\documentclass{article}

\PassOptionsToPackage{numbers, compress}{natbib}
\usepackage[eandd, final]{neurips_2026}
\usepackage{makecell}

\usepackage[utf8]{inputenc} % allow utf-8 input
\usepackage[T1]{fontenc}    % use 8-bit T1 fonts
\usepackage{hyperref}       % hyperlinks
\usepackage{url}            % simple URL typesetting
\usepackage{booktabs}       % professional-quality tables
\usepackage{array}          % extra column types
\usepackage{multirow}       % multi-row table cells
\usepackage{rotating}       % sideways tables for wide reconstruction grids
\usepackage{amsfonts}       % blackboard math symbols
\usepackage{amsmath}        % math environments
\usepackage{amssymb}        % extra math symbols (\uparrow, \downarrow)
\usepackage{nicefrac}       % compact symbols for 1/2, etc.
\usepackage{microtype}      % microtypography
\usepackage{xcolor}         % colors
\usepackage{placeins}       % float barriers
\usepackage{float}          % better float placements

\usepackage{graphicx}       % figures and placeholder boxes
\usepackage{longtable}      % for large metrics table
\usepackage{booktabs}       % for \toprule, \midrule, \bottomrule
\usepackage{placeins}       % prevent longtable mix with figures
\usepackage{wrapfig}
\usepackage{colortbl}   % \cellcolor for the shaded table cells

\definecolor{dmiv}{HTML}{A82F2E}\definecolor{dmiii}{HTML}{D4504C}\definecolor{dmii}{HTML}{EC918B}\definecolor{dmi}{HTML}{F7D0CC}
\definecolor{dzero}{HTML}{ECEBE7}
\definecolor{dpi}{HTML}{CFE1F7}\definecolor{dpii}{HTML}{8FBAEF}\definecolor{dpiii}{HTML}{3D7ED2}\definecolor{dpiv}{HTML}{1C5AA8}
\definecolor{nacell}{HTML}{F4F4F2}                                % not measured
\definecolor{latdisc}{HTML}{F77189}\definecolor{latcont}{HTML}{50B131}  % latent type, as in Fig. 5

\usepackage{caption}
\title{Evaluating Neural Data Tokenizers: A Framework for Assessing Learned Representations of Spiking Activity}
\newcommand{\authsep}{\hspace{0.6em}}% breakable gap between names

\author{%
  \parbox[t]{\dimexpr\textwidth-2\tabcolsep\relax}{%
    \centering
    \normalfont\bfseries\normalsize
    Federico~D'Agostino\textsuperscript{1,2,4,5*}\authsep
    Alex~Gilbert\textsuperscript{3,4,5*}\authsep
    Susanne~Keller\textsuperscript{1,2,*}\authsep
    Jaivardhan~Kapoor\textsuperscript{1,9,\textdagger}\authsep
    Nicolas~Reategui\textsuperscript{1,9,\textdagger}\authsep
    Vedang~Lad\textsuperscript{3,4,5}\authsep
    Matthias~K\"ummerer\textsuperscript{1, 2}\authsep
    Jai~Bhagat\textsuperscript{3,4,5}\authsep
    Emin~Orhan\textsuperscript{8}\authsep
    Hasan~Atakan~Bedel\textsuperscript{3,5}\authsep
    Nikos~Karantzas\textsuperscript{3,5}\authsep
    Fabian~H.~Sinz\textsuperscript{6}\authsep
    Surya~Ganguli\textsuperscript{3}\authsep
    Jakob~H.~Macke\textsuperscript{1,2,9,10}\authsep
    Alexander~S.~Ecker\textsuperscript{6,7}\authsep
    Katrin~Franke\textsuperscript{1,3,5}\authsep
    Sophia~Sanborn\textsuperscript{3,4,5\textdaggerdbl}\authsep
    Matthias~Bethge\textsuperscript{1,2\textdaggerdbl}\authsep
    Andreas~S.~Tolias\textsuperscript{3,4,5\textdaggerdbl}\authsep
    Konstantin~F.~Willeke\textsuperscript{3,4,5\textdaggerdbl}
    \\[1.2ex]
    \normalfont\small
    \textsuperscript{1}University of T\"ubingen \quad
    \textsuperscript{2}T\"ubingen AI Center \quad
    \textsuperscript{3}Stanford University \quad
    \textsuperscript{4}Metamorphic \quad \newline
    \textsuperscript{5}Enigma Project \enspace
    \textsuperscript{6}CIDAS, University of G\"ottingen \enspace
    \textsuperscript{7}MPI for Dynamics and Self-Organization, G\"ottingen \enspace
    \textsuperscript{8}ORNL \quad
    \textsuperscript{9}ELLIS Institute T\"ubingen \quad
    \textsuperscript{10}MPI for Intelligent Systems, T\"ubingen
    \\[0.8ex]
    \normalfont\small
    \textsuperscript{*}Co-first authors \quad
    \textsuperscript{\textdagger}Co-second authors \quad
    \textsuperscript{\textdaggerdbl}Co-senior authors
    \\[0.8ex]
    \normalfont\small
    \texttt{Project page:}\ %
    \href{https://github.com/enigma-brain/ntt}{\texttt{\textcolor[HTML]{2B6FDE}{github.com/enigma-brain/ntt}}}
  }%
}

\begin{document}

\maketitle

\begin{abstract}
Foundation models of brain activity aim to learn representations from large-scale neural recordings that generalize across sessions, subjects, species, and downstream tasks. Transformer-based models have emerged as a promising architecture for this goal and all begin with the same first step -- tokenization: transforming sparse neuronal data into vector representations that downstream networks can consume.
In other areas of machine learning, tokenization is its own design problem, distinct from the downstream foundation model. 
Several mostly implicit tokenization approaches dominate the literature, but their impact on downstream performance has not been systematically evaluated. Here, we propose four explicit desiderata for a neural tokenizer---\textit{reconstruction fidelity}, \textit{downstream utility}, \textit{cross-instance generalization}, and \textit{compression}---and a framework of metrics spanning seven dimensions, including firing statistics, encoding and decoding, and latent structure and robustness. 
We assemble a benchmark of ten public datasets spanning diverse species, brain areas, recording technologies, and behavioral paradigms, and evaluate a representative set of discrete and continuous tokenizers, both single-neuron and population variants. 
Across the strategies evaluated, we show how tokenizers trained only on reconstruction can support downstream tasks. Yet, reconstruction alone is misleading: reconstruction quality and downstream utility can be sharply dissociated. Our multi-axis framework is a response to this failure mode. 
These findings highlight tokenization as a critical and underexplored design axis in brain foundation models, and suggest that further progress will require continued systematic study of tokenization strategies. 
We release the data preparation code, training pipeline, evaluation suite, and trained tokenizers as a community resource.

\end{abstract}

\section{Introduction}

In recent years, advances in neural recording technology have enabled large-scale datasets of brain activity at single-neuron resolution~\cite{iblBrainwideMap2025, siegleSurveySpikingMouse2021, microns2025, perich2018rapid}. These datasets have, in turn, driven growing interest in building brain foundation models that ingest neural activity~\cite{ye2021ndt, ye2023ndt2, azabou2023poyo, zhang2025neds, pandarinath2018lfads, willeke2026omnimouse}. A central promise of such models is that they are trained in a task-agnostic manner, learning representations that generalize across downstream tasks, experimental conditions, and potentially subjects or species. However, in most current approaches, the tokenization of neural activity is not treated as a reusable component, but is instead implicitly defined by task-specific end-to-end training, undermining the goal of learning general-purpose neural representations. This raises a fundamental question: how should we tokenize spiking activity in a way that preserves the information necessary for a broad range of downstream uses? 
% How best to tokenize this modality i.e. mapping sparse spike events into representations suitable for downstream models remains an open question.
%A central challenge is how best to tokenize this modality\textemdash i.e., how to map sparse spike events into representations suitable for downstream models\textemdash remains an open research question.

In other domains such as language and vision, tokenization is a design choice that critically shapes model behavior. Byte-pair encoding variants systematically impact downstream text performance at a fixed scale~\cite{ali2024tokenizer, rust2021tokenizer, bostrom2020bpe}. In vision, patch tokenization and learned codebooks (VQ-VAE, ViT, MAGVIT) play an equivalent role~\cite{vandenoord2017vqvae, dosovitskiy2020vit, yu2023magvit}. By contrast, the impact of tokenization choices on performance in brain foundation models has rarely been examined~\cite[but see e.g.][]{aroraKnowThyselfKnowing2025}.

\begin{figure}[t!]
  \centering
  \includegraphics[width=0.99\linewidth]{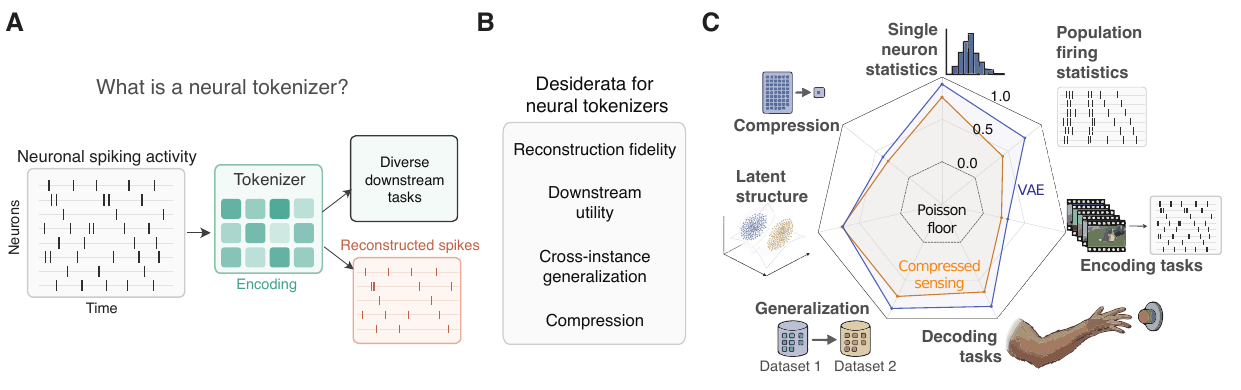}
  \caption{\textbf{The neural spike tokenization problem.} \textbf{A}. A tokenizer maps a population spike train to a latent token sequence and back. Two objectives trade off: the reconstructed spike train should resemble the input, and the latents should expose structure useful to downstream tasks. \textbf{B}. We propose explicit desiderata for neural tokenizers.  \textbf{C}. We introduce a multi-axis evaluation framework providing diagnostic views for different use cases and failure modes in tokenization for neural data, shown here for two example tokenizers (a VAE and compressed sensing architecture) across all axes.}
  \label{fig:problem1}
\end{figure}

The tokenizer dictates the model's data interface; any structure it discards cannot be recovered downstream (Figure~\ref{fig:problem1}). Work in adjacent neural modalities such as EEG \citep{jiang2024labram, barmpas2025neurorvq} and calcium imaging \citep{xuSelfSupervisedFoundationModel2026} has begun to develop standalone tokenizers, but spiking activity differs in ways that prevent direct transfer. Spikes are binary at sub-millisecond resolution, sparse, dominated by Poisson-like single-trial variability, and recorded from neuron sets that change from session to session. Tokenizers for spiking activity must thus be designed and evaluated on their own terms. 

\begin{figure}[b!]
  \centering
  \begin{minipage}[c]{0.70\linewidth}
    \includegraphics[width=\linewidth]{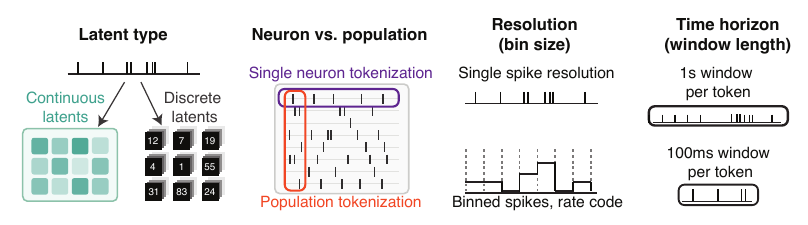}
  \end{minipage}\hfill
  \begin{minipage}[c]{0.28\linewidth}
    \caption{\textbf{Tokenizer design space.} The design space of tokenizers in our evaluation framework covers latent types, population, resolution, and window sizes. }
    \label{fig:tokenizer-design}
  \end{minipage}
\end{figure}

To date, neural tokenizers have not been systematically evaluated as standalone components. A key question is how to evaluate their quality. The naive default is reconstruction fidelity, which measures how well the original spike train can be recovered from the tokens. However, we show that reconstruction alone is insufficient with the tokenizers we tested: tokenizers with near-perfect reconstruction of held-out spike trains can yield latent representations that range from highly informative to nearly useless for downstream decoding and encoding. %Reconstruction alone is therefore an insufficient criterion for evaluating spike tokenizers. 
Here, we introduce an extensive evaluation suite and multi-dataset benchmark for systematic comparison:

%Evaluating spike tokenizers therefore requires multiple axes, including reconstruction, latent structure, and downstream utility, rather than collapsing to a single reconstruction score.

\begin{enumerate}
    \item We define spike tokenization as a modular component with explicit desiderata, decoupling it from the input layer of end-to-end models.
    \item We introduce a multi-axis evaluation framework spanning reconstruction fidelity, single-unit and pairwise temporal statistics, population geometry, and downstream utility. %, with metrics normalized between a rate-matched Poisson floor and a perfect-reconstruction ceiling.
    \item We collate a benchmark of ten public datasets spanning three species, multiple brain areas, three recording technologies, and diverse behavioral paradigms, featuring cross-time, cross-session, cross-subject, and cross-source generalization splits.
    \item We define a zoo of both continuous and discrete tokenizers for single-neuron and population-level tokenization,  exploring a variety of architectures, bottlenecks, and losses, and apply the evaluation framework to this suite (Figure~\ref{fig:tokenizer-design}).
\end{enumerate}

Together, this provides a standardized framework for evaluating spike tokenization, enabling the systematic comparison and development of tokenizers for brain foundation models. Our design takes inspiration from tokenizer evaluation suites in vision, audio, and protein modeling, where tokenizers rank differently across fidelity, downstream utility, robustness, and rate~\citep{yu2023language, yuImageWorth322024, yaoReconstructionVsGeneration2025a, mousaviDASBDiscreteAudio2025, yuanProteinStructureTokenization2025}. Applying the suite to a range of tokenization strategies, we find that, as in those domains, no method is universally optimal: reconstruction quality and downstream utility dissociate, and population-level and single-neuron tokenizers differ in the aspects of neural structure they preserve. These results highlight spike tokenization as an open design problem. Just as tokenization has become a central research question in language and vision, we expect that progress in brain foundation models will hinge on a more systematic study of how neural activity is tokenized, and we hope this evaluation suite will serve the community as a common testbed for that effort.

%We take inspiration from similar tokenizer evaluation suites in vision, audio, and protein modeling, where models rank differently across fidelity, downstream utility, robustness and rate~\citep{yu2023language, yuImageWorth322024, yaoReconstructionVsGeneration2025a, mousaviDASBDiscreteAudio2025, yuanProteinStructureTokenization2025}; Appendix~\ref{app:grounding} maps each desideratum and metric dimension to prior work and to the failure mode it diagnoses.

%%%%%%%%%%%%%%%%%%%%%%

\section{Related work}
\label{sec:related}

\paragraph{Tokenization in models of spiking activity.}
In large-scale models of population spiking activity, fixed binning is the dominant convention for tokenization: spike times are converted into count vectors at a chosen temporal resolution and then passed to a latent-variable model, transformer, or multimodal architecture. This includes NDT \citep{ye2021ndt, ye2023ndt2, zhangUniversalTranslatorNeural2024}, LFADS \citep{pandarinath2018lfads, keshtkaranLargescaleNeuralNetwork2022}, and NEDS \citep{zhang2025neds}-style pipelines, where the bin size, population vectorization, and embedding layers determine the information available to the downstream model. 
% The tokenizer is trained jointly with the rest of the model without measuring its individual contribution separately.
% \paragraph{Event-level and two-stage neural representations.}
Another approach is spike-level tokenization. POYO-like architectures  \citep{azabou2023poyo, azabouMultisessionMultitaskNeural2024, ryooGeneralizableRealtimeNeural2025a} treat individual spikes as events with associated timestamps and neuron identities, preserving temporal information before aggregation by a larger model. This makes the tokenization step explicit, but it is still trained jointly with the downstream objective rather than as a reusable module, so its contribution is not directly assessed. 
Two-stage generative approaches are closer to our setting. LDNS \citep{kapoor2024ldns}, for example, trains an autoencoder before fitting a diffusion model in the inferred latent space, showing that separately learned neural latents can be useful for generation. However, these continuous latents are neither evaluated nor intended as general-purpose tokens. 
% \vspace{-0.1cm}

\paragraph{Discrete neural tokenizers in adjacent modalities.}
The closest precedents for standalone neural tokenization come from non-invasive signals. EEG tokenizers such as LaBraM \citep{jiang2024labram} and NeuroRVQ \citep{barmpas2025neurorvq}, among others \citep{pradeepkumarTokenizingSingleChannelEEG2026, bettinardiBioSerenityE1SelfsupervisedEEG2025}, train discrete codebooks that can be reused by downstream models, and show how tokenizer choice can affect downstream performance. These motivate the modular perspective adopted here, but spiking data differs in ways that make directly transferring insights from other neural data modalities non-trivial: spikes are sparse, binary events observed at millisecond resolution, with variable neuron identities, session structure, firing-rates, and population statistics.

%%%%%%%%%%%%%%%%%%%

\section{Problem formulation}
A neural spike tokenizer maps observed spiking activity into a compressed latent representation that can be reused across modeling  (Figure~\ref{fig:problem1}). Let $x$ denote spike activity over neurons and time. We focus on tokenizers that define an encoder $q_\phi(z \mid x)$ and a decoder $p_\theta(x \mid z)$, with $z$ as the resulting latent representation. Both components are required because foundation models for spiking activity operate in two directions: from spikes to tokens (encoder), supporting downstream decoding into behavioral or external variables, and from tokens back to spikes (decoder network), supporting generation and reconstruction. 
% The use in foundation models implies competing demands on the tokens $z$: They should compress neural data to lighten compute burden and speed up training, while allowing efficient extraction of task-relevant signals~\citep{whitneyEvaluatingRepresentationsComplexity2020,allemanTaskStructureNonlinearity2023}, as well as accurate spike reconstruction.
% Finally, for the foundation model to be universally applicable, the tokenizer must generalize across recordings, subjects, brain regions and even species. These considerations lead to four desiderata for neural tokenizers:
The latent $z$ is used directly by these downstream models. The geometry of $z$ determines whether the tokenizer is useful in practice: it sets how easily downstream models extract task-relevant signal~\citep{whitneyEvaluatingRepresentationsComplexity2020,allemanTaskStructureNonlinearity2023}, whether the same encoder transfers to new sessions and subjects, and whether the representation compresses input efficiently. These considerations motivate four key desiderata:

\label{sec:desiderata}

\textbf{1. Reconstruction fidelity.} The encoder and decoder network preserve the structure of the input without distorting its properties, so that closed-loop computation is possible, and to enable use of the model as a digital twin for neuron populations. Properties that should be preserved include individual spike events, single-unit temporal statistics, and population structure.
%at multiple levels: individual spike events, single-unit temporal statistics, and pairwise and population structure. Reconstruction is the most direct test of whether information has been retained, but as we show in Section~\ref{sec:results}, it is not indicative of downstream performance. %sufficient on its own.

\textbf{2. Downstream utility.} The tokens support downstream tasks either with simple downstream models or by allowing faster training compared to raw data. This has two facets: \emph{cross-task transferability}, where a single frozen tokenizer supports multiple downstream uses (neural decoding, neural encoding, latent analysis) without per-task fine-tuning, and \emph{sample-efficient downstream learning}, where downstream tasks reach competitive performance with few training trials. Two representations can carry identical task-relevant information but differ substantially in how readily a downstream predictor extracts it.

\textbf{3. Cross-instance generalization.} The same tokenizer functions without retraining across sessions, subjects, recording sources, and ideally even brain areas or species. This is the property that makes a tokenizer reusable as a foundation model component, rather than a per-session preprocessing step.

\textbf{4. Compression.} The tokenizer represents the inputs efficiently, as measured, for example, by the number of \textit{neurons} $\times$ \textit{time bins} per token or by the ratio of input bits to output bits (\textit{compression rate}). Other things being equal, a tokenizer that uses one bit per spike on average would be preferable to a tokenizer requiring ten bits per spike to achieve the same reconstruction fidelity. Compression is motivated by at least two practical goals: more efficient model training (for example, requiring less memory or fewer FLOPs) and better generalization. 
%Here, the number of token dimensions is a better proxy than the bits required for each number.

These desiderata are not independent in practice. Compression interacts with reconstruction, since higher compression discards more structure. Cross-instance generalization interacts with downstream utility, since a tokenizer that overfits to a single session is unlikely to support efficient learning on a new one.

\section{Evaluation framework}
\label{sec:evaluation-framework}
% \vspace{-0.1cm}
We provide an evaluation framework for the systematic development of tokenizers on spiking neural data, built on our curated dataset collection (Section~\ref{sec:datasets}). Any candidate tokenizer can be evaluated on the same metrics. The set of evaluation metrics in our current implementation is broad, but can be extended with new metrics as the field develops. On the other hand, the four desiderata above (Section~\ref{sec:desiderata}) are intended as a more stable conceptual framework.

The mapping from desiderata to metrics is not one-to-one: some metrics serve multiple desiderata, and desideratum 3 (cross-instance generalization) applies to every metric. We divide the metrics into multiple groups below and indicate which desideratum each group serves. 
Our metric groups are intended as non-orthogonal diagnostic views covering established properties, use cases and failure modes, rather than independent factors or an exhaustive basis. 
Appendix~\ref{app:metric-corr-analysis} contains metric correlation and dimensionality analyses; Fig.~\ref{fig:metric-category-correlation} shows how the main evaluation dimensions correlate. %With only 12 tokenizer families, these correlations are preliminary. %Comment out to save space
Appendix~\ref{app:metric-details} includes details of metric implementations.

\textbf{Single-neuron reconstruction (SN recon) statistics.} These metrics compare the decoded spike train $\hat{x}$ to the input $x$ at the level of individual units, serving desideratum 1 (reconstruction fidelity) and establishing the per-unit floor: reconstruction is a tokenizer's essential function, and information discarded here is lost for all downstream tasks that follow. Metrics include spike-detection F1 score, temporal precision (single-neuron tokenizers only), mutual information, spike-count correlation, smoothed-rate error, binned rate difference, ISI divergence, ISI CV and CV2 errors, autocorrelogram correlation, firing-rate error, and Fano factor error.

\textbf{Population firing statistics.} These metrics compare $\hat{x}$ to $x$ at the level of joint activity across neurons, also serving desideratum 1: A tokenizer could sufficiently preserve every neuron's marginal statistics while degrading the cross-neuron timing that defines a population code.  Metrics include cross-correlogram correlation, functional connectivity F1, participation ratio error, subspace overlap, and population coupling preservation.%, and binned rate difference (population tokenizers).

\textbf{Downstream decoding.} These metrics test the tokenizers on a canonical downstream task in spike modeling: reading behavior off neural activity. We train both a simple readout (a ridge-regularized linear probe) and a sequence-to-sequence non-linear temporal decoder \citep{azabou2023poyo} on top of the frozen tokenizer, serving desideratum 2 (downstream utility) across both facets. 
We use two protocols: one trains on the full public training set and tests cross-task transferability; another varies the training-set size and tests how sample-efficient a downstream model learns, based on the frozen tokenizer.
%An asymptotic protocol trains on the full training set and tests cross-task transferability; a learning-curve protocol varies training-set size and tests sample-efficient downstream learning. 
Targets include cursor position and velocity ($R^2$) and trial-target and movement-phase classification (accuracy).

\textbf{Downstream encoding.} Neural encoding is the symmetric counterpart to decoding and the second direction any unified neural foundation model must support, where the token sequence is the prediction target rather than the input.
On stimulus-locked data (V4 responses to natural images \citep{cowleyCompactDeepNeural2026}), we train a shared image backbone with session-specific readouts to predict the frozen tokenizer's latent distribution from each image. The decoded conditional-mean response is then scored against held-out PSTHs (see Appendix~\ref{app:downstream-encoding} for details), serving desideratum 2 along both facets.
%We train simple neural encoding models to predict neural activity from images (using a dataset from \citet{cowleyCompactDeepNeural2026})
%These metrics are applied to simple models trained to predict tokens from external variables on held-out frames, also serving desideratum 2 along both facets. 

\textbf{Latent structure and latent robustness.} Latent structure metrics ask whether $z$ itself carries the structure of $x$, independently of whether the tokenizer's decoder network can reconstruct it. Latent geometry is what downstream models actually consume, so probing it directly diagnoses whether the encoder has done useful work independently of the decoder. They are most diagnostic of downstream utility (desideratum 2), with the corruption-robustness sub-metrics also bearing on the stability underlying cross-instance generalization. We measure RSA between input and latent dissimilarity matrices, repeat-trial SNR in the latent space on stimulus-locked datasets, and stability of the latents under input-side corruption, and of the reconstruction under latent-side corruption.

\textbf{Generalization.} Desideratum 3 (cross-instance generalization) applies to every metric above. All metrics can be reported under four held-out splits, defined in Section~\ref{sec:datasets}. However, we include only reconstruction metrics in the aggregate generalization score, since encoding and decoding retrain a downstream network on part of the held-out data.
%: cross-time (held-out time within a session), cross-session (held-out session, same subject), cross-subject (held-out subject, same source), and cross-source (entirely held-out source). 
For each reconstruction metric, across the three held-out datasets, we report a direction-aware retention ratio, comparing to the metrics' mean value across held-out parts of training sessions (clipped to [0, 1]), averaged over metrics into one score (Appendix~\ref{app:generalization-retention}). A value of 1 indicates full score retention on unseen datasets.
%For each reconstruction metric, on each held-out session, we report a direction-aware retention ratio against the cross-time sessions of the same source (clipped to [0, 1]), aggregated as a geometric mean over metrics on the cross-source split (Appendix~\ref{app:generalization-retention}).

\textbf{Compression.} These metrics quantify how efficiently the tokenizer represents the inputs, serving desideratum 4. We report the \textit{compression rate} of tokenizers, as measured by the ratio of input bits to output bits (\textit{i.e.}~number of bits required to represent the input \textit{vs.}~the output). 
% \drafttodo{I think it would be really nice to include the $N\times T \rightarrow 1$ token ``context compression'' rate here as well (we could call the other metric \textit{memory compression rate}), since (1) compression rates in Table 2 are not super convincing and (2) this is computationally highly relevant for transformers as it measures how much tokenizers reduce the quadratic complexity cost of self-attention. I think this should not be hard to add to the tables.}

Reconstruction metrics (single-neuron and population) are normalized between two anchors: a rate-matched Poisson floor and, when applicable, a perfect-reconstruction ceiling. The Poisson floor corresponds to a tokenizer that preserves per-neuron firing rates and discards all other structure (fine timing, pairwise structure, population geometry). Firing-rate error is the one exception: this floor matches it by construction, so it is scored against a reference that gives every neuron the population's mean rate. The remaining categories use their native scales: $R^2$ for decoding and encoding, scores mapped to $[0,1]$ for latent structure and robustness, a retention ratio in $[0,1]$ for generalization, and a bit ratio (input bits/output bits) %natural units
for compression.

Each axis is a family-weighted mean of its constituent metrics (the bin-size variants of one metric together carry weight one) in that axis’s natural units (Poisson-normalized for reconstruction, $R^2$ for encoding/decoding, $[0,1]$-mapped scores such as the soft retention $1/(1+d)$ of the normalized input-corruption distance $d$ for latent robustness, geometric-mean retention for generalization); per-metric Poisson-normalized scores are clipped at $-1$ before averaging to bound outliers. 
Equal weighting of metric families within an axis is our assumption-light default: which metrics matter most will depend on what a tokenizer is for, and we retain all per-metric values so that practitioners can apply their own weighting. 
For the same reason, we deliberately do not collapse to a single composite score. The framework was explicitly designed to expose cross-axis trade-offs (e.g., compression at the cost of population structure).

\section{Dataset suite}
% \vspace{-0.1cm}
\label{sec:datasets}
We compiled a diverse set of public datasets to cover a wide range of settings in which spike tokenizers are typically applied. Each dataset consists of multiple distinct extracellular electrophysiology recordings spanning species and brain regions. Datasets vary along five axes (Figure~\ref{fig:dataset-suite}A): (1) recording technology, (2) species, (3) brain region, (4) behavioral paradigm, (5) population size. 

Datasets vary by 1--3 orders of magnitude on properties that directly affect tokenizer training: median session duration ranges from \textasciitilde10\, to \textasciitilde170\,min, median population size from 5 to \textasciitilde1000\,units, and median per-neuron firing rate from \textasciitilde4\ to \textasciitilde38\,Hz (Figure~\ref{fig:session_stats_boxstrip}). Spiking sparsity has similar variability: at 1\,ms binning, population sparsity ranges from $<1\%$ in dense cortical recordings to $>85\%$ in cerebellar populations, and it decays in a strongly dataset-dependent way as the bin width is coarsened to 10--1000\,ms (Fig.~\ref{fig:sparsity_ladder}). Overall, the corpus contains nearly 20 billion sorted spikes: over 14 billion spikes from public datasets (Table~\ref{tab:datasets}) and an additional 5.5 billion from proprietary unreleased datasets. These recordings enter only the training of the Conv-VAE in Figure~\ref{fig:problem1}C and the largest training-data condition in Figure~\ref{fig:recon-quality}; every tokenizer in Table~\ref{tab:main_results} and in the appendix result tables is trained on public data only.

To probe various levels of generalization, we partition the held-out data into four tiers (Figure~\ref{fig:dataset-suite}B): (1) $xT$: held-out time within a session, (2) $xS$: held-out sessions, same subject, (3) $xSub$: held-out subjects, same dataset, and (4) $xSrc$: held-out datasets, which can also withhold whole species or brain regions. The four lower rows of Table~\ref{tab:datasets} report, per dataset, the fraction of total recording time withheld from training for each tier. To support the community, we release our data preparation code (\href{https://github.com/enigma-brain/ntt}{here})  that builds on existing \texttt{brainsets} pipelines~\citep{azabou2023poyo} to convert each public dataset into a standardized tokenizer-ready format.  
% \vspace{-0.1cm}

\begin{table}[t]
  \centering
  \setlength{\tabcolsep}{3pt}
  \renewcommand{\arraystretch}{1.05}
  \hspace*{-0.1in}
  \resizebox{1.03\linewidth}{!}{%
  \begin{tabular}{@{}l cccccccccc@{}}
    \toprule
      & \makecell{IBL Brain-\\wide Map~\cite{iblBrainwideMap2025}}
      & \makecell{Allen Vis.\\Behavior~\cite{allenVisualBehavior2022}}
      & \makecell{Allen Vis.\\Coding~\cite{siegleSurveySpikingMouse2021}}
      & \makecell{C4 Database~\cite{beauC4Cerebellum2025}}
      & \makecell{Churchland\\Shenoy\\2012~\cite{churchlandNeuralPopulation2012}}
      & \makecell{Flint\\Slutzky\\2012~\cite{flintAccurateDecoder2012}}
      & \makecell{O'Doherty\\Sabes\\2017~\cite{odohertyNonhumanPrimate2017}}
      & \makecell{Perich\\Miller\\2018~\cite{perich2018rapid}}
      & \makecell{Cowley\\Smith\\2026~\cite{cowleyCompactDeepNeural2026}}
      & \makecell{Wingert\\David\\2026~\cite{wingertFerretAuditory2026}}
    \\
    \midrule
    Recording   & Neuropixels & Neuropixels & Neuropixels & Neuropixels & Utah arrays & Utah arrays & Utah arrays & Utah arrays & Utah arrays & Linear arrays \\
    Species     & mouse & mouse & mouse & macaque \& mouse & macaque & macaque & macaque & macaque & macaque & ferret \\
    Brain region  & brainwide & \makecell{visual ctx,\\thal, HPC, MB} & \makecell{visual ctx,\\thal, HPC, MB} & cerebellum & M1, PMd & M1, PMd & M1, S1 & M1, PMd & V4 & A1, PEG \\
    Sessions    & 331 & 103 & 58 & 321 & 10 & 5 & 47 & 111 & 50 & 68 \\
    Units       & 101{,}557 & 82{,}133 & 40{,}010 & 4{,}960 & 1{,}911 & 957 & 16{,}566 & 10{,}410 & 4{,}774 & 3{,}382 \\
    Spikes      & 2.32e9 & 7.01e9 & 3.08e9 & 5.07e8 & 7.39e8 & 7.88e6 & 1.05e8 & 1.11e8 & 2.10e8 & 1.03e8 \\
    \midrule
    $xT$        & 0.10 & 0.10 & 0.10 & - & 0.10 & 0.10 & 0.10 & 0.10 & - & - \\
    $xS$        & 0.04 & 0.10 & 0.07 & - & 0.45 & 0.43 & 0.12 & 0.07 & - & - \\
    $xSub$      & 0.10 & 0.00 & 0.07 & - & 0.00 & 0.00 & 0.41 & 0.10 & - & - \\
    $xSrc$      & 0 & 0 & 0 & 1 & 0 & 0 & 0 & 0 & 1 & 1 \\
    \bottomrule
  \end{tabular}%
  }
  \vspace{0.1cm}
  \caption{\textbf{Public datasets in the pretraining corpus}. Rows are dataset properties, columns are datasets. Spikes are sourced exclusively from single-unit, spike-sorted data, and counts are aggregated across train/val/test splits. The four lower rows give the fraction of total recording \emph{time} withheld from training under each of the generalization regimes defined in Section~\ref{sec:datasets}.
%  : $xT$: held-out time within a session; $xS$: held-out sessions, same subject; $xSub$: held-out subjects, same dataset; and $xSrc$: held-out datasets, which can also withhold whole species or brain regions.
}
    \label{tab:datasets}
  %Brain-region acronyms from: Allen Mouse Brain CCF~\cite{wangAllenCCF2020} (mouse), Felleman-VanEssen parcellation~\cite{fellemanVanEssen1991} (macaque) Bizley-King ferret auditory parcellation~\cite{bizleyFerretAuditory2005} (ferret).
%\vspace{-0.3cm}
\end{table}

\begin{figure}[b!]
  \centering
  \includegraphics[width=0.95\linewidth]{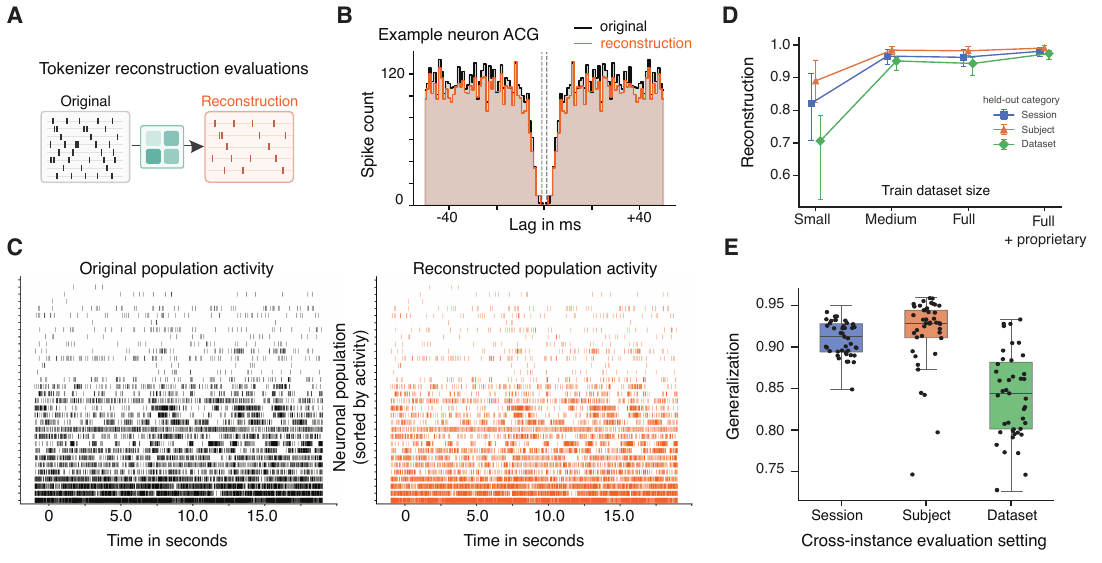}
  \caption{\textbf{Reconstruction results.} \textbf{A}. Schematic of the spike-tokenization reconstruction task. \textbf{B}. Autocorrelogram (ACG) of an example neuron, original (black) and reconstruction (orange), for the Conv-VAE ($100:1$). \textbf{C}. Held-out population spike rasters for the same tokenizer; ground truth (left) and reconstruction (right) align in spike timing over multi-second windows. \textbf{D}. Reconstruction score of a Conv-VAE ($100:1$) trained on increasing amounts of data (Small, Medium: 30\%, 60\% of training sessions), per held-out category; 95\% bootstrap CIs over test sessions. \textbf{E}. Reconstruction retention across the zoo (one dot per tokenizer), relative to held-out-time ($xT$) sessions. Reconstruction generalizes well across sessions and subjects, but degrades across held-out datasets.}
  \label{fig:recon-quality}
\end{figure}

\section{Tokenizer implementations}
\label{sec:tokenizers}

To evaluate tokenizers systematically, we developed a modular framework in which a tokenizer is any model writable as an encoder--decoder pair, the minimal interfaces a foundation model needs to both read spikes in and write spikes out. 
While the interface is minimal, the possible design space behind it is large: our framework, \href{https://github.com/enigma-brain/ntt}{\texttt{ntt}}, exposes each design choice (input type and binning, window length, architecture, bottleneck and loss, among others) as an explicit setting that a tokenizer must declare, which will determine which metrics apply and how they are computed. Our representative zoo spans several of these axes (Figure~\ref{fig:tokenizer-design}).
%: single-neuron and population inputs, continuous and discrete bottlenecks, model architectures, and compression levels.
% For learned, autoencoder-style models, we developed composable bottlenecks, losses, and input transforms. Using this framework, we assembled a zoo spanning several design axes (Figure~\ref{fig:tokenizer-design}): single-neuron/population inputs, continuous/discrete bottlenecks, model architectures, and varying levels of compression (as a function of binning resolution, input window length, and downsampling rates).

We define a \textit{single-neuron} tokenizer as taking as input a 1-D temporal sequence from a single unit; for our exploration in this regime we restrict to binary spike bins at 1\,ms and two temporal downsampling ratios (input bins per token)---$100:1$ and $1000:1$---producing one token per 100\,ms and per 1\,s of activity, respectively. These ratios count input bins per token and are distinct from the bit-level Compression column of Table~\ref{tab:main_results}. A \textit{population} tokenizer instead encodes the joint activity of all simultaneously recorded neurons, taking as input an $N \times T$ matrix of binned firing rates rather than a single-unit sequence. 

As \textit{architectures}, we implement convolutional and transformer-based models. We also provide continuous VAE-style bottlenecks as well as discrete quantization methods---VQ \cite{vandenoord2017vqvae}, FSQ \cite{mentzer2023finite}, BSQ \cite{zhao2024image}, LFQ \cite{yu2023language}, and RVQ \cite{zeghidour2021soundstream}---along with various regularization strategies, for example, a latent straightening penalty~\citep{wang2026temporal} or SIGReg~\citep{balestriero2025lejepa}. 
For the \textit{reconstruction objective}, we explored binary cross-entropy and focal loss~\citep{lin2017focal} for binary spike outputs, Poisson NLL for rate-valued decoders, and MSE, L1, Huber, and Wasserstein losses as alternatives, as well as optional target softening transforms. Finally, we also provide a variety of hand-designed \textit{input-space transforms} based on fixed discrete wavelet families (e.g.\ Haar, \texttt{db2}, \texttt{db4}). Our framework also supports non-learned algorithms, and along these lines, we included a PCA baseline and a compressed-sensing tokenizer. Full architectural and training details of our models are in Appendix~\ref{app:tokenizer-details}.

\section{Results}

We evaluate the framework on our entire tokenizer zoo. Table~\ref{tab:main_results} reports tokenizer scores across the seven evaluation dimensions defined in Section~\ref{sec:evaluation-framework}, with rows grouped by population vs.\ single-neuron tokenizers. %The scores reveal that w
While there are substantial differences across tokenizers, no single tokenizer dominates all dimensions: single-neuron reconstruction, population statistics, encoding, and decoding each order the zoo differently, and no tokenizer holds the same rank across all four.
Therefore, single-axis evaluation would obscure the design trade-offs our framework is built to expose. 

Three claims structure the rest of this section. First, reconstruction quality is high for many tokenizers and scales with training data. Second, reconstruction-trained tokenizers % serve as 
are viable backbones for downstream encoding and decoding. Third, reconstruction alone is not sufficient: some tokenizers achieve high reconstruction with latents that carry little structure a downstream readout can use, making it the most tractable axis to optimize and the most misleading one to optimize alone.

% Generated by ntt/experiments/2026_10_13_camera_ready_figures/main_table_heatmap (r3); do not edit by hand.
% Needs the packages and colours in preamble.tex.
\begin{table*}[t]
\centering
\scriptsize
\setlength{\tabcolsep}{0pt}\renewcommand{\arraystretch}{1.25}
\newcolumntype{G}{>{\centering\arraybackslash}p{4.2mm}}
\newcolumntype{T}{>{\centering\arraybackslash}p{6.2mm}}
\newcolumntype{C}{>{\centering\arraybackslash}p{2.5mm}}
\newcolumntype{R}{>{\centering\arraybackslash}p{6.3mm}}
\newcolumntype{N}{>{\centering\arraybackslash}p{6.5mm}}
\begin{tabular}{@{}lGT@{\hspace{3pt}}CCCCCCCCCCCCR@{\hspace{3pt}}CCCCCR@{\hspace{3pt}}CCCCR@{\hspace{3pt}}CCR@{\hspace{3pt}}NNN@{}}
\toprule
 & \multicolumn{2}{c}{\textit{Design}} & \multicolumn{13}{c}{\textbf{SN Recon}} & \multicolumn{6}{c}{\textbf{Pop.\ Stats}} & \multicolumn{5}{c}{\textbf{Latent Rob.}} & \multicolumn{3}{c}{\shortstack{\textbf{Behav.}\\\textbf{decoding}}} & \multicolumn{3}{c}{} \\
\cmidrule(lr{2pt}){2-3} \cmidrule(lr{2pt}){4-16} \cmidrule(lr{2pt}){17-22} \cmidrule(lr{2pt}){23-27} \cmidrule(lr{2pt}){28-30}
Tokenizer & \rotatebox{90}{Latent} & \rotatebox{90}{Token} & \rotatebox{90}{Spike F1 {\tiny$\times$2}} & \rotatebox{90}{Temporal prec.} & \rotatebox{90}{Firing-rate err.} & \rotatebox{90}{Fano err. {\tiny$\times$5}} & \rotatebox{90}{ISI CV} & \rotatebox{90}{ISI CV$_2$} & \rotatebox{90}{ISI diverg.} & \rotatebox{90}{Count corr. {\tiny$\times$5}} & \rotatebox{90}{Mutual inf. {\tiny$\times$5}} & \rotatebox{90}{Smoothed SE {\tiny$\times$5}} & \rotatebox{90}{Binned rate {\tiny$\times$5}} & \rotatebox{90}{ACG corr. {\tiny$\times$3}} & \rotatebox{90}{\textbf{composite}} & \rotatebox{90}{CCG corr. {\tiny$\times$3}} & \rotatebox{90}{Funct.\ conn.} & \rotatebox{90}{Particip.\ ratio {\tiny$\times$3}} & \rotatebox{90}{Pop.\ coupling {\tiny$\times$2}} & \rotatebox{90}{Subspace {\tiny$\times$3}} & \rotatebox{90}{\textbf{composite}} & \rotatebox{90}{Pop.\ RSA} & \rotatebox{90}{SN RSA} & \rotatebox{90}{Input corr.} & \rotatebox{90}{Latent corr.} & \rotatebox{90}{\textbf{composite}} & \rotatebox{90}{POYO} & \rotatebox{90}{Linear} & \rotatebox{90}{\textbf{composite}} & \rotatebox{90}{\textbf{Stim.\ encoding}} & \rotatebox{90}{\textbf{Generalization}} & \rotatebox{90}{\textbf{Compression}} \\
\midrule
\multicolumn{33}{@{}l}{\textit{Single-neuron tokenizers}} \\[1pt]
C.S. & \textcolor{latcont}{\scalebox{1.5}{$\bullet$}} & {\fontsize{6pt}{7pt}\selectfont 0.1\,s} & \cellcolor{dpiv}\textcolor{white}{} & \cellcolor{dpiv}\textcolor{white}{} & \cellcolor{dpiv}\textcolor{white}{} & \cellcolor{dpiv}\textcolor{white}{} & \cellcolor{dpiv}\textcolor{white}{} & \cellcolor{dpiv}\textcolor{white}{} & \cellcolor{dpiv}\textcolor{white}{} & \cellcolor{dpiv}\textcolor{white}{} & \cellcolor{dpiv}\textcolor{white}{} & \cellcolor{dpiv}\textcolor{white}{} & \cellcolor{dpiv}\textcolor{white}{} & \cellcolor{dpiv}\textcolor{white}{} & \cellcolor{dpiv}\textcolor{white}{{\fontsize{6pt}{7pt}\selectfont 0.93}} & \cellcolor{dpiv}\textcolor{white}{} & \cellcolor{dpiii}\textcolor{white}{} & \cellcolor{dmi} & \cellcolor{dpiv}\textcolor{white}{} & \cellcolor{dpii} & \cellcolor{dpiii}\textcolor{white}{{\fontsize{6pt}{7pt}\selectfont 0.53}} & \cellcolor{dzero} & \cellcolor{dzero} & \cellcolor{dpiii}\textcolor{white}{} & \cellcolor{dpiv}\textcolor{white}{} & \cellcolor{dpii}{\fontsize{6pt}{7pt}\selectfont 0.40} & \cellcolor{dpiv}\textcolor{white}{} & \cellcolor{dzero} & \cellcolor{dpii}{\fontsize{6pt}{7pt}\selectfont 0.43} & \cellcolor{dpii}{\fontsize{6pt}{7pt}\selectfont \textbf{0.35}} & \cellcolor{dpiv}\textcolor{white}{{\fontsize{6pt}{7pt}\selectfont 0.80}} & \cellcolor{dmiii}\textcolor{white}{{\fontsize{6pt}{7pt}\selectfont 0.31$\times$}} \\
Conv-VAE & \textcolor{latcont}{\scalebox{1.5}{$\bullet$}} & {\fontsize{6pt}{7pt}\selectfont 0.1\,s} & \cellcolor{dpiv}\textcolor{white}{} & \cellcolor{dpiv}\textcolor{white}{} & \cellcolor{dpiv}\textcolor{white}{} & \cellcolor{dpiv}\textcolor{white}{} & \cellcolor{dpiv}\textcolor{white}{} & \cellcolor{dpiv}\textcolor{white}{} & \cellcolor{dpiv}\textcolor{white}{} & \cellcolor{dpiv}\textcolor{white}{} & \cellcolor{dpiv}\textcolor{white}{} & \cellcolor{dpiv}\textcolor{white}{} & \cellcolor{dpiv}\textcolor{white}{} & \cellcolor{dpiv}\textcolor{white}{} & \cellcolor{dpiv}\textcolor{white}{{\fontsize{6pt}{7pt}\selectfont \textbf{0.98}}} & \cellcolor{dpiv}\textcolor{white}{} & \cellcolor{dpiv}\textcolor{white}{} & \cellcolor{dpii} & \cellcolor{dpiv}\textcolor{white}{} & \cellcolor{dpiii}\textcolor{white}{} & \cellcolor{dpiii}\textcolor{white}{{\fontsize{6pt}{7pt}\selectfont \textbf{0.72}}} & \cellcolor{dpi} & \cellcolor{dzero} & \cellcolor{dpiii}\textcolor{white}{} & \cellcolor{dpiv}\textcolor{white}{} & \cellcolor{dpii}{\fontsize{6pt}{7pt}\selectfont 0.45} & \cellcolor{dpiv}\textcolor{white}{} & \cellcolor{dpi} & \cellcolor{dpiii}\textcolor{white}{{\fontsize{6pt}{7pt}\selectfont \textbf{0.53}}} & \cellcolor{dpii}{\fontsize{6pt}{7pt}\selectfont 0.33} & \cellcolor{dpiv}\textcolor{white}{{\fontsize{6pt}{7pt}\selectfont 0.88}} & \cellcolor{dmii}{\fontsize{6pt}{7pt}\selectfont 0.39$\times$} \\
Conv-RFSQ & \textcolor{latdisc}{$\blacksquare$} & {\fontsize{6pt}{7pt}\selectfont 0.1\,s} & \cellcolor{dpiv}\textcolor{white}{} & \cellcolor{dpiv}\textcolor{white}{} & \cellcolor{dpiii}\textcolor{white}{} & \cellcolor{dpiv}\textcolor{white}{} & \cellcolor{dpiii}\textcolor{white}{} & \cellcolor{dpiv}\textcolor{white}{} & \cellcolor{dpiv}\textcolor{white}{} & \cellcolor{dpiv}\textcolor{white}{} & \cellcolor{dpiii}\textcolor{white}{} & \cellcolor{dpiii}\textcolor{white}{} & \cellcolor{dpiii}\textcolor{white}{} & \cellcolor{dpiv}\textcolor{white}{} & \cellcolor{dpiv}\textcolor{white}{{\fontsize{6pt}{7pt}\selectfont 0.81}} & \cellcolor{dpiv}\textcolor{white}{} & \cellcolor{dpii} & \cellcolor{dmiv}\textcolor{white}{} & \cellcolor{dpii} & \cellcolor{dmi} & \cellcolor{dpi}{\fontsize{6pt}{7pt}\selectfont 0.14} & \cellcolor{dpii} & \cellcolor{dzero} & \cellcolor{dpiv}\textcolor{white}{} & \cellcolor{dpii} & \cellcolor{dpii}{\fontsize{6pt}{7pt}\selectfont 0.37} & \cellcolor{dpiv}\textcolor{white}{} & \cellcolor{dpi} & \cellcolor{dpiii}\textcolor{white}{{\fontsize{6pt}{7pt}\selectfont 0.52}} & \cellcolor{dpii}{\fontsize{6pt}{7pt}\selectfont 0.32} & \cellcolor{dpiv}\textcolor{white}{{\fontsize{6pt}{7pt}\selectfont 0.81}} & \cellcolor{dpii}{\fontsize{6pt}{7pt}\selectfont 2.5$\times$} \\
Transformer-VQ & \textcolor{latdisc}{$\blacksquare$} & {\fontsize{6pt}{7pt}\selectfont 0.1\,s} & \cellcolor{dpiii}\textcolor{white}{} & \cellcolor{dpiv}\textcolor{white}{} & \cellcolor{dpiv}\textcolor{white}{} & \cellcolor{dpiv}\textcolor{white}{} & \cellcolor{dpiv}\textcolor{white}{} & \cellcolor{dpiv}\textcolor{white}{} & \cellcolor{dpiv}\textcolor{white}{} & \cellcolor{dpii} & \cellcolor{dpii} & \cellcolor{dpii} & \cellcolor{dpi} & \cellcolor{dpiii}\textcolor{white}{} & \cellcolor{dpiii}\textcolor{white}{{\fontsize{6pt}{7pt}\selectfont 0.61}} & \cellcolor{dpii} & \cellcolor{dpi} & \cellcolor{dmi} & \cellcolor{dpii} & \cellcolor{dmi} & \cellcolor{dpi}{\fontsize{6pt}{7pt}\selectfont 0.14} & \cellcolor{dpii} & \cellcolor{dpi} & \cellcolor{dpiii}\textcolor{white}{} & \cellcolor{dpiv}\textcolor{white}{} & \cellcolor{dpiii}\textcolor{white}{{\fontsize{6pt}{7pt}\selectfont 0.53}} & \cellcolor{dpiv}\textcolor{white}{} & \cellcolor{dpi} & \cellcolor{dpiii}\textcolor{white}{{\fontsize{6pt}{7pt}\selectfont \textbf{0.53}}} & \cellcolor{dpi}{\fontsize{6pt}{7pt}\selectfont 0.21} & \cellcolor{dpiv}\textcolor{white}{{\fontsize{6pt}{7pt}\selectfont 0.83}} & \cellcolor{dpiv}\textcolor{white}{{\fontsize{6pt}{7pt}\selectfont 9.1$\times$}} \\
C.S. & \textcolor{latcont}{\scalebox{1.5}{$\bullet$}} & {\fontsize{6pt}{7pt}\selectfont 1\,s} & \cellcolor{dpiv}\textcolor{white}{} & \cellcolor{dpiv}\textcolor{white}{} & \cellcolor{dpiii}\textcolor{white}{} & \cellcolor{dpiv}\textcolor{white}{} & \cellcolor{dpiv}\textcolor{white}{} & \cellcolor{dpiv}\textcolor{white}{} & \cellcolor{dpiv}\textcolor{white}{} & \cellcolor{dpiii}\textcolor{white}{} & \cellcolor{dpiii}\textcolor{white}{} & \cellcolor{dpiii}\textcolor{white}{} & \cellcolor{dpii} & \cellcolor{dpiv}\textcolor{white}{} & \cellcolor{dpiv}\textcolor{white}{{\fontsize{6pt}{7pt}\selectfont 0.80}} & \cellcolor{dpiv}\textcolor{white}{} & \cellcolor{dpii} & \cellcolor{dmiv}\textcolor{white}{} & \cellcolor{dpii} & \cellcolor{dzero} & \cellcolor{dpi}{\fontsize{6pt}{7pt}\selectfont 0.12} & \cellcolor{dzero} & \cellcolor{dzero} & \cellcolor{dpiii}\textcolor{white}{} & \cellcolor{dpiv}\textcolor{white}{} & \cellcolor{dpii}{\fontsize{6pt}{7pt}\selectfont 0.40} & \cellcolor{dpiii}\textcolor{white}{} & \cellcolor{dzero} & \cellcolor{dpii}{\fontsize{6pt}{7pt}\selectfont 0.33$^\dagger$} & \cellcolor{nacell}\textcolor{black!45}{{\fontsize{6pt}{7pt}\selectfont --}} & \cellcolor{dpiv}\textcolor{white}{{\fontsize{6pt}{7pt}\selectfont 0.81}} & \cellcolor{dmi}{\fontsize{6pt}{7pt}\selectfont 0.62$\times$} \\
PCA & \textcolor{latcont}{\scalebox{1.5}{$\bullet$}} & {\fontsize{6pt}{7pt}\selectfont 1\,s} & \cellcolor{nacell}\textcolor{black!45}{\textperiodcentered} & \cellcolor{nacell}\textcolor{black!45}{\textperiodcentered} & \cellcolor{dpiv}\textcolor{white}{} & \cellcolor{dmii} & \cellcolor{nacell}\textcolor{black!45}{\textperiodcentered} & \cellcolor{nacell}\textcolor{black!45}{\textperiodcentered} & \cellcolor{nacell}\textcolor{black!45}{\textperiodcentered} & \cellcolor{dmiii}\textcolor{white}{} & \cellcolor{dmiii}\textcolor{white}{} & \cellcolor{dzero} & \cellcolor{dmii} & \cellcolor{dpii} & \cellcolor{dmi}{\fontsize{6pt}{7pt}\selectfont $-$0.12} & \cellcolor{dmii} & \cellcolor{dmiv}\textcolor{white}{} & \cellcolor{dmiv}\textcolor{white}{} & \cellcolor{dmiv}\textcolor{white}{} & \cellcolor{dmiv}\textcolor{white}{} & \cellcolor{dmiv}\textcolor{white}{{\fontsize{6pt}{7pt}\selectfont $-$0.88}} & \cellcolor{dpiii}\textcolor{white}{} & \cellcolor{dzero} & \cellcolor{dpiv}\textcolor{white}{} & \cellcolor{dpiv}\textcolor{white}{} & \cellcolor{dpiii}\textcolor{white}{{\fontsize{6pt}{7pt}\selectfont \textbf{0.63}}} & \cellcolor{dpiii}\textcolor{white}{} & \cellcolor{dpi} & \cellcolor{dpii}{\fontsize{6pt}{7pt}\selectfont 0.41$^\dagger$} & \cellcolor{nacell}\textcolor{black!45}{{\fontsize{6pt}{7pt}\selectfont --}} & \cellcolor{dpiv}\textcolor{white}{{\fontsize{6pt}{7pt}\selectfont \textbf{0.92}}} & \cellcolor{dpiv}\textcolor{white}{{\fontsize{6pt}{7pt}\selectfont \textbf{62$\times$}}} \\
Conv-VAE & \textcolor{latcont}{\scalebox{1.5}{$\bullet$}} & {\fontsize{6pt}{7pt}\selectfont 1\,s} & \cellcolor{dpii} & \cellcolor{dpiv}\textcolor{white}{} & \cellcolor{dmi} & \cellcolor{dpiii}\textcolor{white}{} & \cellcolor{dzero} & \cellcolor{dmi} & \cellcolor{dmii} & \cellcolor{dmii} & \cellcolor{dmii} & \cellcolor{dmii} & \cellcolor{dmi} & \cellcolor{dmii} & \cellcolor{dzero}{\fontsize{6pt}{7pt}\selectfont $-$0.01} & \cellcolor{dmii} & \cellcolor{dpi} & \cellcolor{dmiv}\textcolor{white}{} & \cellcolor{dmiii}\textcolor{white}{} & \cellcolor{dmiv}\textcolor{white}{} & \cellcolor{dmiii}\textcolor{white}{{\fontsize{6pt}{7pt}\selectfont $-$0.56}} & \cellcolor{dpii} & \cellcolor{dzero} & \cellcolor{dpiv}\textcolor{white}{} & \cellcolor{dpiv}\textcolor{white}{} & \cellcolor{dpiii}\textcolor{white}{{\fontsize{6pt}{7pt}\selectfont 0.55}} & \cellcolor{dpiv}\textcolor{white}{} & \cellcolor{dpii} & \cellcolor{dpiii}\textcolor{white}{{\fontsize{6pt}{7pt}\selectfont 0.55$^\dagger$}} & \cellcolor{nacell}\textcolor{black!45}{{\fontsize{6pt}{7pt}\selectfont --}} & \cellcolor{dpiv}\textcolor{white}{{\fontsize{6pt}{7pt}\selectfont 0.85}} & \cellcolor{dpii}{\fontsize{6pt}{7pt}\selectfont 2.0$\times$} \\
\midrule
\multicolumn{33}{@{}l}{\textit{Population tokenizers}} \\[1pt]
LDNS & \textcolor{latcont}{\scalebox{1.5}{$\bullet$}} & {\fontsize{6pt}{7pt}\selectfont 10\,ms} & \cellcolor{nacell}\textcolor{black!45}{\textperiodcentered} & \cellcolor{nacell}\textcolor{black!45}{\textperiodcentered} & \cellcolor{dpiii}\textcolor{white}{} & \cellcolor{dpiii}\textcolor{white}{} & \cellcolor{nacell}\textcolor{black!45}{\textperiodcentered} & \cellcolor{nacell}\textcolor{black!45}{\textperiodcentered} & \cellcolor{nacell}\textcolor{black!45}{\textperiodcentered} & \cellcolor{dpii} & \cellcolor{dpii} & \cellcolor{dmiv}\textcolor{white}{} & \cellcolor{dpi} & \cellcolor{dpii} & \cellcolor{dpi}{\fontsize{6pt}{7pt}\selectfont 0.22} & \cellcolor{dpii} & \cellcolor{dzero} & \cellcolor{dpii} & \cellcolor{nacell}\textcolor{black!45}{\textperiodcentered} & \cellcolor{dmii} & \cellcolor{dpi}{\fontsize{6pt}{7pt}\selectfont 0.05} & \cellcolor{dpi} & \cellcolor{nacell}\textcolor{black!45}{\textperiodcentered} & \cellcolor{dpiv}\textcolor{white}{} & \cellcolor{dpiv}\textcolor{white}{} & \cellcolor{dpiii}\textcolor{white}{{\fontsize{6pt}{7pt}\selectfont 0.53}} & \cellcolor{dpiv}\textcolor{white}{} & \cellcolor{dpi} & \cellcolor{dpii}{\fontsize{6pt}{7pt}\selectfont \textbf{0.48}} & \cellcolor{dzero}{\fontsize{6pt}{7pt}\selectfont 0.00} & \cellcolor{dpiv}\textcolor{white}{{\fontsize{6pt}{7pt}\selectfont 0.84}} & \cellcolor{dmi}{\fontsize{6pt}{7pt}\selectfont 0.86$\times$} \\
Wavelet-FSQ & \textcolor{latdisc}{$\blacksquare$} & {\fontsize{6pt}{7pt}\selectfont 2\,ms} & \cellcolor{nacell}\textcolor{black!45}{\textperiodcentered} & \cellcolor{nacell}\textcolor{black!45}{\textperiodcentered} & \cellcolor{dpiv}\textcolor{white}{} & \cellcolor{dpiv}\textcolor{white}{} & \cellcolor{nacell}\textcolor{black!45}{\textperiodcentered} & \cellcolor{nacell}\textcolor{black!45}{\textperiodcentered} & \cellcolor{nacell}\textcolor{black!45}{\textperiodcentered} & \cellcolor{dpiv}\textcolor{white}{} & \cellcolor{dpiv}\textcolor{white}{} & \cellcolor{dpiv}\textcolor{white}{} & \cellcolor{dpiv}\textcolor{white}{} & \cellcolor{dpiv}\textcolor{white}{} & \cellcolor{dpiv}\textcolor{white}{{\fontsize{6pt}{7pt}\selectfont 0.90}} & \cellcolor{dpiv}\textcolor{white}{} & \cellcolor{dpiii}\textcolor{white}{} & \cellcolor{dpiii}\textcolor{white}{} & \cellcolor{dpiii}\textcolor{white}{} & \cellcolor{dpii} & \cellcolor{dpiii}\textcolor{white}{{\fontsize{6pt}{7pt}\selectfont 0.67}} & \cellcolor{dpi} & \cellcolor{nacell}\textcolor{black!45}{\textperiodcentered} & \cellcolor{dpiii}\textcolor{white}{} & \cellcolor{dpiii}\textcolor{white}{} & \cellcolor{dpiii}\textcolor{white}{{\fontsize{6pt}{7pt}\selectfont 0.51}} & \cellcolor{dpiii}\textcolor{white}{} & \cellcolor{dzero} & \cellcolor{dpii}{\fontsize{6pt}{7pt}\selectfont 0.37} & \cellcolor{dpi}{\fontsize{6pt}{7pt}\selectfont 0.14} & \cellcolor{dpiv}\textcolor{white}{{\fontsize{6pt}{7pt}\selectfont 0.86}} & \cellcolor{dpiii}\textcolor{white}{{\fontsize{6pt}{7pt}\selectfont \textbf{3.2$\times$}}} \\
Wavelet-AE & \textcolor{latcont}{\scalebox{1.5}{$\bullet$}} & {\fontsize{6pt}{7pt}\selectfont 2\,ms} & \cellcolor{nacell}\textcolor{black!45}{\textperiodcentered} & \cellcolor{nacell}\textcolor{black!45}{\textperiodcentered} & \cellcolor{dpiv}\textcolor{white}{} & \cellcolor{dpiv}\textcolor{white}{} & \cellcolor{nacell}\textcolor{black!45}{\textperiodcentered} & \cellcolor{nacell}\textcolor{black!45}{\textperiodcentered} & \cellcolor{nacell}\textcolor{black!45}{\textperiodcentered} & \cellcolor{dpiv}\textcolor{white}{} & \cellcolor{dpiv}\textcolor{white}{} & \cellcolor{dpiv}\textcolor{white}{} & \cellcolor{dpiv}\textcolor{white}{} & \cellcolor{dpiv}\textcolor{white}{} & \cellcolor{dpiv}\textcolor{white}{{\fontsize{6pt}{7pt}\selectfont \textbf{0.96}}} & \cellcolor{dpiv}\textcolor{white}{} & \cellcolor{dpiii}\textcolor{white}{} & \cellcolor{dpiv}\textcolor{white}{} & \cellcolor{dpiv}\textcolor{white}{} & \cellcolor{dpii} & \cellcolor{dpiv}\textcolor{white}{{\fontsize{6pt}{7pt}\selectfont \textbf{0.77}}} & \cellcolor{dpi} & \cellcolor{nacell}\textcolor{black!45}{\textperiodcentered} & \cellcolor{dpiii}\textcolor{white}{} & \cellcolor{dpiv}\textcolor{white}{} & \cellcolor{dpiii}\textcolor{white}{{\fontsize{6pt}{7pt}\selectfont 0.60}} & \cellcolor{dpiv}\textcolor{white}{} & \cellcolor{dpi} & \cellcolor{dpii}{\fontsize{6pt}{7pt}\selectfont 0.44} & \cellcolor{dpi}{\fontsize{6pt}{7pt}\selectfont \textbf{0.18}} & \cellcolor{dpiv}\textcolor{white}{{\fontsize{6pt}{7pt}\selectfont \textbf{0.93}}} & \cellcolor{dmii}{\fontsize{6pt}{7pt}\selectfont 0.50$\times$} \\
LDNS-Perceiver & \textcolor{latcont}{\scalebox{1.5}{$\bullet$}} & {\fontsize{6pt}{7pt}\selectfont 10\,ms} & \cellcolor{nacell}\textcolor{black!45}{\textperiodcentered} & \cellcolor{nacell}\textcolor{black!45}{\textperiodcentered} & \cellcolor{dpiv}\textcolor{white}{} & \cellcolor{dpiii}\textcolor{white}{} & \cellcolor{nacell}\textcolor{black!45}{\textperiodcentered} & \cellcolor{nacell}\textcolor{black!45}{\textperiodcentered} & \cellcolor{nacell}\textcolor{black!45}{\textperiodcentered} & \cellcolor{dpiii}\textcolor{white}{} & \cellcolor{dpii} & \cellcolor{dmi} & \cellcolor{dpi} & \cellcolor{dpii} & \cellcolor{dpii}{\fontsize{6pt}{7pt}\selectfont 0.35} & \cellcolor{dpiii}\textcolor{white}{} & \cellcolor{dpii} & \cellcolor{dzero} & \cellcolor{nacell}\textcolor{black!45}{\textperiodcentered} & \cellcolor{dmi} & \cellcolor{dpi}{\fontsize{6pt}{7pt}\selectfont 0.18} & \cellcolor{dpii} & \cellcolor{nacell}\textcolor{black!45}{\textperiodcentered} & \cellcolor{dpiv}\textcolor{white}{} & \cellcolor{dpiv}\textcolor{white}{} & \cellcolor{dpiii}\textcolor{white}{{\fontsize{6pt}{7pt}\selectfont \textbf{0.74}}} & \cellcolor{dpiv}\textcolor{white}{} & \cellcolor{dpi} & \cellcolor{dpii}{\fontsize{6pt}{7pt}\selectfont 0.46} & \cellcolor{dpi}{\fontsize{6pt}{7pt}\selectfont 0.12} & \cellcolor{dpiv}\textcolor{white}{{\fontsize{6pt}{7pt}\selectfont 0.86}} & \cellcolor{dmi}{\fontsize{6pt}{7pt}\selectfont 0.86$\times$} \\
\bottomrule
\end{tabular}
\caption{\textbf{Evaluating tokenizers across all seven evaluation framework dimensions (Section~\ref{sec:evaluation-framework}).} \emph{Design}: latent type (\textcolor{latcont}{$\bullet$}~continuous, \textcolor{latdisc}{$\blacksquare$}~discrete) and the time one token covers. Composites average the metric families to their left; reconstruction is normalized against a rate-matched Poisson baseline (representing $0$, see Appendix~\ref{app:poisson-baseline}). \emph{Behav.\ decoding}: cursor $R^2$ of a POYO-style decoder and of linear probes; \emph{Stim.\ encoding}: noise-corrected $R^2$. Blue is better, red worse; bold: best per group. $^\dagger$1\,s tokens hold a whole decoding window, so their score is not directly comparable. Details in Appendix~\ref{sec:metrics-details}.}
\label{tab:main_results}
\vspace{-0.2cm}
\end{table*}

\subsection{Reconstruction quality \& generalization}

Many tokenizers achieve near-perfect reconstruction. Figure~\ref{fig:recon-quality} illustrates this for the Conv-VAE ($100:1$), the best single-neuron tokenizer in Table~\ref{tab:main_results}, evaluated on a held-out session: reconstructed spike trains closely match held-out originals over multi-second windows (Fig.~\ref{fig:recon-quality}C), and single-unit temporal statistics are recovered with high fidelity (Fig.~\ref{fig:recon-quality}B; per-metric values in Table~\ref{tab:results-recon}). Reconstruction quality scales with training data, with diminishing returns reached early (Fig.~\ref{fig:recon-quality}D): the cross-source regime starts substantially lower at a small scale and closes most of the gap already with the medium corpus, and only the substantially larger corpus with proprietary recordings improves it further. Across the full zoo (Fig.~\ref{fig:recon-quality}E), reconstruction generalizes well across sessions and subjects of the training datasets, but loses quality on datasets unseen in training: the generalization score, the retention on those datasets, averages 0.84 (s.d.\ 0.05) over 47 tokenizers.

\subsection{Downstream utility of tokenizers}\label{sec:downstream-utility}

\begin{figure}[t]
    \centering
    \includegraphics[width=\linewidth]{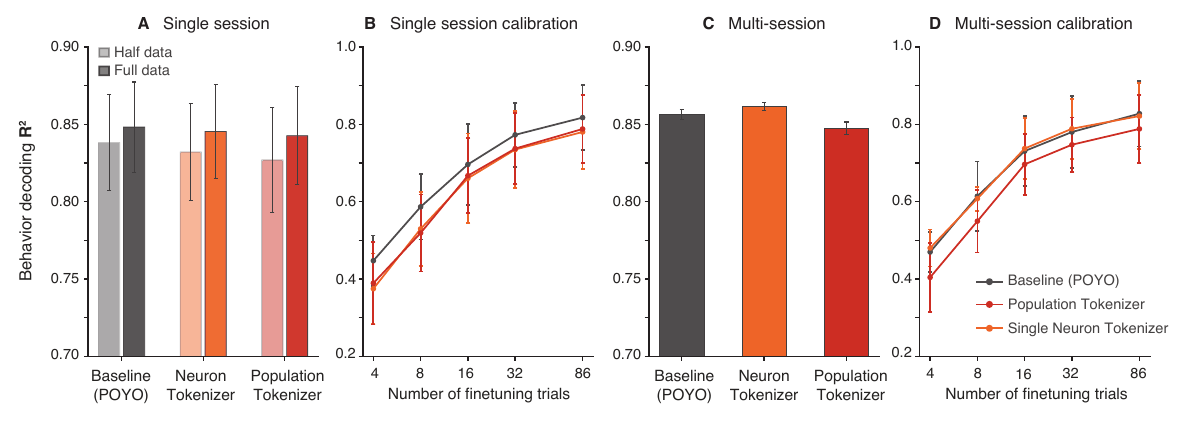}
    \caption{\textbf{Tokenized data supports POYO~\citep{azabou2023poyo}-style temporal decoding.}
    We compare the native spike-time baseline, a frozen single-unit tokenizer, and a frozen population-level tokenizer on reaching sessions from the Perich--Miller dataset~\citep{perich2018rapid}.  Error bars indicate sample standard deviations across runs.
    \textbf{A.} Single-session scratch decoding with mean held-out test \(R^2\) over eight sessions and two seeds, training on half (light bars) or all (dark bars) of each session's training trials.
    \textbf{B.} Single-session calibration from two base sessions to held-out sessions over different budgets of target-session trials.
    \textbf{C.} Multi-session training with four sessions over two seeds.
    \textbf{D.} Calibration to held-out sessions over different budgets of target-session trials.}
    \label{fig:poyo-evals}
    \vspace{-0.2cm}
\end{figure}

Reconstruction quality shows whether a tokenizer preserves the information about its input, but whether a downstream model can use that information is a separate question.
%but downstream utility determines its value in foundation model pipelines.
We test %this directly by evaluating
whether frozen tokenizer outputs can replace native spike-time inputs in a POYO~\citep{azabou2023poyo}-style temporal decoder (Figure~\ref{fig:poyo-evals}). This setting is more general than a linear probe, using frozen tokenizer representations as the neural input interface for an attention-based behavioral decoder. 
Across single-session training, multi-session base training, and cross-session calibrations with different trial budgets, 
tokenizers perform competitively with the native spike-time POYO baseline. 
% the single-unit tokenizer performs competitively with the native spike-time POYO baseline. 
% The population-level Perceiver-LDNS tokenizer trails both representations, suggesting that the population bottleneck preserves useful structure but may require longer training or a larger bottleneck to match performance. 
We note that POYO was designed around spike-time inputs, so the ceiling for tokenized inputs could likely be raised with architectural adaptations. For more details and the plotted values, see Appendix~\ref{app:downstream-decoding-poyo} and Table~\ref{tab:poyo-plotted-means}; decoding scores across the zoo are in Table~\ref{tab:results-decoding}.

A parallel evaluation reverses the direction, asking whether stimuli can be used to predict into the tokenizer's latent space and whether the frozen decoder can then recover PSTH-like responses. Encoding results in Table~\ref{tab:main_results} (full breakdown in Table~\ref{tab:results-encoding}) show substantial variation across tokenizers despite similar reconstruction quality (Fig.~\ref{fig:dissociation}B), also motivating our dissociation analysis.

One trend is that population tokenizers tend to come with lower encoding and decoding scores, suggesting either that their mixing across units loses downstream-relevant information, or that they work better with downstream architectures other than ours.% our downstream architectures are not sufficiently optimized for them.

Tokenizers differ widely on the latent-structure proxies (Table~\ref{tab:latent_metrics}), and tokens generally are less robust and share less structure with the input, than 100ms-binned Poisson rates. Preliminary evidence suggests a relationship between latent-structure scores and decoding performance (App.~\ref{app:latent_vs_downstream}).
%On proxy tasks that evaluate latent space alignment with stimulus structure and robustness, most tokenizers underperform Poisson rates (Table~\ref{tab:latent_metrics}), showing the difficulty of learning complex neural tokenizers while maintaining structural fidelity to the data.

\subsection{Reconstruction quality is necessary but not sufficient}

A tokenizer that achieves high reconstruction need not produce latents that are useful downstream, structured or linearly accessible. Compressed sensing illustrates this. %makes the point sharply. 
As a fixed random projection, it has no learned structure, yet its decoder reconstructs spike trains with fidelity competitive with the best learned tokenizers. 
It even performs well on our nonlinear encoding and POYO decoding benchmarks. Its latent space itself, however, carries little structure: 
its representational similarity to the input collapses, and linear probes recover little behavioral information from it.

\begin{figure}[t!]
  \centering
  \hspace*{-0.2in}\includegraphics[width=1.05\linewidth]{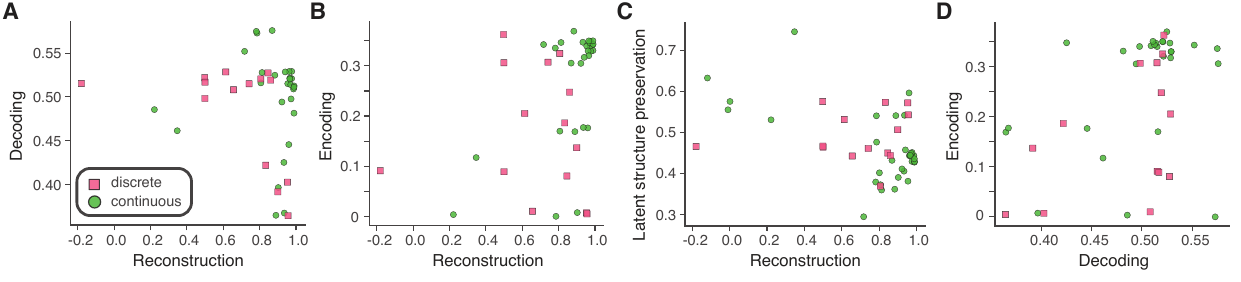}
  \caption{\textbf{Tokenizer dissociations.} Each point is a tokenizer in the zoo, colored by bottleneck type (discrete, pink; continuous, green); axes are the composite scores of Table~\ref{tab:main_results}, and A, B, D omit the 1\,s-token tokenizers. \textbf{A}. Reconstruction quality alone does not determine behavioral decoding. \textbf{B}. Likewise, high-fidelity reconstruction does not guarantee latents that support neural encoding. \textbf{C}. Latent structure preservation is not highest for the best single-neuron reconstructions, suggesting that fine spike-level fidelity and population geometry are distinct objectives. \textbf{D}. Tokenizers with strong decoding span nearly the full range of encoding performance.}
  \label{fig:dissociation}
  \vspace{-0.2cm}
\end{figure}

Across the full model zoo, reconstruction quality and downstream utility dissociate along several axes (Figure~\ref{fig:dissociation}). 
Reconstruction fidelity shows no clear monotonic relationship with decoding performance (Fig.~\ref{fig:dissociation}A), and %only a moderate one with encoding $R^2$ (Fig.~\ref{fig:dissociation}B) that does not hold up across model families.
neither with encoding $R^2$ (Fig.~\ref{fig:dissociation}B).
Models that reconstruct well are not necessarily the best at preserving latent structure (Fig.~\ref{fig:dissociation}C). Decoding and encoding performance are not interchangeable either: tokenizers with strong decoding span nearly the full range of encoding performance (Fig.~\ref{fig:dissociation}D). These trends suggest that the tokenizer’s full profile across axes is more diagnostic than any single metric.

% Across the full zoo, reconstruction quality and downstream utility decouple along several axes (Figure~\ref{fig:dissociation}). Single-neuron reconstruction fidelity (ACG correlation) is uncorrelated with encoding $R^2$ and with linear decoding accuracy, so a tokenizer that recovers single-unit timing well gives no guarantee of supporting downstream readouts. Population structure preservation is anti-correlated with single-neuron reconstruction, revealing a trade-off between preserving fine timing and preserving population geometry, and it is positively correlated with linear decoding accuracy, suggesting that population-level structure rather than single-unit timing is what linear downstream readouts use. The shape of a tokenizer's profile across axes, not its summary reconstruction score, is what carries diagnostic information.

\section{Discussion}
We introduce a framework for evaluating spike tokenization as a modular component, and use it to compare a range of tokenization strategies systematically. A dissociation emerges between reconstruction fidelity and downstream utility: tokenizers with near-perfect reconstruction can yield latent representations that are either highly informative or nearly useless for tasks the tokenizer was never trained for. This reveals a fundamental open problem. Reconstruction loss is the natural unsupervised objective for a task-agnostic tokenizer, yet it is not guaranteed to capture the structure that downstream models need. 

This points to a deeper question that the field has not yet answered: what self-supervised objective should a standalone spike tokenizer be trained with? In language, next-token prediction turned out to capture surprisingly universal structure \citep{radford2019language, brown2020language, radford2018improving}. In vision, masked reconstruction \citep{bao2021beit, heMAE2022} and contrastive objectives \citep{caronEmerging2021, moco2020} transfer broadly across tasks.
The analogous question for spikes, i.e., what unsupervised task over spike trains forces a model to capture universally useful structure, remains open. %, and its answer is potentially coupled to the downstream architecture. 
What our study makes explicit is the axes that are combined in tokenization:
%What tokenizers do is make explicit which axes survive as units of computation. 
Single-neuron tokenizers keep neurons as separate units and compress time, leaving the downstream model to combine neurons, e.g., via unit embeddings re-learned for each new population~\citep{azabou2023poyo}; population tokenizers compress units and keep finer time, but must then deal with changing unit identities themselves, similar to problems in multivariate time-series forecasting \citep{nieTimeSeriesWorth2022, liuITransformerInvertedTransformers2023}.
%Across our tokenizer zoo, collapsing information across the unit dimension comes with lowered encoding and decoding, indicating that mixing here loses information valuable for downstream performance. %Todo this would better fit into results, how to formulate a take-away instead? 
Whatever a frozen tokenizer mixes, the downstream model no longer has to learn, but can no longer undo. % - across our tokenizer zoo, mixing the unit dimension appears to happen suboptimally for downstream performance. 
The tradeoff is also on efficiency: Reducing sequence length lowers attention cost quadratically, and collapsing a dimension removes a dimension that the downstream model has to process.
%In short, whatever a frozen tokenizer mixes, the downstream model no longer has to learn, but can no longer undo. 
Our common interface (Section~\ref{sec:tokenizers}) makes the trade-off testable by swapping tokenizers under a fixed downstream architecture; the right objective, by contrast, is a broader question that tokenization shapes but does not settle.

%PREVIOUS CLOSER:
%The analogous question for spikes, i.e., what prediction task over spike trains forces a model to capture universally useful structure, remains open. Our evaluation framework provides the diagnostic tool to measure progress toward that goal by starting with systematic evaluation of the first component of the foundation model stack, but identifying the right objective is still the central open problem for the broader field.
% \drafttodo{As promised in the rebuttal, we should  expand the discussion of the tokenizer's interface with foundation models. Touching on population vs single neuron, analogies in vision etc. Also discussing mixing in the tokenizer vs mixing in the downstream model. Can replace parts of this discussion section. See linear for more notes.}

\paragraph{Practical insights from building the tokenizer zoo.}
% Building the zoo also taught us which design choices make a tokenizer work in practice. The most reliable recipe we found is a modest one: a continuous convolutional VAE over short (100\,ms) windows, with SIGReg~\cite{balestriero2025lejepa} as latent regularizer. Together, this implementation combines reconstruction fidelity with broad downstream use in improved encoding, decoding and reconstruction. Frozen and invertible, such a tokenizer already rivals POYO's jointly trained input layer (Fig.~\ref{fig:poyo-evals}), and it is the one we use in our own follow-up work. 
Building the zoo also taught us which design choices make a tokenizer work in practice. The most reliable recipe we found is a modest one: a continuous convolutional VAE over short (100\,ms) windows, with SIGReg~\cite{balestriero2025lejepa} as latent regularizer. Together, this implementation combines reconstruction fidelity with promising results for downstream tasks such as encoding and decoding. Despite being frozen for downstream tasks and invertible, our VAE tokenizer already rivals POYO's jointly trained input layer (Fig.~\ref{fig:poyo-evals}), and it is the one we use in our own follow-up work. 
Moving away from this recipe toward what long-context, decoder-only models prefer came at a cost at our data scale. Discrete bottlenecks were markedly harder to train stably and scored lower on most axes, and lengthening the window to 1\,s to lower the token rate drastically degrades reconstruction quality and training stability. Population tokenizers, in turn, offer direct access to cross-neuron structure but scored lower on encoding (and, for the LDNS variants, on reconstruction), and required much more design iteration. Once again, we read our results less as a ranking than as a map of trade-offs: because metrics dissociate, the downstream task rather than any single score should decide which tokenizer to use and which functionality it should support.

These recommendations also carry a caveat about scope: we built the zoo to exercise the evaluation framework and not to exhaust a vast design space that spans architectures, bottlenecks, losses, and compression levels. A practitioner who targets one region of that space for a specific use can undoubtedly train stronger tokenizers than ours. 

%We introduce a framework for evaluating spike tokenization as a modular component, and use it to systematically compare a range of tokenization strategies. Two observations emerge. First, reconstruction fidelity is a poor proxy for representation quality: tokenizers with near-perfect reconstruction can yield latent representations that are either highly informative or nearly useless for downstream readouts. Second, the choice between population-level and single-neuron tokenization reflects a fundamental trade-off, prioritizing either population geometry or fine temporal structure. Taken together, these results suggest that tokenization is not merely a preprocessing step, but a core representational decision that shapes what information is available to downstream models. This reinforces the need to treat tokenization as a critical design axis in brain foundation models.

% \paragraph{Relationship to existing benchmarks.}
% NLB \citep{pei2021nlb}, FALCON \citep{karpowicz2024falcon}, and CtDB \citep{versteeg2025ctdb} fix a representation and evaluate models built on top of it. Brain-Score \citep{schrimpf2018brainscore} made model-to-brain comparison a community-driven enterprise by standardizing the protocol. Our framework plays the analogous role one layer upstream, scoring the tokenizer itself against desiderata that are independent of any specific downstream model. The two lines of work compose.

\paragraph{Limitations.}
The framework's predictive validity for full foundation-model performance remains a hypothesis: our POYO and encoding experiments test it only partially, for independent decoding-only and encoding-only models. Predictive relationships may depend on architecture, scale, and data regime.  The metric set is not exhaustive and is designed to grow through community contributions. The current suite covers electrophysiological spikes only, not calcium imaging data or LFP. The evaluation pipeline assumes fixed-rate tokenization, excluding variable-length and event-based designs.

% The framework opens several directions beyond the tokenizers evaluated here. Hybrid tokenization schemes that combine single-neuron and population-level representations can be evaluated within the same framework, turning the trade-off we identify into a design space rather than a binary choice. The dissociation between reconstruction and downstream utility suggests that alternative training objectives—targeting latent structure or downstream alignment—may yield more useful representations than reconstruction alone. Finally, integration with benchmarks such as NLB and FALCON will test whether tokenizer-axis scores predict end-to-end foundation model performance. By releasing the dataset pipeline, evaluation suite, and tokenizer interface, we aim to enable systematic comparison and extension of spike tokenization methods.

\paragraph{Future directions.}
The framework opens several directions beyond the tokenizers evaluated here. On the tokenizer design side, hybrid spike tokenizers that compress both units and time at high fidelity with tractable cost remain to be developed; the dissociation between reconstruction and downstream utility also suggests further exploring training objectives that target latent structure or downstream alignment beyond reconstruction only. On the evaluation side, tokenizer metrics are cheap relative to training a foundation model per tokenizer (Appendix~\ref{app:compute}), so their value lies in predicting costlier measurements. As such, the next step is to further explore holding downstream architectures fixed and swapping tokenizers, as we do here with the POYO decoding downstream task. This inverts the task of benchmarks like FALCON~\citep{karpowicz2024falcon}, which fixes the input and varies the decoder. Ultimately, only properly ablating tokenizer choices in a large foundation model that, at minimum, both encodes and decodes \citep[such as NEDS; ][]{zhang2025neds}, would test whether tokenizer rankings and the predictions of our framework carry through at full scale.

% ORIGINAL SUBMISSION VERSION:
%The framework opens several directions beyond the tokenizers evaluated here. Hybrid tokenization schemes that combine single-neuron and population-level representations can be evaluated within the same framework, turning the trade-off between fine timing and population geometry into a design space rather than a forced choice. The dissociation between reconstruction and downstream utility suggests alternative training objectives\textemdash targeting latent structure or downstream alignment\textemdash may yield more useful representations than reconstruction alone. Finally, integration with benchmarks such as NLB and FALCON will test whether tokenizer-axis scores predict downstream foundation model performance. By releasing the dataset pipeline, evaluation suite, and tokenizer interface, we aim to enable systematic comparison and extension of spike tokenization methods.

\FloatBarrier
\newpage
\section*{Acknowledgments}
This work was generously supported by The James Fickel Enigma Project Fund.
The authors thank the International Max Planck Research School for Intelligent Systems (IMPRS-IS) for supporting FD.
This work was supported by the German Research Foundation (DFG): SFB 1233, Robust Vision: Inference Principles and Neural Mechanisms, project number: 276693517.
SK was supported by the German Research Foundation (DFG): SFB 1233, Robust Vision: Inference Principles and Neural Mechanisms, TP B3, project number: 276693517.
NR and JK were supported by the German Research Foundation (DFG) through Germany’s Excellence Strategy (EXC-Number 2064/1, PN 390727645) and SFB1233 (PN 276693517).
MK was supported by the Tübingen AI Center.
FHS and ASE were supported by the European Research Council (ERC) under the European Union’s
Horizon Europe research and innovation programme (Grant agreement No. 101171526 and grant agreement No. 101041669).
FHS, KF, ASE and MB are supported by the German Federal Ministry of Research, Technology and Space (BMFTR) via the project PRISMA-V (FKZ 01GQ2501A).
FHS and ASE acknowledge the support of the Ministry of Science and Culture of Lower Saxony through funds from the program zukunft.niedersachsen of the Volkswagen Foundation for the “CAIMed – Lower Saxony Center for Artificial Intelligence and Causal Methods in Medicine” project (grant no. ZN4257).
FHS acknowledges the support of the Else-Kröner-Fresenius Center for Optogenetic Therapies, University Medical Center Göttingen, Göttingen, Germany.

{
\small
\bibliographystyle{abbrvnat}
\bibliography{references}

@article{wang2025foundation,
  title={Foundation model of neural activity predicts response to new stimulus types},
  author={Wang, Eric Y and Fahey, Paul G and Ding, Zhuokun and Papadopoulos, Stelios and Ponder, Kayla and Weis, Marissa A and Chang, Andersen and Muhammad, Taliah and Patel, Saumil and Ding, Zhiwei and others},
  journal={Nature},
  volume={640},
  number={8058},
  pages={470--477},
  year={2025},
  publisher={Nature Publishing Group UK London}
}

@article{willeke2026omnimouse,
  title={OmniMouse: Scaling properties of multi-modal, multi-task Brain Models on 150B Neural Tokens},
  author={Willeke, Konstantin F and Turishcheva, Polina and Gilbert, Alex and Chakrabarty, Goirik and Bedel, Hasan A and Fahey, Paul G and Qiu, Yongrong and Weis, Marissa A and Vystr{\v{c}}ilov{\'a}, Michaela and Muhammad, Taliah and others},
  journal={arXiv preprint arXiv:2604.18827},
  year={2026}
}

@article{sahani2002linear,
  title={How linear are auditory cortical responses?},
  author={Sahani, Maneesh and Linden, Jennifer},
  journal={Advances in neural information processing systems},
  volume={15},
  year={2002}
}

@article{vanrossum2001novel,
  author  = {van Rossum, Mark C. W.},
  title   = {A novel spike distance},
  journal = {Neural Computation},
  volume  = {13},
  number  = {4},
  pages   = {751--763},
  year    = {2001},
  doi     = {10.1162/089976601300014321}
}

@inproceedings{aroraKnowThyselfKnowing2025,
  title = {Know {{Thyself}} by {{Knowing Others}}: {{Learning Neuron Identity}} from {{Population Context}}},
  shorttitle = {Know {{Thyself}} by {{Knowing Others}}},
  booktitle = {The {{Thirty-ninth Annual Conference}} on {{Neural Information Processing Systems}}},
  author = {Arora, Vinam and Lachi, Divyansha and Knight, Ian Jarratt and Azabou, Mehdi and Richards, Blake Aaron and Hurwitz, Cole Lincoln and Siegle, Josh and Dyer, Eva L.},
  year = 2025,
  month = oct,
  urldate = {2026-04-29},
  langid = {english}
}

@article{gerstner1997neural,
  title={Neural codes: firing rates and beyond},
  author={Gerstner, Wulfram and Kreiter, Andreas K and Markram, Henry and Herz, Andreas VM},
  journal={Proceedings of the National Academy of Sciences},
  volume={94},
  number={24},
  pages={12740--12741},
  year={1997},
  publisher={The National Academy of Sciences of the USA}
}

@article{panzeri2001unified,
  title={A unified approach to the study of temporal, correlational, and rate coding},
  author={Panzeri, Stefano and Schultz, Simon R},
  journal={Neural Computation},
  volume={13},
  number={6},
  pages={1311--1349},
  year={2001},
  publisher={MIT Press One Rogers Street, Cambridge, MA 02142-1209, USA journals-info~…}
}

@article{victor1996nature,
  title={Nature and precision of temporal coding in visual cortex: a metric-space analysis},
  author={Victor, Jonathan D and Purpura, Keith P},
  journal={Journal of neurophysiology},
  volume={76},
  number={2},
  pages={1310--1326},
  year={1996},
  publisher={American Physiological Society Bethesda, MD}
}

@article{azouz1999cellular,
  title={Cellular mechanisms contributing to response variability of cortical neurons in vivo},
  author={Azouz, Rony and Gray, Charles M},
  journal={Journal of Neuroscience},
  volume={19},
  number={6},
  pages={2209--2223},
  year={1999},
  publisher={Society for Neuroscience}
}

@article{churchland2010stimulus,
  title={Stimulus onset quenches neural variability: a widespread cortical phenomenon},
  author={Churchland, Mark M and Yu, Byron M and Cunningham, John P and Sugrue, Leo P and Cohen, Marlene R and Corrado, Greg S and Newsome, William T and Clark, Andrew M and Hosseini, Paymon and Scott, Benjamin B and others},
  journal={Nature neuroscience},
  volume={13},
  number={3},
  pages={369--378},
  year={2010},
  publisher={Nature Publishing Group US New York}
}

@inproceedings{allemanTaskStructureNonlinearity2023,
  title = {Task Structure and Nonlinearity Jointly Determine Learned Representational Geometry},
  booktitle = {The {{Twelfth International Conference}} on {{Learning Representations}}},
  author = {Alleman, Matteo and Lindsey, Jack and Fusi, Stefano},
  year = 2024,
  urldate = {2026-05-02},
  langid = {english}
}

@article{whitneyEvaluatingRepresentationsComplexity2020,
  title = {Evaluating Representations by the Complexity of Learning Low-Loss Predictors},
  author = {Whitney, William F. and Song, Min Jae and Brandfonbrener, David and Altosaar, Jaan and Cho, Kyunghyun},
  year = 2020,
  month = oct,
  urldate = {2026-05-02},
  langid = {english}
}

@inproceedings{
    azabou2023poyo,
    title={A Unified, Scalable Framework for Neural Population Decoding},
    author={Mehdi Azabou and Vinam Arora and Venkataramana Ganesh and Ximeng Mao and Santosh Nachimuthu and Michael Mendelson and Blake Richards and Matthew Perich and Guillaume Lajoie and Eva L. Dyer},
    booktitle={Thirty-seventh Conference on Neural Information Processing Systems},
    year={2023},
}

@misc{xuSelfSupervisedFoundationModel2026,
  title = {Self-{{Supervised Foundation Model}} for {{Calcium-imaging Population Dynamics}}},
  author = {Xu, Xinhong and Zhang, Yimeng and Qian, Qichen and Zhang, Yuanlong},
  year = 2026,
  month = apr,
  number = {arXiv:2604.04958},
  eprint = {2604.04958},
  primaryclass = {q-bio},
  publisher = {arXiv},
  doi = {10.48550/arXiv.2604.04958},
  urldate = {2026-04-08},
  archiveprefix = {arXiv}
}

@inproceedings{pei2021nlb,
  title     = {Neural Latents Benchmark '21: Evaluating Latent Variable Models of Neural Population Activity},
  author    = {Pei, Felix and Ye, Joel and Zoltowski, David and Wu, Anqi and Chowdhury, Raeed H. and Sohn, Hansem and O'Doherty, Joseph E. and Shenoy, Krishna V. and Kaufman, Matthew T. and Churchland, Mark M. and Jazayeri, Mehrdad and Miller, Lee E. and Pillow, Jonathan and Park, Il Memming and Dyer, Eva L. and Pandarinath, Chethan},
  booktitle = {Advances in Neural Information Processing Systems Track on Datasets and Benchmarks},
  year      = {2021}
}

@inproceedings{karpowicz2024falcon,
  title     = {Few-shot Algorithms for Consistent Neural Decoding ({FALCON}) Benchmark},
  author    = {Karpowicz, Brianna M. and Ye, Joel and Fan, Chaofei and Tostado-Marcos, Pablo and Rizzoglio, Fabio and Washington, Clayton B. and Scodeler, Thiago and de Lucena, Diogo S. and Nason-Tomaszewski, Samuel R. and Mender, Matthew and Ma, Xuan and Arneodo, Ezequiel Matias and Hochberg, Leigh and Chestek, Cynthia and Henderson, Jaimie M. and Gentner, Timothy Q. and Gilja, Vikash and Miller, Lee E. and Rouse, Adam G. and Gaunt, Robert and Pandarinath, Chethan},
  booktitle = {Advances in Neural Information Processing Systems Track on Datasets and Benchmarks},
  year      = {2024}
}

@article{versteeg2024ctdb,
  title   = {Computation-through-Dynamics Benchmark: Simulated Datasets and Quality Metrics for Dynamical Models of Neural Activity},
  author  = {Versteeg, Christopher and McCart, Jonathan D. and Ostrow, Mitchell and Zoltowski, David M. and Washington, Clayton B. and Driscoll, Laura and Codol, Olivier and Michaels, Jonathan A. and Linderman, Scott W. and Sussillo, David and Pandarinath, Chethan},
  journal = {bioRxiv},
  year    = {2025},
  doi     = {10.1101/2025.02.07.637062}
}

@misc{balestriero2025lejepa,
  title         = {{LeJEPA}: Provable and Scalable Self-Supervised Learning Without the Heuristics},
  author        = {Balestriero, Randall and LeCun, Yann},
  year          = {2025},
  eprint        = {2511.08544},
  archivePrefix = {arXiv},
  primaryClass  = {cs.LG}
}

@misc{ali2024tokenizer,
  title         = {Tokenizer Choice for {LLM} Training: Negligible or Crucial?},
  author        = {Ali, Mehdi and Fromm, Michael and Thellmann, Klaudia and Rutmann, Richard and L{\"u}bbering, Max and Leveling, Johannes and Klug, Katrin and Ebert, Jan and Doll, Niclas and Buschhoff, Jasper Schulze and Jain, Charvi and Weber, Alexander Arno and Jurkschat, Lena and Abdelwahab, Hammam and John, Chelsea and Suarez, Pedro Ortiz and Ostendorff, Malte and Weinbach, Samuel and Sifa, Rafet and Kesselheim, Stefan and Flores-Herr, Nicolas},
  year          = {2024},
  eprint        = {2310.08754},
  archivePrefix = {arXiv},
  primaryClass  = {cs}
}

@article{kapoor2024ldns,
  title   = {Latent Diffusion for Neural Spiking Data},
  author  = {Kapoor, Jaivardhan and Schulz, Auguste and Vetter, Julius and Pei, Felix and Gao, Richard and Macke, Jakob H.},
  journal = {Advances in Neural Information Processing Systems},
  volume  = {37},
  pages   = {118119--118154},
  year    = {2024}
}

@inproceedings{rust2021tokenizer,
  title     = {How Good is Your Tokenizer? On the Monolingual Performance of Multilingual Language Models},
  author    = {Rust, Phillip and Pfeiffer, Jonas and Vuli{\'c}, Ivan and Ruder, Sebastian and Gurevych, Iryna},
  booktitle = {Proceedings of the 59th Annual Meeting of the Association for Computational Linguistics and the 11th International Joint Conference on Natural Language Processing (Volume 1: Long Papers)},
  year      = {2021},
  pages     = {3118--3135}
}

@inproceedings{bostrom2020bpe,
  title     = {Byte Pair Encoding is Suboptimal for Language Model Pretraining},
  author    = {Bostrom, Kaj and Durrett, Greg},
  booktitle = {Findings of the Association for Computational Linguistics: EMNLP 2020},
  year      = {2020},
  pages     = {4617--4624}
}

@inproceedings{vandenoord2017vqvae,
  title     = {Neural Discrete Representation Learning},
  author    = {van den Oord, Aaron and Vinyals, Oriol and Kavukcuoglu, Koray},
  booktitle = {Advances in Neural Information Processing Systems},
  year      = {2017}
}

@inproceedings{dosovitskiy2020vit,
  title     = {An Image is Worth 16x16 Words: Transformers for Image Recognition at Scale},
  author    = {Dosovitskiy, Alexey and Beyer, Lucas and Kolesnikov, Alexander and Weissenborn, Dirk and Zhai, Xiaohua and Unterthiner, Thomas and Dehghani, Mostafa and Minderer, Matthias and Heigold, Georg and Gelly, Sylvain and Uszkoreit, Jakob and Houlsby, Neil},
  booktitle = {International Conference on Learning Representations},
  year      = {2021}
}

@inproceedings{yu2023magvit,
  title     = {{MAGVIT}: Masked Generative Video Transformer},
  author    = {Yu, Lijun and Cheng, Yong and Sohn, Kihyuk and Lezama, Jos{\'e} and Zhang, Han and Chang, Huiwen and Hauptmann, Alexander G. and Yang, Ming-Hsuan and Hao, Yuan and Essa, Irfan and Jiang, Lu},
  booktitle = {Proceedings of the IEEE/CVF Conference on Computer Vision and Pattern Recognition},
  year      = {2023}
}

@article{microns2025,
  title   = {Functional Connectomics Spanning Multiple Areas of Mouse Visual Cortex},
  author  = {{The MICrONS Consortium}},
  journal = {Nature},
  volume  = {640},
  pages   = {435--447},
  year    = {2025},
  doi     = {10.1038/s41586-025-08790-w}
}

@article{perich2018rapid,
  title   = {A Neural Population Mechanism for Rapid Learning},
  author  = {Perich, Matthew G. and Gallego, Juan A. and Miller, Lee E.},
  journal = {Neuron},
  volume  = {100},
  number  = {4},
  pages   = {964--976.e7},
  year    = {2018},
  doi     = {10.1016/j.neuron.2018.09.030}
}

@article{ye2021ndt,
  title   = {Representation Learning for Neural Population Activity with Neural Data Transformers},
  author  = {Ye, Joel and Pandarinath, Chethan},
  journal = {Neurons, Behavior, Data Analysis, and Theory},
  year    = {2021},
  doi     = {10.51628/001c.27358}
}

@inproceedings{ye2023ndt2,
  title     = {Neural Data Transformer 2: Multi-context Pretraining for Neural Spiking Activity},
  author    = {Ye, Joel and Collinger, Jennifer L. and Wehbe, Leila and Gaunt, Robert},
  booktitle = {Advances in Neural Information Processing Systems},
  year      = {2023}
}

@inproceedings{zhang2025neds,
  title     = {Neural Encoding and Decoding at Scale},
  author    = {Zhang, Yizi and Wang, Yanchen and Azabou, Mehdi and Andre, Alexandre and Wang, Zixuan and Lyu, Hanrui and {International Brain Laboratory} and Dyer, Eva L. and Paninski, Liam and Hurwitz, Cole},
  booktitle = {International Conference on Machine Learning},
  year      = {2025},
  eprint    = {2504.08201},
  archivePrefix = {arXiv}
}

@article{pandarinath2018lfads,
  title   = {Inferring Single-Trial Neural Population Dynamics Using Sequential Auto-Encoders},
  author  = {Pandarinath, Chethan and O'Shea, Daniel J. and Collins, Jasmine and Jozefowicz, Rafal and Stavisky, Sergey D. and Kao, Jonathan C. and Trautmann, Eric M. and Kaufman, Matthew T. and Ryu, Stephen I. and Hochberg, Leigh R. and Henderson, Jaimie M. and Shenoy, Krishna V. and Abbott, L. F. and Sussillo, David},
  journal = {Nature Methods},
  volume  = {15},
  number  = {10},
  pages   = {805--815},
  year    = {2018},
  doi     = {10.1038/s41592-018-0109-9}
}

@inproceedings{jiang2024labram,
  title     = {Large Brain Model for Learning Generic Representations with Tremendous {EEG} Data in {BCI}},
  author    = {Jiang, Wei-Bang and Zhao, Li-Ming and Lu, Bao-Liang},
  booktitle = {The Twelfth International Conference on Learning Representations},
  year      = {2024}
}

@misc{barmpas2025neurorvq,
  title         = {{NeuroRVQ}: Multi-Scale {EEG} Tokenization for Generative Large Brainwave Models},
  author        = {Barmpas, Konstantinos and Lee, Na and Koliousis, Alexandros and Panagakis, Yannis and Adamos, Dimitrios A. and Laskaris, Nikolaos and Zafeiriou, Stefanos},
  year          = {2025},
  eprint        = {2510.13068},
  archivePrefix = {arXiv},
  primaryClass  = {cs.LG}
}

@article{cowleyCompactDeepNeural2026,
  title = {Compact Deep Neural Network Models of the Visual Cortex},
  author = {Cowley, Benjamin R. and Stan, Patricia L. and Pillow, Jonathan W. and Smith, Matthew A.},
  year = 2026,
  month = feb,
  journal = {Nature},
  pages = {1--8},
  publisher = {Nature Publishing Group},
  issn = {1476-4687},
  doi = {10.1038/s41586-026-10150-1},
  urldate = {2026-02-26},
  langid = {english}
}

@article{candes2006near,
  author={Candes, Emmanuel J. and Tao, Terence},
  journal={IEEE Transactions on Information Theory}, 
  title={Near-Optimal Signal Recovery From Random Projections: Universal Encoding Strategies?}, 
  year={2006},
  volume={52},
  number={12},
  pages={5406-5425},
  doi={10.1109/TIT.2006.885507}
}

@article{donoho2006compressed,
  author={Donoho, D.L.},
  journal={IEEE Transactions on Information Theory}, 
  title={Compressed sensing}, 
  year={2006},
  volume={52},
  number={4},
  pages={1289-1306},
  doi={10.1109/TIT.2006.871582}
}

@misc{donoho2010counting,
  title = {Counting the {{Faces}} of {{Randomly-Projected Hypercubes}} and {{Orthants}}, with {{Applications}}},
  author = {Donoho, David L. and Tanner, Jared},
  year = 2008,
  month = jul,
  number = {arXiv:0807.3590},
  eprint = {0807.3590},
  primaryclass = {math},
  publisher = {arXiv},
  doi = {10.48550/arXiv.0807.3590},
  urldate = {2026-05-06},
  archiveprefix = {arXiv}
}

@article{beck2009fast,
author = {Beck, Amir and Teboulle, Marc},
title = {A Fast Iterative Shrinkage-Thresholding Algorithm for Linear Inverse Problems},
journal = {SIAM Journal on Imaging Sciences},
volume = {2},
number = {1},
pages = {183-202},
year = {2009},
doi = {10.1137/080716542},
URL = { 
        https://doi.org/10.1137/080716542
},
eprint = { 
        https://doi.org/10.1137/080716542
}
,
}

@inproceedings{vaswani2017attention,
  title = {Attention Is {{All}} You {{Need}}},
  booktitle = {Advances in {{Neural Information Processing Systems}}},
  author = {Vaswani, Ashish and Shazeer, Noam and Parmar, Niki and Uszkoreit, Jakob and Jones, Llion and Gomez, Aidan N and Kaiser, {\L}ukasz and Polosukhin, Illia},
  year = 2017,
  volume = {30},
  publisher = {Curran Associates, Inc.},
  urldate = {2026-05-06}
}

@article{iblBrainwideMap2025,
  title   = {A brain-wide map of neural activity during complex behaviour},
  author  = {{International Brain Laboratory}},
  journal = {Nature},
  year    = {2025},
  volume  = {645},
  number  = {8079},
  pages   = {177--191},
  doi     = {10.1038/s41586-025-09235-0},
  url     = {https://www.nature.com/articles/s41586-025-09235-0},
  note    = {Preprint posted on bioRxiv in 2023 (doi:10.1101/2023.07.04.547681)},
}

@techreport{allenVisualBehavior2022,
  title       = {Visual Behavior Neuropixels Technical Whitepaper},
  author      = {{Allen Institute for Brain Science}},
  institution = {Allen Institute for Brain Science},
  year        = {2022},
  type        = {Technical Whitepaper},
  url         = {https://portal.brain-map.org/circuits-behavior/visual-behavior-neuropixels},
}

@article{siegleSurveySpikingMouse2021,
  title   = {Survey of spiking in the mouse visual system reveals functional hierarchy},
  author  = {Siegle, Joshua H. and Jia, Xiaoxuan and Durand, S{\'e}verine and Gale, Sam and Bennett, Corbett and Graddis, Nile and Heller, Greggory and Ramirez, Tamina K. and Choi, Hannah and Luviano, Jennifer A. and Groblewski, Peter A. and Ahmed, Ruweida and Arkhipov, Anton and Bernard, Amy and Billeh, Yazan N. and Brown, Dillan and Buice, Michael A. and Cain, Nicholas and Caldejon, Shiella and Casal, Linzy and Cho, Andrew and Chvilicek, Maggie and Cox, Timothy C. and Dai, Kael and Denman, Daniel J. and de Vries, Saskia E. J. and Dietzman, Roald and Esposito, Luke and Farrell, Colin and Feng, David and Galbraith, John and Garrett, Marina and Gelfand, Emily C. and Hancock, Nile and Harris, Julie A. and Howard, Robert and Hu, Brian and Hytnen, Ross and Iyer, Ramakrishnan and Jessett, Erika and Johnson, Katelyn and Kato, India and Kiggins, Justin and Lambert, Sophie and Lecoq, Jerome and Ledochowitsch, Peter and Lee, Jung Hoon and Leon, Arielle and Li, Yang and Liang, Elizabeth and Long, Fuhui and Mace, Kyla and Melchior, Jose and Millman, Daniel and Mollenkopf, Tyler and Nayan, Chelsea and Ng, Lydia and Ngo, Kiet and Nguyen, Thuyahn and Nicovich, Philip R. and North, Kat and Ocker, Gabriel Koch and Ollerenshaw, Doug and Oliver, Michael and Pachitariu, Marius and Perkins, Jed and Reding, Melissa and Reid, David and Robertson, Miranda and Ronellenfitch, Kara and Seid, Sam and Slaughterbeck, Cliff and Stoecklin, Michelle and Sullivan, David and Sutton, Ben and Swapp, Jackie and Thompson, Carol and Turner, Kyle and Wakeman, Wayne and Whitesell, Jennifer D. and Williams, Derric and Williford, Ali and Young, Rob and Zeng, Hongkui and Naylor, Sarah and Phillips, John W. and Reid, R. Clay and Mihalas, Stefan and Olsen, Shawn R. and Koch, Christof},
  journal = {Nature},
  year    = {2021},
  volume  = {592},
  number  = {7852},
  pages   = {86--92},
  doi     = {10.1038/s41586-020-03171-x},
  url     = {https://www.nature.com/articles/s41586-020-03171-x},
}

@article{beauC4Cerebellum2025,
  title   = {A deep-learning strategy to identify cell types across species from high-density extracellular recordings},
  author  = {Beau, Maxime and Herzfeld, David J. and Naveros, Francisco and Hemelt, Marie E. and D'Agostino, Federico and Oostland, Marlies and Sanchez-Lopez, Alvaro and Chung, Young Yoon and Maibach, Michael and Kyranakis, Stephen and Stabb, Hannah N. and Martinez Lopera, M. Gabriela and Lajko, Agoston and Zedler, Marie and Ohmae, Shogo and Hall, Nathan J. and Clark, Beverley A. and Cohen, Dana and Lisberger, Stephen G. and Kostadinov, Dimitar and Hull, Court and H{\"a}usser, Michael and Medina, Javier F.},
  journal = {Cell},
  year    = {2025},
  volume  = {188},
  number  = {8},
  pages   = {2218--2234.e22},
  doi     = {10.1016/j.cell.2025.01.041},
  url     = {https://www.cell.com/cell/abstract/S0092-8674(25)00110-2},
}

@article{churchlandNeuralPopulation2012,
  title   = {Neural population dynamics during reaching},
  author  = {Churchland, Mark M. and Cunningham, John P. and Kaufman, Matthew T. and Foster, Justin D. and Nuyujukian, Paul and Ryu, Stephen I. and Shenoy, Krishna V.},
  journal = {Nature},
  year    = {2012},
  volume  = {487},
  number  = {7405},
  pages   = {51--56},
  doi     = {10.1038/nature11129},
  url     = {https://www.nature.com/articles/nature11129},
}

@article{flintAccurateDecoder2012,
  title   = {Accurate decoding of reaching movements from field potentials in the absence of spikes},
  author  = {Flint, Robert D. and Lindberg, Eric W. and Jordan, Luke R. and Miller, Lee E. and Slutzky, Marc W.},
  journal = {Journal of Neural Engineering},
  year    = {2012},
  volume  = {9},
  number  = {4},
  pages   = {046006},
  doi     = {10.1088/1741-2560/9/4/046006},
  url     = {https://iopscience.iop.org/article/10.1088/1741-2560/9/4/046006},
}

@misc{odohertyNonhumanPrimate2017,
  title        = {Nonhuman Primate Reaching with Multichannel Sensorimotor Cortex Electrophysiology},
  author       = {O'Doherty, Joseph E. and Cardoso, Mariana M. B. and Makin, Joseph G. and Sabes, Philip N.},
  year         = {2017},
  publisher    = {Zenodo},
  doi          = {10.5281/zenodo.3854034},
  url          = {https://zenodo.org/records/3854034},
  howpublished = {Zenodo dataset},
}

@inproceedings{ryooGeneralizableRealtimeNeural2025a,
  title = {Generalizable, Real-Time Neural Decoding with Hybrid State-Space Models},
  booktitle = {The {{Thirty-ninth Annual Conference}} on {{Neural Information Processing Systems}}},
  author = {Ryoo, Avery Hee-Woon and Krishna, Nanda H. and Mao, Ximeng and Azabou, Mehdi and Dyer, Eva L. and Perich, Matthew G. and Lajoie, Guillaume},
  year = 2025,
  month = oct,
  urldate = {2026-05-06},
  langid = {english}
}

@inproceedings{azabouMultisessionMultitaskNeural2024,
  title = {Multi-Session, Multi-Task Neural Decoding from Distinct Cell-Types and Brain Regions},
  booktitle = {The {{Thirteenth International Conference}} on {{Learning Representations}}},
  author = {Azabou, Mehdi and Pan, Krystal Xuejing and Arora, Vinam and Knight, Ian Jarratt and Dyer, Eva L. and Richards, Blake Aaron},
  year = 2025,
  urldate = {2025-05-17},
  langid = {english}
}

@inproceedings{zhangUniversalTranslatorNeural2024,
  title = {Towards a "{{Universal Translator}}" for {{Neural Dynamics}} at {{Single-Cell}}, {{Single-Spike Resolution}}},
  booktitle = {The {{Thirty-eighth Annual Conference}} on {{Neural Information Processing Systems}}},
  author = {Zhang, Yizi and Wang, Yanchen and {Jim{\'e}nez-Benet{\'o}}, Donato M. and Wang, Zixuan and Azabou, Mehdi and Richards, Blake Aaron and Tung, Renee and Winter, Olivier and {International Brain Laboratory} and Dyer, Eva L. and Paninski, Liam and Hurwitz, Cole Lincoln},
  year = 2024,
  month = nov,
  urldate = {2025-04-02},
  langid = {english}
}

@article{keshtkaranLargescaleNeuralNetwork2022,
  title = {A Large-Scale Neural Network Training Framework for Generalized Estimation of Single-Trial Population Dynamics},
  author = {Keshtkaran, Mohammad Reza and Sedler, Andrew R. and Chowdhury, Raeed H. and Tandon, Raghav and Basrai, Diya and Nguyen, Sarah L. and Sohn, Hansem and Jazayeri, Mehrdad and Miller, Lee E. and Pandarinath, Chethan},
  year = 2022,
  month = dec,
  journal = {Nature Methods},
  volume = {19},
  number = {12},
  pages = {1572--1577},
  publisher = {Nature Publishing Group},
  issn = {1548-7105},
  doi = {10.1038/s41592-022-01675-0},
  urldate = {2026-05-06},
  langid = {english}
}

@article{wan2025wan,
      title={Wan: Open and Advanced Large-Scale Video Generative Models},
      author={Team Wan and Ang Wang and Baole Ai and Bin Wen and Chaojie Mao and Chen-Wei Xie and Di Chen and Feiwu Yu and Haiming Zhao and Jianxiao Yang and Jianyuan Zeng and Jiayu Wang and Jingfeng Zhang and Jingren Zhou and Jinkai Wang and Jixuan Chen and Kai Zhu and Kang Zhao and Keyu Yan and Lianghua Huang and Mengyang Feng and Ningyi Zhang and Pandeng Li and Pingyu Wu and Ruihang Chu and Ruili Feng and Shiwei Zhang and Siyang Sun and Tao Fang and Tianxing Wang and Tianyi Gui and Tingyu Weng and Tong Shen and Wei Lin and Wei Wang and Wei Wang and Wenmeng Zhou and Wente Wang and Wenting Shen and Wenyuan Yu and Xianzhong Shi and Xiaoming Huang and Xin Xu and Yan Kou and Yangyu Lv and Yifei Li and Yijing Liu and Yiming Wang and Yingya Zhang and Yitong Huang and Yong Li and You Wu and Yu Liu and Yulin Pan and Yun Zheng and Yuntao Hong and Yupeng Shi and Yutong Feng and Zeyinzi Jiang and Zhen Han and Zhi-Fan Wu and Ziyu Liu},
      journal = {arXiv preprint arXiv:2503.20314},
      year={2025}
}

@article{radford2018improving,
  title={Improving language understanding by generative pre-training},
  author={Radford, Alec and Narasimhan, Karthik and Salimans, Tim and Sutskever, Ilya and others},
  year={2018},
  publisher={San Francisco, CA, USA}
}

@inproceedings{bao2021beit,
  title={{BEiT}: {BERT} Pre-Training of Image Transformers},
  author={Bao, Hangbo and Dong, Li and Piao, Songhao and Wei, Furu},
  booktitle={International Conference on Learning Representations},
  year={2022}
}

@article{brown2020language,
  title={Language models are few-shot learners},
  author={Brown, Tom and Mann, Benjamin and Ryder, Nick and Subbiah, Melanie and Kaplan, Jared D and Dhariwal, Prafulla and Neelakantan, Arvind and Shyam, Pranav and Sastry, Girish and Askell, Amanda and others},
  journal={Advances in neural information processing systems},
  volume={33},
  pages={1877--1901},
  year={2020}
}

@article{radford2019language,
  title={Language models are unsupervised multitask learners},
  author={Radford, Alec and Wu, Jeffrey and Child, Rewon and Luan, David and Amodei, Dario and Sutskever, Ilya and others},
  journal={OpenAI blog},
  volume={1},
  number={8},
  pages={9},
  year={2019}
}

@inproceedings{yu2023language,
  title={Language Model Beats Diffusion -- Tokenizer is Key to Visual Generation},
  author={Yu, Lijun and Lezama, Jos{\'e} and Gundavarapu, Nitesh Bharadwaj and Versari, Luca and Sohn, Kihyuk and Minnen, David and Cheng, Yong and Gupta, Agrim and Gu, Xiuye and Hauptmann, Alexander G. and Gong, Boqing and Yang, Ming-Hsuan and Essa, Irfan and Ross, David A. and Jiang, Lu},
  booktitle={International Conference on Learning Representations},
  year={2024}
}

@article{zeghidour2021soundstream,
  title={Soundstream: An end-to-end neural audio codec},
  author={Zeghidour, Neil and Luebs, Alejandro and Omran, Ahmed and Skoglund, Jan and Tagliasacchi, Marco},
  journal={IEEE/ACM Transactions on Audio, Speech, and Language Processing},
  volume={30},
  pages={495--507},
  year={2021},
  publisher={IEEE}
}

@inproceedings{zhao2024image,
  title={Image and Video Tokenization with Binary Spherical Quantization},
  author={Zhao, Yue and Xiong, Yuanjun and Kr{\"a}henb{\"u}hl, Philipp},
  booktitle={International Conference on Learning Representations},
  year={2025}
}

@article{wingertFerretAuditory2026,
  title   = {Convolutional neural network models describe the encoding subspace of local circuits in auditory cortex},
  author  = {Wingert, Jereme C. and Parida, Satyabrata and Norman-Haignere, Sam V. and David, Stephen V.},
  journal = {Nature Neuroscience},
  year    = {2026},
  note    = {In press},
  doi     = {10.1038/s41593-026-02216-0},
}

@inproceedings{jaegle2021perceiver,
  author    = {Andrew Jaegle and Felix Gimeno and Andy Brock and Oriol Vinyals and Andrew Zisserman and Jo{\~{a}}o Carreira},
  editor    = {Marina Meila and Tong Zhang},
  title     = {Perceiver: General Perception with Iterative Attention},
  booktitle = {Proceedings of the 38th International Conference on Machine Learning, {ICML} 2021, 18-24 July 2021, Virtual Event},
  series    = {Proceedings of Machine Learning Research},
  volume    = {139},
  pages     = {4651-4664},
  publisher = {PMLR},
  year      = {2021},
  url       = {http://proceedings.mlr.press/v139/jaegle21a.html},
  bibsource = {dblp computer science bibliography, https://dblp.org}
}

@article{agarwal2025cosmos,
  title={Cosmos world foundation model platform for physical ai},
  author={Agarwal, Niket and Ali, Arslan and Bala, Maciej and Balaji, Yogesh and Barker, Erik and Cai, Tiffany and Chattopadhyay, Prithvijit and Chen, Yongxin and Cui, Yin and Ding, Yifan and others},
  journal={arXiv preprint arXiv:2501.03575},
  year={2025}
}

@inproceedings{mentzer2023finite,
  title={Finite Scalar Quantization: {VQ-VAE} Made Simple},
  author={Mentzer, Fabian and Minnen, David and Agustsson, Eirikur and Tschannen, Michael},
  booktitle={International Conference on Learning Representations},
  year={2024}
}

@inproceedings{he2016deep,
  title={Deep residual learning for image recognition},
  author={He, Kaiming and Zhang, Xiangyu and Ren, Shaoqing and Sun, Jian},
  booktitle={Proceedings of the IEEE conference on computer vision and pattern recognition},
  pages={770--778},
  year={2016}
}

@book{mackay2003information,
  author = {MacKay, David J. C.},
  title = {Information Theory, Inference, and Learning Algorithms},
  year = {2003},
  isbn = {0521642981},
  publisher = {Cambridge University Press},
  address = {Cambridge, UK},
  url = {http://www.inference.org.uk/itila/book.html}
}

@article{wang2026temporal,
  title={Temporal straightening for latent planning},
  author={Wang, Ying and Bounou, Oumayma and Zhou, Gaoyue and Balestriero, Randall and Rudner, Tim GJ and LeCun, Yann and Ren, Mengye},
  journal={arXiv preprint arXiv:2603.12231},
  year={2026}
}

@misc{pradeepkumarTokenizingSingleChannelEEG2026,
  title = {Tokenizing {{Single-Channel EEG}} with {{Time-Frequency Motif Learning}}},
  author = {Pradeepkumar, Jathurshan and Piao, Xihao and Chen, Zheng and Sun, Jimeng},
  year = 2026,
  month = feb,
  number = {arXiv:2502.16060},
  eprint = {2502.16060},
  primaryclass = {cs},
  publisher = {arXiv},
  doi = {10.48550/arXiv.2502.16060},
  urldate = {2026-05-07},
  archiveprefix = {arXiv}
}

@misc{bettinardiBioSerenityE1SelfsupervisedEEG2025,
  title = {{{BioSerenity-E1}}: A Self-Supervised {{EEG}} Model for Medical Applications},
  shorttitle = {{{BioSerenity-E1}}},
  author = {Bettinardi, Ruggero G. and Rahmouni, Mohamed and Gimenez, Ulysse},
  year = 2025,
  month = mar,
  number = {arXiv:2503.10362},
  eprint = {2503.10362},
  primaryclass = {q-bio},
  publisher = {arXiv},
  doi = {10.48550/arXiv.2503.10362},
  urldate = {2026-05-07},
  archiveprefix = {arXiv}
}

@inproceedings{heMAE2022,
  author={He, Kaiming and Chen, Xinlei and Xie, Saining and Li, Yanghao and Dollár, Piotr and Girshick, Ross},
  booktitle={2022 IEEE/CVF Conference on Computer Vision and Pattern Recognition (CVPR)}, 
  title={Masked Autoencoders Are Scalable Vision Learners}, 
  year={2022},
  volume={},
  number={},
  pages={15979-15988},
  doi={10.1109/CVPR52688.2022.01553}}

@inproceedings{caronEmerging2021,
  author={Caron, Mathilde and Touvron, Hugo and Misra, Ishan and Jegou, Hervé and Mairal, Julien and Bojanowski, Piotr and Joulin, Armand},
  booktitle={2021 IEEE/CVF International Conference on Computer Vision (ICCV)}, 
  title={Emerging Properties in Self-Supervised Vision Transformers}, 
  year={2021},
  volume={},
  number={},
  pages={9630-9640},
  doi={10.1109/ICCV48922.2021.00951}}

@inproceedings{moco2020,
  author={He, Kaiming and Fan, Haoqi and Wu, Yuxin and Xie, Saining and Girshick, Ross},
  booktitle={2020 IEEE/CVF Conference on Computer Vision and Pattern Recognition (CVPR)}, 
  title={Momentum Contrast for Unsupervised Visual Representation Learning}, 
  year={2020},
  volume={},
  number={},
  pages={9726-9735},
  doi={10.1109/CVPR42600.2020.00975}}

@article{bjorck1973angles,
  author = {{\AA}ke Bj\"orck and Gene H. Golub},
  title = {Numerical methods for computing angles between linear subspaces},
  journal = {Mathematics of Computation},
  volume = {27},
  number = {123},
  pages = {579--594},
  year = {1973}
}

@article{okun2015populationcoupling,
  author = {Okun, M. and Steinmetz, N. A. and Cossell, L. and Iacaruso, M. F. and Ko, H. and Bart\'o, P. and Moore, T. and Hofer, S. B. and Mrsic-Flogel, T. D. and Carandini, M. and Harris, K. D.},
  title = {Diverse coupling of neurons to populations in sensory cortex},
  journal = {Nature},
  volume = {521},
  number = {7553},
  pages = {511--515},
  month = may,
  year = {2015}
}

@article {Gao214262,
	author = {Gao, Peiran and Trautmann, Eric and Yu, Byron and Santhanam, Gopal and Ryu, Stephen and Shenoy, Krishna and Ganguli, Surya},
	title = {A theory of multineuronal dimensionality, dynamics and measurement},
	elocation-id = {214262},
	year = {2017},
	doi = {10.1101/214262},
	publisher = {Cold Spring Harbor Laboratory},
	URL = {https://www.biorxiv.org/content/early/2017/11/12/214262},
	eprint = {https://www.biorxiv.org/content/early/2017/11/12/214262.full.pdf},
	journal = {bioRxiv}
}

@article{kriegeskorte2008rsa,
  author    = {Kriegeskorte, Nikolaus and Mur, Marieke and Bandettini, Peter A.},
  title     = {Representational similarity analysis: connecting the branches of systems neuroscience},
  journal   = {Frontiers in Systems Neuroscience},
  volume    = {2},
  pages     = {4},
  year      = {2008},
  doi       = {10.3389/neuro.06.004.2008},
  issn      = {1662-5137}
}

@article{diedrichsen2017representational,
  author    = {Diedrichsen, J{\"o}rn and Kriegeskorte, Nikolaus},
  title     = {Representational models: A common framework for understanding encoding, pattern-component, and representational-similarity analysis},
  journal   = {PLOS Computational Biology},
  volume    = {13},
  number    = {4},
  pages     = {e1005508},
  year      = {2017},
  doi       = {10.1371/journal.pcbi.1005508},
  publisher = {Public Library of Science}
}

@misc{yuImageWorth322024,
  title = {An {{Image}} Is {{Worth}} 32 {{Tokens}} for {{Reconstruction}} and {{Generation}}},
  author = {Yu, Qihang and Weber, Mark and Deng, Xueqing and Shen, Xiaohui and Cremers, Daniel and Chen, Liang-Chieh},
  year = 2024,
  month = jun,
  number = {arXiv:2406.07550},
  eprint = {2406.07550},
  primaryclass = {cs.CV},
  publisher = {arXiv},
  doi = {10.48550/arXiv.2406.07550},
  urldate = {2026-07-27},
  archiveprefix = {arXiv}
}

@misc{yaoReconstructionVsGeneration2025a,
  title = {Reconstruction vs. {{Generation}}: {{Taming Optimization Dilemma}} in {{Latent Diffusion Models}}},
  shorttitle = {Reconstruction vs. {{Generation}}},
  author = {Yao, Jingfeng and Yang, Bin and Wang, Xinggang},
  year = 2025,
  month = mar,
  number = {arXiv:2501.01423},
  eprint = {2501.01423},
  primaryclass = {cs.CV},
  publisher = {arXiv},
  doi = {10.48550/arXiv.2501.01423},
  urldate = {2026-07-27},
  archiveprefix = {arXiv}
}

@article{mousaviDASBDiscreteAudio2025,
  title = {{{DASB}} - {{Discrete Audio}} and {{Speech Benchmark}}},
  author = {Mousavi, Pooneh and Duret, Jarod and Petermann, Darius and Ploujnikov, Artem and Libera, Luca Della and Kuznetsova, Anastasia and Subakan, Cem and Ravanelli, Mirco},
  year = 2025,
  month = oct,
  journal = {Transactions on Machine Learning Research},
  issn = {2835-8856},
  urldate = {2026-10-26},
  langid = {english}
}

@inproceedings{yuanProteinStructureTokenization2025,
  title = {Protein {{Structure Tokenization}}: {{Benchmarking}} and {{New Recipe}}},
  shorttitle = {Protein {{Structure Tokenization}}},
  booktitle = {Forty-Second {{International Conference}} on {{Machine Learning}}},
  author = {Yuan, Xinyu and Wang, Zichen and Collins, Marcus D. and Rangwala, Huzefa},
  year = 2025,
  month = jun,
  urldate = {2026-10-26},
  langid = {english}
}

@article{averbeckNeuralCorrelationsPopulation2006b,
  title = {Neural Correlations, Population Coding and Computation},
  author = {Averbeck, Bruno B. and Latham, Peter E. and Pouget, Alexandre},
  year = 2006,
  month = may,
  journal = {Nature Reviews Neuroscience},
  volume = {7},
  number = {5},
  pages = {358--366},
  issn = {1471-003X, 1471-0048},
  doi = {10.1038/nrn1888},
  urldate = {2024-09-24},
  langid = {english}
}

@article{schneidmanWeakPairwiseCorrelations2006,
  title = {Weak Pairwise Correlations Imply Strongly Correlated Network States in a Neural Population},
  author = {Schneidman, Elad and Berry, Michael J. and Segev, Ronen and Bialek, William},
  year = 2006,
  month = apr,
  journal = {Nature},
  volume = {440},
  number = {7087},
  pages = {1007--1012},
  publisher = {Nature Publishing Group},
  issn = {1476-4687},
  doi = {10.1038/nature04701},
  urldate = {2025-05-24},
  langid = {english}
}

@article{dicarloHowDoesBrain2012,
  title = {How {{Does}} the {{Brain Solve Visual Object Recognition}}?},
  author = {DiCarlo, James J. and Zoccolan, Davide and Rust, Nicole C.},
  year = 2012,
  month = feb,
  journal = {Neuron},
  volume = {73},
  number = {3},
  pages = {415--434},
  publisher = {Elsevier},
  issn = {0896-6273},
  doi = {10.1016/j.neuron.2012.01.010},
  urldate = {2026-10-26},
  langid = {english},
  pmid = {22325196}
}

@article{mainenReliabilitySpikeTiming1995,
  title = {Reliability of {{Spike Timing}} in {{Neocortical Neurons}}},
  author = {Mainen, Zachary F. and Sejnowski, Terrence J.},
  year = 1995,
  month = jun,
  journal = {Science},
  volume = {268},
  number = {5216},
  pages = {1503--1506},
  publisher = {American Association for the Advancement of Science},
  doi = {10.1126/science.7770778},
  urldate = {2026-10-26}
}

@article{holtComparisonDischargeVariability1996,
  title = {Comparison of Discharge Variability in Vitro and in Vivo in Cat Visual Cortex Neurons},
  author = {Holt, G. R. and Softky, W. R. and Koch, C. and Douglas, R. J.},
  year = 1996,
  month = may,
  journal = {Journal of Neurophysiology},
  volume = {75},
  number = {5},
  pages = {1806--1814},
  publisher = {American Physiological Society},
  issn = {0022-3077},
  doi = {10.1152/jn.1996.75.5.1806},
  urldate = {2026-10-26}
}

@article{cohenMeasuringInterpretingNeuronal2011b,
  title = {Measuring and Interpreting Neuronal Correlations},
  author = {Cohen, Marlene R and Kohn, Adam},
  year = 2011,
  month = jul,
  journal = {Nature Neuroscience},
  volume = {14},
  number = {7},
  pages = {811--819},
  issn = {1097-6256, 1546-1726},
  doi = {10.1038/nn.2842},
  urldate = {2024-09-24},
  langid = {english}
}

@article{pillowSpatiotemporalCorrelationsVisual2008b,
  title = {Spatio-Temporal Correlations and Visual Signalling in a Complete Neuronal Population},
  author = {Pillow, Jonathan W. and Shlens, Jonathon and Paninski, Liam and Sher, Alexander and Litke, Alan M. and Chichilnisky, E. J. and Simoncelli, Eero P.},
  year = 2008,
  month = aug,
  journal = {Nature},
  volume = {454},
  number = {7207},
  pages = {995--999},
  issn = {0028-0836, 1476-4687},
  doi = {10.1038/nature07140},
  urldate = {2024-09-24},
  langid = {english}
}

@article{mackeGeneratingSpikeTrains2009,
  title = {Generating {{Spike Trains}} with {{Specified Correlation Coefficients}}},
  author = {Macke, Jakob H. and Berens, Philipp and Ecker, Alexander S. and Tolias, Andreas S. and Bethge, Matthias},
  year = 2009,
  month = feb,
  journal = {Neural Computation},
  volume = {21},
  number = {2},
  pages = {397--423},
  issn = {0899-7667},
  doi = {10.1162/neco.2008.02-08-713},
  urldate = {2026-10-26}
}

@inproceedings{molano-mazonSynthesizingRealisticNeural2018a,
  title = {Synthesizing Realistic Neural Population Activity Patterns Using {{Generative Adversarial Networks}}},
  booktitle = {International {{Conference}} on {{Learning Representations}}},
  author = {{Molano-Mazon}, Manuel and Onken, Arno and Piasini, Eugenio and Panzeri, Stefano},
  year = 2018,
  urldate = {2026-10-26},
  langid = {english}
}

@article{naselarisEncodingDecodingFMRI2011,
  title = {Encoding and Decoding in {{fMRI}}},
  author = {Naselaris, Thomas and Kay, Kendrick N. and Nishimoto, Shinji and Gallant, Jack L.},
  year = 2011,
  month = may,
  journal = {NeuroImage},
  series = {Multivariate {{Decoding}} and {{Brain Reading}}},
  volume = {56},
  number = {2},
  pages = {400--410},
  issn = {1053-8119},
  doi = {10.1016/j.neuroimage.2010.07.073},
  urldate = {2026-10-26}
}

@article{wuCompleteFunctionalCharacterization2006,
  title = {Complete {{Functional Characterization}} of {{Sensory Neurons}} by {{System Identification}}},
  author = {Wu, Michael C.-K. and David, Stephen V. and Gallant, Jack L.},
  year = 2006,
  journal = {Annual Review of Neuroscience},
  volume = {29},
  pages = {477--505},
  publisher = {Annual Reviews},
  issn = {1545-4126},
  doi = {10.1146/annurev.neuro.29.051605.113024},
  urldate = {2026-10-26}
}

@article{cunninghamDimensionalityReductionLargescale2014,
  title = {Dimensionality Reduction for Large-Scale Neural Recordings},
  author = {Cunningham, John P. and Yu, Byron M.},
  year = 2014,
  month = nov,
  journal = {Nature Neuroscience},
  volume = {17},
  number = {11},
  pages = {1500--1509},
  publisher = {Nature Publishing Group},
  issn = {1546-1726},
  doi = {10.1038/nn.3776},
  urldate = {2026-10-26},
  langid = {english}
}

@inproceedings{huhStraighteningOutStraightthrough2023,
  title = {Straightening out the Straight-through Estimator: Overcoming Optimization Challenges in Vector Quantized Networks},
  shorttitle = {Straightening out the Straight-through Estimator},
  booktitle = {Proceedings of the 40th {{International Conference}} on {{Machine Learning}}},
  author = {Huh, Minyoung and Cheung, Brian and Agrawal, Pulkit and Isola, Phillip},
  year = 2023,
  month = jul,
  series = {{{ICML}}'23},
  volume = {202},
  pages = {14096--14113},
  publisher = {JMLR.org},
  address = {Honolulu, Hawaii, USA},
  urldate = {2026-10-27}
}

@inproceedings{blauRethinkingLossyCompression2019,
  title = {Rethinking {{Lossy Compression}}: {{The Rate-Distortion-Perception Tradeoff}}},
  shorttitle = {Rethinking {{Lossy Compression}}},
  booktitle = {International {{Conference}} on {{Machine Learning}}},
  author = {Blau, Y. and Michaeli, T.},
  year = 2019,
  month = jan,
  urldate = {2026-10-27}
}

@article{degenhartStabilizationBrainComputer2020,
  title = {Stabilization of a Brain--Computer Interface via the Alignment of Low-Dimensional Spaces of Neural Activity},
  author = {Degenhart, Alan D. and Bishop, William E. and Oby, Emily R. and {Tyler-Kabara}, Elizabeth C. and Chase, Steven M. and Batista, Aaron P. and Yu, Byron M.},
  year = 2020,
  month = jul,
  journal = {Nature Biomedical Engineering},
  volume = {4},
  number = {7},
  pages = {672--685},
  publisher = {Nature Publishing Group},
  issn = {2157-846X},
  doi = {10.1038/s41551-020-0542-9},
  urldate = {2026-10-27},
  langid = {english}
}

@article{gallegoLongtermStabilityCortical2020,
  title = {Long-Term Stability of Cortical Population Dynamics Underlying Consistent Behavior},
  author = {Gallego, Juan A. and Perich, Matthew G. and Chowdhury, Raeed H. and Solla, Sara A. and Miller, Lee E.},
  year = 2020,
  month = feb,
  journal = {Nature Neuroscience},
  volume = {23},
  number = {2},
  pages = {260--270},
  publisher = {Nature Publishing Group},
  issn = {1546-1726},
  doi = {10.1038/s41593-019-0555-4},
  urldate = {2026-10-27},
  langid = {english}
}

@inproceedings{liuITransformerInvertedTransformers2023,
  title = {{{iTransformer}}: {{Inverted Transformers Are Effective}} for {{Time Series Forecasting}}},
  shorttitle = {{{iTransformer}}},
  booktitle = {The {{Twelfth International Conference}} on {{Learning Representations}}},
  author = {Liu, Yong and Hu, Tengge and Zhang, Haoran and Wu, Haixu and Wang, Shiyu and Ma, Lintao and Long, Mingsheng},
  year = 2024,
  urldate = {2026-10-28},
  langid = {english}
}

@inproceedings{nieTimeSeriesWorth2022,
  title = {A {{Time Series}} Is {{Worth}} 64 {{Words}}: {{Long}}-Term {{Forecasting}} with {{Transformers}}},
  shorttitle = {A {{Time Series}} Is {{Worth}} 64 {{Words}}},
  booktitle = {The {{Eleventh International Conference}} on {{Learning Representations}}},
  author = {Nie, Yuqi and Nguyen, Nam H. and Sinthong, Phanwadee and Kalagnanam, Jayant},
  year = 2023,
  urldate = {2026-10-28},
  langid = {english}
}

@incollection{miller1955note,
  title     = {Note on the Bias of Information Estimates},
  author    = {Miller, George A.},
  booktitle = {Information Theory in Psychology: Problems and Methods},
  editor    = {Quastler, Henry},
  publisher = {Free Press},
  address   = {Glencoe, IL},
  year      = {1955},
  pages     = {95--100}
}

@article{pospisil2021unbiased,
  title   = {The Unbiased Estimation of the Fraction of Variance Explained by a Model},
  author  = {Pospisil, Dean A. and Bair, Wyeth},
  journal = {PLOS Computational Biology},
  volume  = {17},
  number  = {8},
  pages   = {e1009212},
  year    = {2021},
  doi     = {10.1371/journal.pcbi.1009212}
}

@inproceedings{ryali2023hiera,
  title     = {Hiera: A Hierarchical Vision Transformer without the Bells-and-Whistles},
  author    = {Ryali, Chaitanya and Hu, Yuan-Ting and Bolya, Daniel and Wei, Chen and Fan, Haoqi and Huang, Po-Yao and Aggarwal, Vaibhav and Chowdhury, Arkabandhu and Poursaeed, Omid and Hoffman, Judy and Malik, Jitendra and Li, Yanghao and Feichtenhofer, Christoph},
  booktitle = {Proceedings of the 40th International Conference on Machine Learning},
  series    = {Proceedings of Machine Learning Research},
  volume    = {202},
  pages     = {29441--29454},
  year      = {2023}
}

@inproceedings{gu2022efficiently,
  title     = {Efficiently Modeling Long Sequences with Structured State Spaces},
  author    = {Gu, Albert and Goel, Karan and R{\'e}, Christopher},
  booktitle = {International Conference on Learning Representations},
  year      = {2022}
}

@inproceedings{lipman2023flow,
  title     = {Flow Matching for Generative Modeling},
  author    = {Lipman, Yaron and Chen, Ricky T. Q. and Ben-Hamu, Heli and Nickel, Maximilian and Le, Matt},
  booktitle = {International Conference on Learning Representations},
  year      = {2023}
}

@inproceedings{lin2017focal,
  title     = {Focal Loss for Dense Object Detection},
  author    = {Lin, Tsung-Yi and Goyal, Priya and Girshick, Ross and He, Kaiming and Doll{\'a}r, Piotr},
  booktitle = {Proceedings of the IEEE International Conference on Computer Vision},
  pages     = {2980--2988},
  year      = {2017}
}

@misc{wang2020vqpytorch,
  title        = {vector-quantize-pytorch},
  author       = {Wang, Phil},
  howpublished = {GitHub repository, \url{https://github.com/lucidrains/vector-quantize-pytorch}},
  year         = {2020}
}
}

\newpage
\appendix

\setcounter{figure}{0}
\renewcommand{\thefigure}{S\arabic{figure}}                                 
\setcounter{table}{0}                                                       
\renewcommand{\thetable}{S\arabic{table}}
\renewcommand{\theHfigure}{S\arabic{figure}}
\renewcommand{\theHtable}{S\arabic{table}}

\section{Extended related work and grounding for the desiderata}
\label{app:grounding}

This section relates each desideratum (Section~\ref{sec:desiderata}) and each metric dimension (Section~\ref{sec:evaluation-framework}) to prior work, and expands on the failure mode each dimension is designed to expose. Our seven evaluation metric groups / design dimensions are non-orthogonal diagnostic views rather than independent factors or a complete basis: their empirical correlations, which depend on our current cohort of tokenizers, are reported in Appendix~\ref{sec:appdx_metric_pairwise}. 
These dimensions are intended to be expanded by practitioners and the broader community depending on the question or modeling problem a tokenizer is used to tackle and practical considerations. As an example of the latter, computational efficiency, which none of our original dimensions captured, is reported separately in Appendix~\ref{app:inference-cost}. % <- label TBD by MOD-1701

\paragraph{Desiderata.}
The four desiderata follow from the functional requirements of a tokenizer meant to be reused across models: it should preserve the relevant structure of its input (reconstruction fidelity), expose that structure to downstream computation (downstream utility), keep doing so across the recordings it is meant to serve (cross-instance generalization), and do so at tractable representational cost (compression). Tokenizer evaluation in other modalities is organized along the same lines, and for the same reason: tokenizers rank differently across these criteria. In vision, reconstruction fidelity, downstream generative quality and token rate are reported separately and can be dissociated in controlled comparisons~\citep{yu2023language, yuImageWorth322024, yaoReconstructionVsGeneration2025a}. Audio and protein-structure tokenizer benchmarks likewise separate fidelity, bitrate or codebook utilization, sensitivity or robustness, and downstream effectiveness, and find that no single tokenizer leads on all of them~\citep{mousaviDASBDiscreteAudio2025, yuanProteinStructureTokenization2025}. 
The case for multiple axes is therefore not that they are independent, but that none of them reliably determines the others, and that all represent useful and desirable qualitative properties for downstream modeling.

\paragraph{Single-neuron reconstruction.}
The baseline failure mode is a tokenizer that preserves each neuron's firing rate and discards everything else about precise spike-timing. This is the motivation for our rate-matched Poisson floor, against which reconstruction metrics are normalized. Cortical neurons can fire with millisecond-scale reliability~\citep{mainenReliabilitySpikeTiming1995}, and spike timing, interval statistics and variability carry information beyond rate~\citep{victor1996nature, panzeri2001unified, gerstner1997neural}. Our metrics therefore build upon standard spike-train distances~\citep{victor1996nature, vanrossum2001novel} and variability measures such as CV2~\citep{holtComparisonDischargeVariability1996} for the comparison of $x$ and $\hat{x}$.
 
\paragraph{Population firing statistics.}
Single-neuron metrics are computed independently on each unit's spike train, so they cannot see dependencies between units: a tokenizer can match every neuron's marginal statistics while decorrelating the population. Pairwise correlations and higher-order population structure carry information absent from single-neuron statistics~\citep{averbeckNeuralCorrelationsPopulation2006b, schneidmanWeakPairwiseCorrelations2006, cohenMeasuringInterpretingNeuronal2011b}, and modeling coupling between cells has been shown to improve the prediction of population responses~\citep{pillowSpatiotemporalCorrelationsVisual2008b}. Generative models of spike trains have accordingly been judged by whether they reproduce such statistics~\citep{mackeGeneratingSpikeTrains2009, molano-mazonSynthesizingRealisticNeural2018a, kapoor2024ldns}; we apply the same criterion to tokenizer reconstructions, using measures such as population coupling~\citep{okun2015populationcoupling} and participation ratio~\citep{Gao214262}.
 
\paragraph{Downstream decoding.}
Reconstruction is computed through the tokenizer's own decoder, so it does not establish that task variables are accessible from the latents a downstream (linear or nonlinear) model actually receives. Information can be present but entangled, in which case readout requires a complex or data-hungry decoder~\citep{dicarloHowDoesBrain2012, whitneyEvaluatingRepresentationsComplexity2020}. Decoding behavior from population activity is also the canonical evaluation of neural population models~\citep{pei2021nlb, karpowicz2024falcon, azabou2023poyo}. We therefore probe frozen latents with simple readouts and with learning curves.
 
\paragraph{Downstream encoding.}
Encoding uses the representation in the opposite direction, as a prediction target rather than an input~\citep{naselarisEncodingDecodingFMRI2011, wuCompleteFunctionalCharacterization2006}. A tokenizer whose latents support decoding may still produce a target that is hard to predict from stimuli due to the structure of the latent space. Decoding-centric evaluation would not be able to detect this, yet it matters for any model that must generate neural activity, as recent encoding-oriented foundation models do~\citep{wang2025foundation}.

\paragraph{Latent structure and robustness.}
A sufficiently expressive decoder can reconstruct well from latents whose geometry bears little relation to that of the input, so we also evaluate $z$ directly. Representational similarity~\citep{kriegeskorte2008rsa, diedrichsen2017representational} tests whether the latent space preserves population geometry, which is the object of much systems-neuroscience analysis~\citep{cunninghamDimensionalityReductionLargescale2014}. Corruption tests check stability under perturbations that would resemble natural trial-to-trial variability~\citep{azouz1999cellular, churchland2010stimulus}. Codebook statistics are the usual check for the collapse and under-utilization common in quantized tokenizers~\citep{huhStraighteningOutStraightthrough2023, mentzer2023finite, yuanProteinStructureTokenization2025}; our current suite does not include such a metric.
 
\paragraph{Generalization.}
All other dimensions are first measured in distribution. A tokenizer can score well on all of them while overfitting to unit identities, electrode geometry or firing statistics of its training sessions. This is the central obstacle to tokenizer reuse: recorded neurons change across days, subjects and labs, even when the underlying latent dynamics are stable~\citep{gallegoLongtermStabilityCortical2020, degenhartStabilizationBrainComputer2020}. Cross-session transfer is also a primary target of recent decoding benchmarks and multi-session models~\citep{karpowicz2024falcon, ye2023ndt2, azabou2023poyo}. In our evaluation the score uses the cross-source split, because it matches how a tokenizer is meant to be used: trained once on a large corpus, then applied to new datasets that it has most likely never seen (Appendix~\ref{app:generalization-retention}).
 
\paragraph{Compression.}
Without a rate constraint, the identity map could be the optimal tokenizer on every reconstruction metric. Compression makes this trade-off explicit, following rate--distortion analysis~\citep{mackay2003information, blauRethinkingLossyCompression2019}. The number and size of tokens also set downstream cost, which, for a hypothetical downstream transformer model, would grow quadratically with sequence length under self-attention~\citep{vaswani2017attention}. For this reason visual tokenizers are increasingly compared at matched token budgets~\citep{yuImageWorth322024}.

\paragraph{Benchmarks for neural data modeling.}
Existing neural benchmarks evaluate models built on a fixed input representation. NLB \citep{pei2021nlb} measures how well latent-variable models recover firing rates of held-out neurons (co-smoothing) and decode behavior from population activity. FALCON \citep{karpowicz2024falcon} measures how well decoders transfer across sessions and days with minimal recalibration data. CtDB \citep{versteeg2024ctdb} measures how well dynamical models recover ground-truth dynamics on synthetic data. 
In each case, the input representation of the data is held fixed and the downstream model is the experimental variable. Our framework works one layer upstream: it fixes the evaluations and compares the tokenizers that decide what any of those models can see.

\section{Additional metric details}
\label{app:metric-details}

Here we describe the metrics referenced in
the main text. Throughout, $x$ denotes the original spike train (or rate trace, when the input is rate-valued) and $\hat{x}$ the tokenizer
reconstruction. 
Per-unit metrics are reported as the average over neurons within a session; population-level metrics return a single scalar per session. Metrics may return missing values per unit when the underlying statistic is not defined (e.g.\ a neuron with fewer than two spikes for ISI-based metrics, or a contingency table too sparse for the bias correction in mutual information); these units are excluded from per-session averages. We describe the metrics in detail in the following sections. Appendix~\ref{app:metric-corr-analysis} analyzes 39 per-metric scores, which count metric variants (e.g.\ strict spike-detection F1, individual input-corruption types and downstream targets) separately.

\subsection{Single-neuron reconstruction statistics}
These metrics compare $\hat{x}$ to $x$ for each unit independently and
serve as the per-unit floor: information discarded here is lost for every
downstream task that follows.

% \begin{description}
\setlength{\itemsep}{2pt}

\subsubsection{Spike-detection F1}
Raw mean-squared error is not very meaningful for spiking data because of the sparsity of the data. As a more balanced, spike-level metric, 
we can consider the reconstruction task as a hit-or-miss distinction per spike: For each spike in the original data, we obtain a true positive ($\mathrm{TP}$) if one reconstruction spike exists within a tolerance window of $\pm 1ms$ of the original spike. If no spike is within range, we have a false negative ($\mathrm{FN}$), and each unmatched reconstruction spike is a false positive ($\mathrm{FP}$). Spikes can only be matched once and are matched greedily. From this we can compute precision $P$ and recall $R$:
\[
P = \frac{\mathrm{TP}}{\mathrm{TP} + \mathrm{FP}},
\qquad
R = \frac{\mathrm{TP}}{\mathrm{TP} + \mathrm{FN}}.
\]
The F1 score is the harmonic mean of precision and recall:
\[
\mathrm{F1} = \frac{2PR}{P + R}
= \frac{2\,\mathrm{TP}}{2\,\mathrm{TP} + \mathrm{FP} + \mathrm{FN}}.
\]
The range is $[0, 1]$, higher is better. We report this metric at two tolerances: the default $\pm 1$\,ms (spike-detection F1) and a stricter $\pm 0.5$\,ms (strict F1).

\subsubsection{Temporal precision}
Using the same matching method as in spike-detection F1, we compute the median absolute timing error in milliseconds for matched spikes. We use a wider matching tolerance (default $5$~ms) than the F1 metric so that the value remains informative on data quantized to a few-millisecond bin size. Range $[0, \infty)$~ms, lower is better; reported only for single-neuron tokenizers, whose outputs retain spike-time resolution. 

\subsubsection{Firing-rate error}
Per-unit absolute error of the mean firing rate in Hz,
$|\bar{r}_{x} - \bar{r}_{\hat{x}}|$. A rate-matched Poisson tokenizer matches this metric by
construction (its error is only sampling noise), so it is the one reconstruction metric whose
normalization does not use the Poisson floor (Appendix~\ref{sec:metrics-details}).

\subsubsection{Binned rate difference}
Per-unit mean absolute difference between binned rates (Hz) of $x$ and $\hat{x}$ at a configurable bin width (default: average over scores at $1, 5, 10, 25, 50$~ms). Used in place of spike-time--based metrics for population tokenizers whose outputs do not preserve sub-bin timing; it provides a count-level fidelity score that does not require spike matching. Range $[0, \infty)$, lower is better.

\subsubsection{Smoothed squared error}
We compute the normalized summed squared error between Gaussian-smoothed traces of $x$ and $\hat{x}$, with default kernel widths of $\sigma \in 1, 5, 10, 50, 100$~ms. We average over scores at different smoothing kernel widths.  %(the implementation also exposes a causal exponential kernel, recovering the
%Van Rossum spike-train distance up to   scale,~\citealp{vanrossum2001novel}). 
The smoothing kernel is $L_{2}$-normalized so that a single missed or
spurious spike contributes $\sim\!1$ unit of error; per-unit values are
divided by the original spike count to give a normalized ``fraction of
mis-localized spikes''. Up to scale, this metric relates to the Van Rossum spike-train distance~\citep{vanrossum2001novel}, where we use Gaussian instead of exponential smoothing.  Range $[0, \infty)$, lower is better.

\subsubsection{Spike count correlation}
To quantify the preservation of temporal fluctuations in spiking activity, we compute the Pearson correlation between binned spike-count
sequences. 
Pearson correlation between $x$ and $\hat{x}$ binned at multiple timescales ($\{1, 5, 10, 25, 50\}$~ms), per unit. We average over scores at the different timescales. Range $[-1, 1]$, higher is better.

\subsubsection{Mutual information}
Normalized mutual information between binned spike counts of $x$ and
$\hat{x}$ at multiple timescales (default bin sizes
$\{1, 5, 10, 25, 50\}$~ms). The joint distribution is built as a contingency table over integer counts, and the plug-in MI is corrected with the Miller--Madow estimator~\citep{miller1955note}. The score is entropy-normalized as
\begin{equation}
    \mathrm{NMI} \;=\; \frac{\mathrm{MI}(x, \hat{x})}{H(x)} \;\in\; [0, 1],
\end{equation}
per bin size, and we average over scores at the different timescales. The estimator is marked unreliable
(NaN) when the sample-to-occupied-cell ratio of the contingency table
falls below five, indicating the bias correction itself has high variance.

\subsubsection{ISI divergence}
$1$-Wasserstein distance between the empirical inter-spike-interval (ISI) distributions of $x$ and $\hat{x}$, per unit. Range $[0, \infty)$, lower is better. Requires fine-resolution spike data.

\subsubsection{ISI CV error}
Absolute error in the global coefficient of variation
$\mathrm{CV} = \mathrm{std}(\mathrm{ISI})/\mathrm{mean}(\mathrm{ISI})$:
$|\mathrm{CV}_{x} - \mathrm{CV}_{\hat{x}}|$, per unit.

\subsubsection{ISI CV$_{2}$ error}
Absolute error in the local irregularity statistic
\begin{equation}
    \mathrm{CV}_{2}
    \;=\;
    \frac{1}{N - 1}
    \sum_{i=1}^{N-1}
    \frac{2\,|\mathrm{ISI}_{i+1} - \mathrm{ISI}_{i}|}
         {\mathrm{ISI}_{i+1} + \mathrm{ISI}_{i}},
\end{equation}
per unit. Unlike the global CV, $\mathrm{CV}_{2}$ is insensitive to slow rate fluctuations and isolates fine-grained ISI irregularity.

\subsubsection{Fano factor error}

This is the per-unit absolute error of the Fano factor $F = \mathrm{Var}(c) / \mathbb{E}[c]$ of binned spike counts $c$ at a window size of $ 1, 5, 10, 25, 50$~ms.  We average over scores at different window sizes. $F = 1$ for a Poisson process, $F < 1$ for sub-Poisson regularity, $F > 1$ for over-dispersed or bursty firing; the metric quantifies whether the reconstruction preserves this variability regime.

%% Single-unit statistics:
\subsubsection{Autocorrelogram (ACG) correlation}

Pearson correlation between the per-unit autocorrelograms of $x$ and
$\hat{x}$, computed by discrete cross-correlation of binned counts at $1$~ms, $5$~ms and $10$~ms resolution, out to $\pm 100$~ms $\times$ its bin size ($\pm 100, \pm 500, \pm1000$~ms each). Range $[-1, 1]$, higher is better.

%%%
\subsection{Population firing statistics}
These metrics compare $\hat{x}$ to $x$ at the level of joint activity
across neurons. A tokenizer that preserves marginals while degrading
cross-neuron timing receives a low score here, even when single-neuron
metrics look good.

\subsubsection{Subspace overlap}

Chordal overlap of the principal angles between the top-$K$ PCA subspaces
of $x$ and $\hat{x}$ (default $K = 20$, population bin $1, 5$ and $10$~ms, we average over scores at different bin sizes). For
PCA bases $V_{1}, V_{2} \in \mathbb{R}^{N \times K}$ the principal angles
follow from the SVD $V_{1}^{\top} V_{2} = U \Sigma V^{\top}$ as
$\theta_{i} = \arccos \sigma_{i}$, and the metric is
$1 - \sqrt{\tfrac{1}{K}\sum_{i} \sin^{2} \theta_{i}}$ (in place of the mean cosine $\tfrac{1}{K}\sum_{i} \cos \theta_{i}$, which saturates near perfect overlap)~\citep{bjorck1973angles}. Range
$[0, 1]$, higher is better.

\subsubsection{Cross-correlogram (CCG) correlation}
Pearson correlation between original and reconstructed pairwise
cross-correlograms, for a sampled set of neuron pairs ($50$ pairs), computed in $1, 5$ and $10$~ms bins with a $\pm 100 \times$ bin size lag window. We average the per-pair CCG correlation per session. We then average over sessions, and then over scores at different bin sizes.  The range is $[-1, 1]$, higher is better.

\subsubsection{Functional-connectivity F1}

This metric is also based on the CCG between the units of the original spiking data. A unit pair is labeled \emph{functionally connected} if its CCG contains a peak with $|z|$-score above a threshold (we use the value $3$). The metric is the F1 score between the connected-pair sets identified from $x$ and from $\hat{x}$. Population-level scalar in $[0, 1]$, higher is better.

\subsubsection{Population coupling preservation}

Following \citep{okun2015populationcoupling}, the population coupling of neuron $u$ is its Pearson correlation with the population-summed activity excluding itself. For each session this yields a coupling vector $c \in \mathbb{R}^{U}$ from $x$ and $\hat{c}$ from $\hat{x}$; the metric is the concordance correlation coefficient between $c$ and $\hat{c}$. We compute this for spikes binned at $1$ and $5$~ms resolution, averaging scores over both resolutions. Range $[-1, 1]$, higher is better.

\subsubsection{Participation ratio error}

Absolute error of the participation ratio of the population covariance,
\begin{equation}
    \mathrm{PR}
    \;=\;
    \frac{(\sum_{i} \lambda_{i})^{2}}{\sum_{i} \lambda_{i}^{2}},
\end{equation}
where $\{\lambda_{i}\}$ are the eigenvalues of the population activity covariance at a population bin width of $1, 5$ and $10$~ms by default. We average over scores at the three bin sizes, after averaging over sessions. PR is a continuous, scale-invariant estimate of the effective linear dimensionality of the population  code~\citep{Gao214262}.

\subsection{Downstream decoding}
Details about the downstream decoding tasks and metrics are in Appendix~\ref{app:downstream-decoding}. Code released at~\href{https://github.com/enigma-brain/ntt-downstream}{\texttt{ntt-downstream}} (decoding entry point: \texttt{src/ntt\_downstream/decoding/}).

\subsection{Downstream encoding}
Details about the downstream encoding task and metrics are in Appendix~\ref{app:downstream-encoding}. Code released at~\href{https://github.com/enigma-brain/ntt-downstream}{\texttt{ntt-downstream}} (encoding entry point: \texttt{src/ntt\_downstream/encoding/}).

\subsection{Compression}
\textbf{Compression rate.} We measure the compression rate as the ratio of input bits to output bits. Input bits are calculated as $NT\log_2(\Delta+1)$, where $N\times T$ is the tokenizer's fixed ``patch size'' (with $N$ denoting the number of neurons and $T$ denoting the number of time bins in a single patch), and $\Delta$ is the bin size in ms (we assume a maximum of 1 spike/ms). For discrete tokenizers, output bits are calculated as $\log_2(V)$ for a single patch, where $V$ is the tokenizer's vocabulary size; for continuous tokenizers, on the other hand, output bits are calculated as $16D$ (again for a single patch), where $D$ is the tokenizer's latent dimensionality (\textit{i.e.}~we assume that latents are 16-bit vectors, \textit{e.g.}~\texttt{bf16} vectors). Compression rates for a number of selected tokenizers are shown in Table~\ref{tab:main_results} (rightmost column) in the main text. For comparison, we estimate the near-optimal lossless compression rate for a variable-length tokenizer using \textit{arithmetic coding} \citep[Ch. 6]{mackay2003information} to be $\sim 2.1\times$ in the same setting as the single-neuron tokenizers in Table~\ref{tab:main_results}.

\subsection{Generalization Retention Metric} %Or just "generalization"?
\label{app:generalization-retention}

We quantify out-of-distribution generalization by measuring how much evaluation
performance is retained when moving from the within-session temporal split to
held-out splits that require progressively stronger transfer. The procedure
takes as input the per-session reports produced by the evaluation
pipeline. Each report contains one scalar value per metric, together with the
metric direction, indicating whether higher or lower values are better.

\paragraph{Tier assignment.}
Each per-session report is assigned to one of four evaluation tiers using the
dataset selections in the evaluation configuration. The within-session temporal
split, denoted $xT$, is used as the in-distribution reference tier.
The held-out tiers are:
\begin{itemize}
    \item $xS$: held-out sessions from subjects present during
    training;
    \item $xSub$: held-out subjects from sources present during
    training;
    \item $xSrc$: held-out data sources absent from training.
\end{itemize}

\paragraph{Metrics.}
The score uses the reconstruction metrics only (single-unit statistics and population structure), which every dataset provides. Latent-quality
metrics are excluded: they are not reconstruction metrics but candidate proxies for downstream utility (desideratum 2; Section~\ref{sec:desiderata}).
%they are signed RSA correlations near zero, so their ratio to a baseline is close to a coin flip. 
Downstream metrics are excluded because 
encoding and decoding train downstream networks on part of the held out data at evaluation time. %Also, decoding and encoding are only evaluated for datasets with behavior and repeats, respectively, and would weigh the score towards them.
%decoding exists only for datasets with behavior and would weight the score
%towards them.

\paragraph{Direction-aware, dataset-matched retention.}
Let $x_{m,s}$ be the value of metric $m$ on held-out session $s$, and let
$b_m(s)$ be the mean of $m$ over the $xT$ sessions of the same dataset
as $s$. If metric $m$ is higher-is-better, retention is
\begin{equation}
    r_{m,s} = \frac{x_{m,s}}{b_m(s)};
\end{equation}
if it is lower-is-better, retention is $r_{m,s} = b_m(s)/x_{m,s}$. An
$xSrc$ session has no $xT$ session from the same dataset, so
its baseline $b_m(s)$ is the mean over all $xT$ sessions. A session is
left out when its baseline is not positive. Comparing each session with its own
dataset keeps retention independent of which datasets a tier happens to
contain: per-session metric values scale with firing rate, which differs
several-fold between datasets. We clip retention to the interval $[0,1]$,
\begin{equation}
    \tilde{r}_{m,s} = \min\left(1, \max(0, r_{m,s})\right),
\end{equation}
which treats improvements over the in-distribution reference as full retention
and prevents them from compensating for degradation on other metrics. The tier
term for metric $m$ is the mean over the tier's sessions,
\begin{equation}
    \tilde{r}_{m,t}
    =
    \frac{1}{|\mathcal{S}_t(m)|}
    \sum_{s \in \mathcal{S}_t(m)} \tilde{r}_{m,s},
\end{equation}
where $\mathcal{S}_t(m)$ is the set of sessions in tier $t$ for which metric
$m$ is available.

\paragraph{Score.}
The generalization score is the geometric mean over the cross-source terms,
\begin{equation}
    G
    =
    \exp\left(
        \frac{1}{|\mathcal{M}|}
        \sum_{m \in \mathcal{M}}
        \log \max(\tilde{r}_{m,xSrc}, \epsilon)
    \right),
\end{equation}
where $\mathcal{M}$ is the set of reconstruction metrics available for the
tokenizer and $\epsilon=10^{-3}$. The floor $\epsilon$ keeps a single
exactly-zero term from collapsing the geometric mean to zero, while still
penalizing that term strongly. The geometric mean is used because retention is
multiplicative: a value of $0.5$ means that only half of the reference
performance is retained, regardless of the metric's original units. It also
penalizes severe degradation on any one metric more strongly than the
arithmetic mean. $G$ lies in $[\epsilon,1]$; a value near one indicates that
the tokenizer retains its in-distribution reconstruction quality on datasets
unseen in training.

\paragraph{Why only the cross-source tier.}
We still report the $xS$ and $xSub$ retention per tier
and per metric, but only the cross-source tier enters the score, because it
matches how a tokenizer is meant to be used: trained once on a large corpus and
then applied to new datasets that it has most likely never seen, and whose
recording setups, brain areas and firing statistics it has to accommodate.

\subsection{Robustness and signal fidelity of tokens}

\subsubsection{Signal-to-noise ratio (SNR) based on stimulus repeats}
Given datasets with repeated stimuli and associated sequences of latents $(z^t_{i,r})_{t=1...T}$, we can define signal-driven variance as the variance across stimulus means:
\[
\sigma_{\mathrm{sig,raw}}^2
=
\frac{1}{T}\sum_{t=1}^T
\operatorname{Var}_{i}\!\left(\bar z^t_i\right), 
\]
where $\bar z^t_i = \frac{1}{R}\sum_{r=1}^R z^t_{i,r}$
is the mean token sequence across repeats.
Similarly, the noise variance is the average within-stimulus variance across repeats:
\[
\sigma_{\mathrm{noise}}^2
=
\frac{1}{S}\sum_{i=1}^S
\left[
\frac{1}{T}\sum_{t=1}^T
\operatorname{Var}\!\left(z^t_{i}\right)
\right].
\]
The signal-to-noise ratio is then
\[
\mathrm{SNR}
=
\frac{\sigma_{\mathrm{sig}}^2}{\sigma_{\mathrm{noise}}^2}.
\]
Because $\bar z^t_i$ is estimated from finitely many repeats, and is biased, we subtract the finite-repeat bias~\cite{sahani2002linear}
\[
\sigma_{\mathrm{sig}}^2
=
\max\!\left(
0,\,
\sigma_{\mathrm{sig,raw}}^2
-
\sigma_{\mathrm{noise}}^2
\cdot
\frac{1}{S}\sum_{i=1}^S \frac{1}{R_i}
\right).
\]
We use a single session, session PE\_201018 of the dataset by \citet{cowleyCompactDeepNeural2026}, where images were shown to monkeys for $100$ ms at a time. Each image was shown $11-12$ times. We chunk the neural data into chunks from stimulus start to at most 200ms past stimulus end, or until the start of the next stimulus, whichever is earlier. (Image presentations are interleaved by gaps of at least $100$ ms each.) We tokenize from the start up to the largest possible multiple of token bin size that fits into each stimulus window, drop outlier windows that are very small, and then cap per-window token sequences to the shortest remaining sequence length before comparing them.

\subsubsection{Input-corruption robustness}
Although we argue for single-millisecond precision in our evaluations because the true nature of the neural code is presently not known, it is still desirable for tokens to be smooth under small input perturbations, as some level of neural variability is expected~\cite{churchland2010stimulus, azouz1999cellular}. As corruptions, we use tokenizer window shifts of up to 3 ms in either direction, jittering 20\% of spikes randomly by up to 1ms, and adding and removing 2\% of spikes randomly.
The score is the normalized squared distance between the latent corresponding to uncorrupted vs corrupted input, averaged across 10 randomly sampled corruptions per type, and across the first 5 minutes of each of four test sessions, one each from IBL (\texttt{ibl\_witten\_17\_20200723\_ibl\_witten\_17}), the C4 cerebellum dataset (\texttt{hausser\_18-08-30\_YC001\_probe1}), the dataset by Cowley et al.\ (\texttt{PE\_201014}) and the dataset by Perich, Miller et al.\ from 2018 (\texttt{t\_20130830\_random\_target\_reaching}). We normalize by the token variance across the entire evaluated part of the session. 

\paragraph{Normalization} Normalization of mean squared Euclidean token distances is session-specific. % TODO Formula
Let $z_0$ be a clean latent sequence and $z_c$ the latent sequence obtained
from a corrupted input. The jointly normalized input-corruption robustness score is
\[
\mathrm{input\_corruption}
=
\frac{
\frac{1}{W}\sum_{w=1}^{W}
\left(
\frac{1}{T_w}\sum_{t=1}^{T_w}
\|z^{t}_{w,c} - z^{t}_{w,0}\|_2^2
\right)
}{
\sigma_{\mathrm{joint}}^2
},
\]
where $W$ is the number of evaluated windows, $T_w$ is the number of latent
tokens in window $w$, and $\sigma_{\mathrm{joint}}^2$ is the variance of clean
tokens computed jointly over all clean tokens from all evaluated windows
and all latent dimensions:
\[
\sigma_{\mathrm{joint}}^2
=
\frac{1}{\sum_{w=1}^{W} T_w}
\sum_{w=1}^{W}\sum_{t=1}^{T_w}
\left\| z_{w,0}^t - \bar{z}_0 \right\|_2^2.
\]
% \[
% \mathrm{input\_corruption}
% =
% \frac{
% \frac{1}{T}\sum_{t=1}^T \|z_c^t - z_0^t\|_2^2
% }{
% \sigma_{\mathrm{joint}}^2
% },
% \]
% where $\sigma_{\mathrm{joint}}^2$ is the variance of clean latent values
% computed jointly over all clean tokens and all latent dimensions.

\subsubsection{Latent-space corruption robustness}
We evaluate how strongly small changes in latent space affect reconstruction by perturbing latents (pre-discretization for discrete tokenizers) by a small fraction of their per-session standard deviation. We apply Gaussian noise with standard deviation $\sigma_{i,corr} = 0.1 \times \sigma_{i,tok}$, where $\sigma_{i,tok}$ is the standard deviation of the $i$-th token dimension, computed across the session. Reconstruction perturbation can be measured by any reconstruction metric; for simplicity, we choose to report the ACG correlation between unperturbed and perturbed reconstruction. We compute ACGs for up to 300 1\,s reconstructed windows and average them before computing the correlation between uncorrupted and corrupted reconstruction. % todo per window and averaged? It should be. - EDIT: No it was not.... so another round of removing M2 scores. Up to May 5th, 18:18 german time.
We use the same four sessions as in input-corruption robustness.

\subsubsection{Latent-Space Structure Preservation}
\label{app:eval-metrics-rsa}

Two complementary metrics quantify whether the relational geometry of the
input data survives encoding into the tokenizer's latent space. Both are
based on representational similarity analysis~\citep{kriegeskorte2008rsa}:
We construct a representational dissimilarity matrix (RDM) in the input
space and one in the latent space, and report the rank correlation between
their upper triangles.

\paragraph{RDM construction}
Given a matrix $X \in \mathbb{R}^{N \times D}$ whose rows are entities
(time windows or neurons, depending on the variant), the RDM
$M \in \mathbb{R}^{N \times N}$ is the symmetric matrix of pairwise
distances $M_{ij} = d(X_{i,:}, X_{j,:})$ under a chosen metric $d$.
We compute one RDM in the input space, $M_{x}$, and one in the latent
space, $M_{z}$, and score their structural agreement by the Spearman rank
correlation between the upper triangles,
\begin{equation}
    \mathrm{RSA}
    \;=\;
    \mathrm{corr}_{\text{Spearman}}\!\bigl(\,
        \mathrm{vec}_{i<j}(M_{x}),\;
        \mathrm{vec}_{i<j}(M_{z})
    \,\bigr)
    \;\in\; [-1, 1].
\end{equation}
The metric is rank-based by design, so it is invariant to monotonic
rescaling of either RDM and does not require the input and latent metrics
to be on the same scale. Defaults: input metric is correlation distance
$d_{x}(u, v) = 1 - \mathrm{corr}(u, v)$, latent metric is Euclidean,
comparison method is Spearman. Direction: higher is better, range
$[-1, 1]$.

\paragraph{Single-neuron RSA}

Each row of $X$ corresponds to one neuron: in the input space, that
neuron's activity across all windows and within-window time bins is
flattened into a single vector, and analogously for the latent. The RDM
is therefore $N \times N$, where $N$ is the number of neurons. This
variant aims to answer: \emph{do neurons whose temporal activity patterns were
similar in input space remain similar in the latent space?} It applies
only to per-neuron tokenizers (e.g.\ the compressed-sensing tokenizer in
\texttt{per\_neuron} mode and the CalM-style (Transformer-VQ) tokenizer, whose
tokens stay per neuron despite its cross-neuron attention layers), where the latent has a distinguishable
neuron axis with shape $[N_{w}, N, m]$. For population tokenizers whose
latent does not factor across neurons, the metric returns NaN.

\paragraph{Population RSA}

Each row of $X$ is a single time window: in the input space, the spike
counts of all neurons in that window are concatenated into one vector; in
the latent space, the per-window latent is flattened into one vector. The
RDM is therefore $N_{w} \times N_{w}$, where $N_{w}$ is the number of
windows. This variant aims at answering: \emph{do windows that were similar across the
whole population in input space remain similar in the latent space?} It
applies to any tokenizer that produces a per-window latent and is the
primary RSA metric reported in the main text.

\paragraph{Smoothing of the input}

At the $1$~ms resolution used in this paper, single-neuron spike trains
are very sparse and their pairwise distances under any of the supported
metrics are dominated by the shared zeros. To avoid degenerate, nearly
uniform input RDMs, the input matrix is Gaussian-smoothed along the time
axis with $\sigma = 25$ bins (equivalent to $25$~ms at $1$~kHz) before the
RDM is computed; for the single-neuron variant the smoothing is applied
within each neuron independently, and for the population variant it is
applied within each neuron's segment so as not to bleed across neuron
boundaries. The latent RDM is computed on the raw latents without
smoothing.

\paragraph{Subsampling and reliability}
Both RDMs are $O(N^{2})$ in storage and computation, so when the number
of entities exceeds $N_{\max} = 200$ a fixed-seed random subset of size
$N_{\max}$ is drawn before the RDM is built. Sessions with fewer than
$10$ windows return NaN. The pipeline also returns the underlying RDMs
and a $k$-nearest-neighbor neighborhood-overlap curve as metadata, which
we use for diagnostic plots but not for scalar scoring.

\section{\texorpdfstring{Table~\ref{tab:main_results}}{Table} Metrics Details}
\label{sec:metrics-details}

This section describes how each column in Table~\ref{tab:main_results} is computed from the underlying evaluation results.

\paragraph{Poisson normalization (SN Recon, Pop.\ Stats).}
Both reconstruction columns use a shared normalization scheme.
For each scalar metric $m$, we compute a Poisson-normalized score
\begin{equation}
    \tilde{m} = \frac{m_\text{model} - m_\text{Poisson}}{m_\text{perfect} - m_\text{Poisson}},
\end{equation}
where $m_\text{Poisson}$ is the value obtained by a rate-matched Poisson baseline tokenizer (firing rates estimated at 10\,ms resolution, new spikes drawn from a Poisson process with those rates; see Appendix~\ref{app:poisson-baseline}) and $m_\text{perfect}$ is the value under perfect reconstruction.
By construction, $\tilde{m}=0$ corresponds to Poisson-level performance and $\tilde{m}=1$ to perfect reconstruction; values below zero indicate worse-than-Poisson performance.
Firing-rate error is the one exception: the Poisson baseline keeps each neuron's mean rate, so its error is only sampling noise and the ratio above would be dominated by it. For this metric the floor is a reference that gives every unit the mean rate of its population, $m_\text{ref} = \frac{1}{N}\sum_{u} |\bar{r}_u - \frac{1}{N}\sum_{u'} \bar{r}_{u'}|$, aggregated over segments and sessions like the model's error, so that $\tilde{m} = 1 - m_\text{model}/m_\text{ref}$.
Because Poisson normalization is unbounded below for lower-is-better metrics whose floor is near zero, a single pathologically-bad metric can dominate the arithmetic mean.
We therefore clip each per-metric normalized score at $-1$ before averaging, bounding the influence of any single metric to at most one unit of negative contribution.

\paragraph{SN Recon.}
The \emph{SN Recon} score is the family-weighted mean of the normalized scores above (Poisson-normalized, except firing-rate error) over the single-neuron metrics: spike detection (F1, strict F1, temporal precision), spike-count statistics (firing rate error, Fano factor error at 1--50\,ms, ISI CV, ISI CV$_2$, ISI divergence), temporal fidelity (spike-count correlation and mutual information at 1--50\,ms, smoothed squared error at 1--100\,ms, binned rate difference) and ACG correlation (1, 5, 10\,ms).
Each metric family has weight one, whatever its number of bin sizes. Spike-detection and ISI metrics need spike times, so they are omitted for tokenizers with rate outputs.

\paragraph{Pop.\ Stats.}
The \emph{Pop.\ Stats} score is the family-weighted mean of Poisson-normalized scores over the population-level metrics: CCG correlation (1, 5, 10\,ms), functional-connectivity F1, participation ratio error (1, 5, 10\,ms), population coupling preservation (1, 5\,ms, where available) and subspace overlap (1, 5, 10\,ms).

\paragraph{Decoding.}
The \emph{POYO} cell is the mean cursor-velocity $R^2$ on the held-out test split of 12 Perich--Miller sessions (six from monkey C, Chewie, and six from monkey T; center-out and random-target reaching). For each session, a POYO-style~\cite{azabou2023poyo} transformer is trained on the frozen tokenizer's latent representations (Appendix~\ref{app:downstream-decoding-poyo}); tokenizers whose encoders read neurons by position (LDNS and the Wavelet tokenizers) receive every unit of the session (Appendix~\ref{app:poyo-tokenizer-zoo}). The \emph{Linear} cell is the mean $R^2$ of the ridge probes (Appendix~\ref{app:downstream-decoding-linear-probes}) for cursor velocity and cursor position on the Perich--Miller sessions of the reconstruction evaluation set (monkeys C and J); for the $1000:1$ tokenizers only velocity enters, because their latents cannot be grouped into 200\,ms segments. The \emph{composite} is the mean of the POYO and Linear cells, weighted equally. For the $1000:1$ tokenizers each 1\,s decoding window holds a single latent token, so their values ($^\dagger$) are not directly comparable.

\paragraph{Encoding.}
We report the mean noise-corrected $R^2$ ($R^2_{\mathrm{NC}}$) of the image-to-response encoding model (Appendix~\ref{app:downstream-encoding}).
It measures how well responses to held-out images, predicted in the tokenizer's latent space and decoded by its frozen decoder, match the recorded responses.

\paragraph{Latent Robustness.}
The column averages up to four members, each mapped to $[0,1]$: (i) population RSA and (ii) single-neuron RSA (per-neuron tokenizers only), each as $\max(\rho,0)$ (Appendix~\ref{app:eval-metrics-rsa}); (iii) input-corruption retention
\begin{equation}
    s = \frac{1}{1 + d},
\end{equation}
where $d$ is the mean normalized latent distance over three corruptions (window shifts of up to 3\,ms, 1\,ms jitter of 20\% of spikes, adding and removing 2\% of spikes), so that $s=1$ means no degradation; and (iv) latent-noise robustness, the ACG correlation between reconstructions from clean and perturbed latents, floored at 0.

\paragraph{Generalization.}
We report the geometric mean, over reconstruction metrics, of the retention on the cross-source tier (datasets unseen in training).
Each held-out session is compared with the held-out-time sessions of its own dataset, or, for cross-source sessions, with the mean over all held-out-time sessions. The ratio is direction-aware and clipped to $[0,1]$, so a score of 1.0 indicates no degradation.
Cross-session and cross-subject retention are reported separately but not scored (Appendix~\ref{app:generalization-retention}).

\paragraph{Compression.}
We report the ratio of input bits to output bits, computed as
\begin{equation}
    \text{compression} = \frac{B_\text{in}}{B_\text{out}},
\end{equation}
where $B_\text{in}$ is the number of bits required to represent the raw spike counts in the input window (at 1\,ms resolution) and $B_\text{out}$ is the number of bits in the tokenizer's output (16 bits per latent dimension for continuous latents, $\log_2 V$ per discrete token for a codebook of size $V$; Appendix~\ref{app:metric-details}).
A ratio greater than 1 indicates compression; ratios below 1 indicate expansion.

\section{Public dataset details}
\label{app:public-datasets}

Here we provide additional public dataset details, expounding upon those introduced in Table~\ref{tab:datasets}. All public datasets are converted from their source formats into standardized representations via brainsets~\citep{azabou2023poyo} data-preparation code released alongside the paper (\url{https://github.com/enigma-brain/ntt}). Each dataset is used under its original license. 

\begin{figure}[h]
  \centering
  \includegraphics[width=\linewidth]{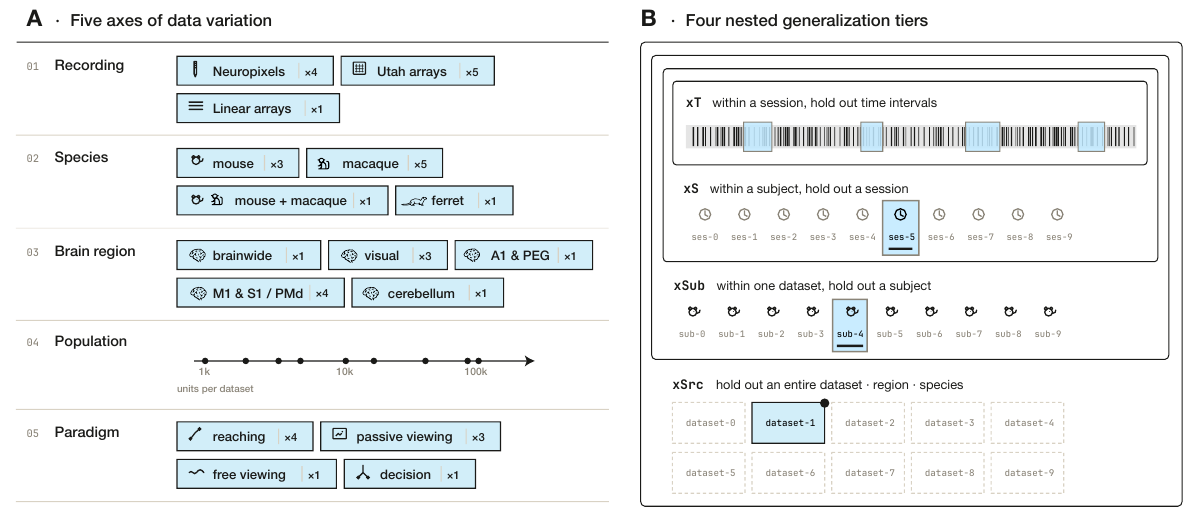}
  \caption{\textbf{Dataset suite at a glance.} \emph{(A)}~Suite breadth across the five axes of variation; counts ($\times n$) tally how many of the public datasets fall in each category, and population size is shown on a log axis. \emph{(B)}~The four nested held-out tiers used by the evaluation framework ($xT$, $xS$, $xSub$, $xSrc$); per-dataset values for each tier appear in the lower rows of Table~\ref{tab:datasets}.}   
  \label{fig:dataset-suite}
\end{figure}

\begin{figure}[h]
  \centering
  \includegraphics[width=\linewidth]{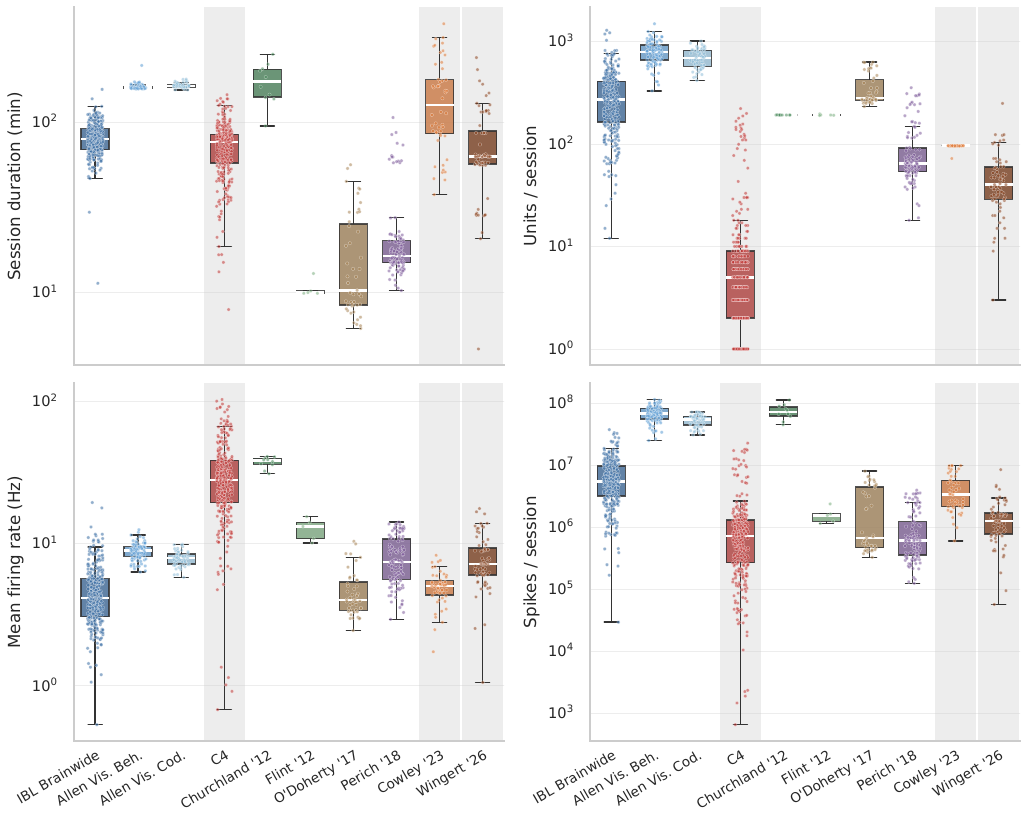}
  \caption{\textbf{Per-dataset session statistics.} Distributions over per-session values for four properties: session duration (min), units recorded per session, mean per-neuron firing rate (Hz), and total spikes per session. Boxes span the interquartile range with the median as a white line; jittered dots show individual sessions. All vertical axes are log-scaled. Light-gray shading marks the three $xSrc$ datasets (C4, Cowley, Wingert) that are held entirely out of training. Across the corpus, datasets span 1--3 orders of magnitude on each property, motivating cross-dataset evaluation rather than tuning to any single recording.}
  \label{fig:session_stats_boxstrip}
\end{figure}

\begin{figure}[h]
  \centering
  \includegraphics[width=\linewidth]{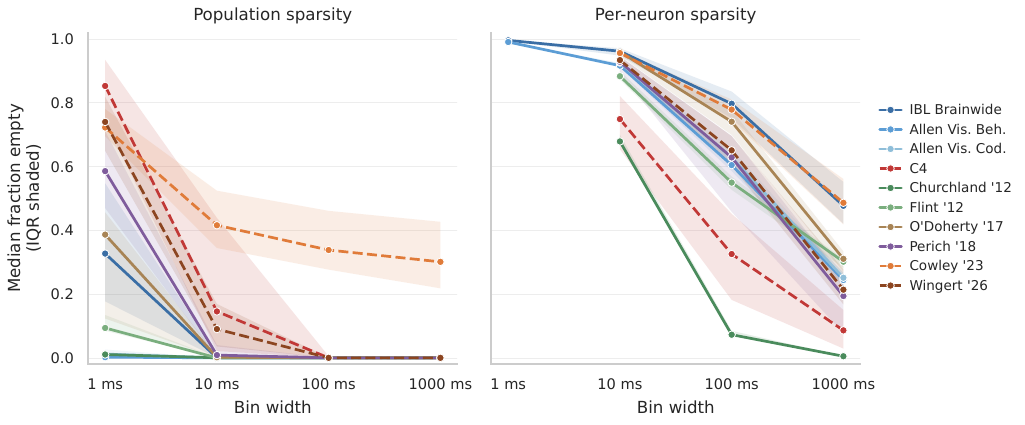}
  \caption{\textbf{Per-dataset spike-grid sparsity vs.\ binning resolution.} Median fraction of empty bins as a function of bin width, on a log axis (1, 10, 100, 1000\,ms). \emph{Left:} \emph{population-level} sparsity --- the fraction of bins in which the entire recorded population is silent (relevant for window-aggregated tokenizers). \emph{Right:} \emph{per-neuron} sparsity --- the fraction of (neuron, bin) cells with zero spikes (relevant for per-neuron tokenizers). Lines show per-dataset medians across sessions; shaded bands show the IQR. $xSrc$ datasets (C4, Cowley, Wingert) are drawn dashed. Two patterns are visible: (i)~at fine resolution (1\,ms) population sparsity ranges from $<1\%$ in dense visual-cortex recordings (Allen Vis.\ Beh., Allen Vis.\ Cod.) to $>85\%$ in cerebellum (C4); and (ii)~the rate at which sparsity decays with coarser binning is itself dataset-dependent, with only the V4 population (Cowley) remaining sparse at the population level even at 1\,s windows, while the cerebellar population (C4) is rarely silent for bins of 100\,ms or wider. The decay slope is the central trade-off any spike tokenizer must navigate.}
  \label{fig:sparsity_ladder}
\end{figure}

Figure~\ref{fig:dataset-suite} sketches the suite's coverage along the five qualitative axes of variation and the four nested generalization tiers, and Figures~\ref{fig:session_stats_boxstrip} and~\ref{fig:sparsity_ladder} characterize the per-session quantitative spread that motivates evaluating tokenizers across varied datasets.

\FloatBarrier
\newpage

\section{Tokenizer and training details}
\label{app:tokenizer-details}
\subsection{Single-neuron tokenizers}

Single-neuron tokenizers take as input a \textit{1-D temporal sequence of spiking activity from a single unit}. In this work, we restrict the design space here to binary spike bins with 1\,ms resolution, where the input is a binary sequence $\mathbf{x} \in \{0,1\}^{l_w}$, with window length $l_w$ in 1\,ms bins. We also consider two temporal downsampling ratios, $100:1$ and $1000:1$ (input bins per token), where we produce a single token for every 100\,ms and 1000\,ms of input, respectively. We explored two architecture families in the single-neuron setting, \textit{convolutional} and \textit{transformer-based}, along with a variety of bottlenecks and losses.

\paragraph{Convolutional} In our convolutional architecture an encoder composed of 1-D residual blocks~\citep{he2016deep} with strided convolutions compresses input bins to $l_z$ features (where $\frac{l_w}{l_z}$ is the temporal downsampling ratio). This is passed to a bottleneck producing either continuous or quantized latents, $\mathbf{z} \in \mathbb{R}^{d_z}$ (the latter have associated token IDs). A symmetric decoder reconstructs logits $\hat{\mathbf{x}} \in \mathbb{R}^{l_w}$ at the original resolution, and a sigmoid maps these to spiking probabilities in $[0,1]$. Appendix~\ref{app:wanneural} describes the implementation (WanNeural) behind the Conv-VAE and Conv-RFSQ rows of Table~\ref{tab:main_results}.

\paragraph{Transformer-based} The transformer-based tokenizer, inspired by recent work with calcium imaging data, CalM~\citep{xuSelfSupervisedFoundationModel2026}, uses a causal transformer encoder with patch-based temporal tokenization. The input is partitioned into $l_z = l_w / p$ non-overlapping patches of $p$ consecutive bins; each patch is linearly projected to a $d_z$-dimensional embedding before encoding, yielding latents $\mathbf{z} \in \mathbb{R}^{l_z \times d_z}$. We explored only discrete bottlenecks for this architecture. The encoder can optionally append attention layers across neurons after the per-neuron temporal stack; with zero such layers it is a purely per-neuron tokenizer. The Transformer-VQ model in Table~\ref{tab:main_results} uses two, so its tokens also depend on the other recorded units.

\paragraph{Bottlenecks.}
The bottleneck is the primary site at which structure is imposed on the latent representation; a tokenizer that minimizes reconstruction loss alone need not produce latents that are useful downstream. We explored one continuous and five discrete bottleneck families. The standard VAE bottleneck uses a diagonal Gaussian posterior with KL regularization, optionally augmented with additional regularization, for example a straightening penalty~\cite{wang2026temporal} or SIGReg~\cite{balestriero2025lejepa}. For discrete latents, we explored a variety of vector quantization techniques, including standard vector quantization (VQ) \cite{vandenoord2017vqvae}, finite scalar (FSQ) \cite{mentzer2023finite}, binary spherical (BSQ) \cite{zhao2024image}, label-free (LFQ) \cite{yu2023language}, and residual vector (RVQ) \cite{zeghidour2021soundstream}.

\paragraph{Training losses.}
Because spikes are sparse binary events, the reconstruction loss choice is consequential. We explored binary cross-entropy and focal loss~\citep{lin2017focal} for binary spike outputs; Poisson NLL for rate-valued reconstructions (where the decoder predicts expected counts); and MSE, Huber, L1, and Wasserstein losses as alternatives. These can be combined arbitrarily via a multi-loss wrapper with per-component weights and optional target transforms---for instance, pairing a spike-level BCE with a smoothed-rate MSE. An auxiliary latent regularizer \citep[e.g.\ SIGReg][]{balestriero2025lejepa}, can be added to push the aggregate latent distribution toward an isotropic Gaussian without modifying the reconstruction target.

\paragraph{Non-learned baselines.}
We include two non-learned baselines to anchor the zoo: a PCA tokenizer (Appendix~\ref{app:pca-baseline}) and a compressed sensing (CS) tokenizer (Appendix~\ref{app:cs-tokenizer}). Both are training-free and serve as a floor against which learned tokenizers can be measured; the CS baseline in particular demonstrates empirically that high reconstruction scores do not imply useful latent representations.

\subsection{Population tokenizers}

Population tokenizers encode the joint activity of all recorded neurons, giving them direct access to pairwise correlations and shared latent dynamics at the cost of requiring mechanisms for variable-size neuron sets across heterogeneous recordings. Table~\ref{tab:main_results} contains four population tokenizers, two of which are our constructions: LDNS-Perceiver (Appendix~\ref{app:perceiver-ldns}), a continuous-latent tokenizer operating at 10\,ms bins that extends the LDNS autoencoder~\citep{kapoor2024ldns} to variable neural populations via Perceiver-style cross-attention; and Wavelet-FSQ (Appendix~\ref{app:cosmos-2d}), a discrete tokenizer adapted from the Cosmos video tokenizer~\citep{agarwal2025cosmos} that operates at 1\,ms resolution and models joint activity across a fixed population of neurons; Wavelet-AE is its continuous counterpart. The fourth, LDNS, is a fixed-population baseline.

\paragraph{LDNS.}
The LDNS baseline is the autoencoder of LDNS~\citep{kapoor2024ldns} without its diffusion stage, applied to a fixed population of $N = 128$ neurons. The input is a 2.5\,s window of spike counts in 10\,ms bins. An S4-based encoder (6 blocks, hidden width 256) maps the neurons at each time bin to a 32-dimensional continuous latent, and a 2-block decoder predicts Poisson firing rates for each neuron. Training minimizes the Poisson negative log-likelihood with an $L_2$ and a temporal-difference penalty on the latents.

\paragraph{LDNS-Perceiver.}
Our main continuous-latent population tokenizer, LDNS-Perceiver, extends the autoencoder in LDNS~\citep{kapoor2024ldns} to variable neural populations. The input is a 4s window of binned spike counts with a 10\,ms bin size, together with precomputed per-neuron embeddings that serve as descriptors for each recorded unit. The encoder first mixes information across $N$ neurons and time using neuron-space attention and S4 temporal layers. A Perceiver-style \cite{jaegle2021perceiver} cross-attention bottleneck then compresses the variable-size recorded population into a fixed set of $M<N$ latent tokens at each time bin, each of fixed dimension $d$. The decoder performs the inverse operation using neuron-specific queries, initialized from the same precomputed neuron embeddings, and predicts Poisson firing rates for each unit. During training, a random subset of neurons is masked from the encoder, while the decoder reconstructs all neurons, encouraging the latent representations to capture population-level structure rather than simply copying observed units. For architectural and training details, see Appendix~\ref{app:perceiver-ldns}.

\paragraph{Wavelet-FSQ.}
Wavelet-FSQ is a discrete high-resolution population tokenizer, adapted from
the Cosmos video tokenizer~\citep{agarwal2025cosmos} to binned neural population
activity. The input is a $1$\,s window of $1$\,ms-binned spike counts for a
fixed population of $N = 32$ neurons, treated as a single-channel 2D image
of shape $(N, T)$ with $T = 1000$. A 2D wavelet patcher applies a
Daubechies-4 (\texttt{db4}) DWT with $\text{patch}_h = 4$ over neurons and
$\text{patch}_w = 2$ over time, and a three-stage 2D convolutional encoder
with channel widths $[64, 128, 128]$ further compresses the result to a
latent of shape $[T_\text{lat} = 500,\,N_\text{lat} = 1]$ that fully
collapses the neuron axis; convolutions are causal along the time axis. The
bottleneck is a finite-scalar quantizer
(FSQ)~\citep{mentzer2023finite} with two latent dimensions of 32 levels
each (1024 codes, 10 bits per latent position). The decoder
mirrors the encoder and is followed by an inverse DWT unpatcher; we train by
minimizing the Poisson negative log-likelihood between predicted and binned
per-neuron rates at $1$\,ms resolution. This places it
between the two main design extremes in our zoo: like single-neuron
tokenizers, it preserves millisecond-scale input resolution; like population
tokenizers, it models joint activity across simultaneously recorded units.
For architectural and training details, see Appendix~\ref{app:cosmos-2d}.

\subsection{Poisson Baseline Tokenizer}
\label{app:poisson-baseline}

\begin{table}[H]
\centering
\small
\begin{tabular}{ll}
\toprule
\textbf{Property} & \textbf{Value} \\
\midrule
Operates over        & per-neuron (independent across units) \\
Latent type          & continuous (firing rates in Hz) \\
Output type          & non-negative integer spikes (Poisson samples) \\
Window length        & variable \\
Neuron count         & variable \\
Trainable parameters & none \\
\bottomrule
\end{tabular}
\end{table}

The Poisson baseline tokenizer is the lower anchor used to normalize
reconstruction and latent-structure metrics. It is training-free and
deliberately simple: it preserves per-neuron firing rates over a
chosen rate-estimation window and discards millisecond
timing, cross-neuron correlations, and fine-grained population geometry.
We note that this is the implicit tokenizer chosen when spiking data is binned before being fed into a downstream model.

\paragraph{Encoding.}
Let $\mathbf{x} \in \mathbb{Z}_{\geq 0}^{T}$ be the binned spike train of a
single neuron at fine resolution $\Delta = 1~\mathrm{ms}$ (so
$T = W/\Delta$ for a window of duration $W$). The tokenizer chooses a
coarser rate-estimation bin $\Delta_{r}$ that is an integer multiple of
$\Delta$, partitions the window into $T' = \lfloor T/r \rfloor$ disjoint
chunks of $r = \Delta_{r}/\Delta$ fine bins each, and computes the empirical
firing rate per chunk:
\begin{equation}
    \hat{\lambda}_{t'}
    \;=\;
    \frac{1}{\Delta_{r}}
    \sum_{k=1}^{r} x_{(t'-1)r + k}
    \quad [\mathrm{Hz}],
    \qquad t' = 1, \ldots, T'.
\end{equation}
Each neuron is encoded independently, so a window with $N$ neurons yields a
latent of shape $[T', N]$ containing one rate per (rate-bin, neuron) pair.
There is no cross-neuron mixing.

\paragraph{Decoding and reconstruction.}
The encoded rates are expanded back to the fine grid by repeating each
$\hat{\lambda}_{t'}$ across the $r$ fine bins it covers and converting to an
expected count per fine bin:
\begin{equation}
    \mu_{t}
    \;=\;
    \hat{\lambda}_{\lceil t/r \rceil} \cdot \Delta.
\end{equation}
The reconstruction is then a single independent Poisson draw per fine bin,
\begin{equation}
    \tilde{x}_{t}
    \;\sim\;
    \mathrm{Poisson}(\mu_{t}),
\end{equation}
with a fixed seed for reproducibility. By construction, the expected total
count per rate-bin is matched to the observed total count, while the
within-rate-bin temporal pattern is replaced by an inhomogeneous Poisson
process with constant rate over each rate-bin.

\paragraph{Role as a normalization anchor.}
The single hyperparameter $\Delta_{r}$ trades off how much temporal structure
the floor retains. At $\Delta_{r} = \Delta$ the floor becomes a near-perfect
copy of the input; at $\Delta_{r} \gg W$ it collapses to a window-wide mean
rate. The intermediate setting of 10 ms used throughout the paper preserves slow rate modulation but destroys fine-timing information (ISI distributions, bursts). Reconstruction metrics (all but firing-rate error) are
normalized between this Poisson floor and a perfect-reconstruction ceiling
(when the metric admits one), so a normalized score of $0$ means a tokenizer
matches the floor and a normalized score of $1$ means it saturates the
ceiling. Because the floor is rate-matched per neuron, any normalized score
above zero is attributable to structure beyond per-neuron firing rates.

\FloatBarrier
\subsection{PCA Baseline Tokenizer}
\label{app:pca-baseline}

\begin{table}[H]
\centering
\small
\begin{tabular}{ll}
\toprule
\textbf{Property} & \textbf{Value} \\
\midrule
Operates over        & per-neuron (independent across units) \\
Latent type          & continuous \\
Output type          & non-negative firing rates (linear reconstruction clamped at a small floor) \\
Window length        & fixed ($T = 1000$ bins, $1$\,s at $1$\,ms) \\
Neuron count         & variable (single projection shared across units) \\
Trainable parameters & none (closed-form fit on running statistics) \\
\bottomrule
\end{tabular}
\end{table}

The PCA tokenizer is a closed-form, non-learned linear baseline. For each
neuron independently, a $1$-second window of $1$\,ms-binned activity is
projected onto the top-$k$ principal components of the empirical covariance
estimated from the training set. Running mean and outer-product statistics are
accumulated over a single pass through the training data; after the first
training epoch the projection matrix $\mathbf{W} \in \mathbb{R}^{T \times k}$
is set to the leading eigenvectors of the covariance via
\texttt{torch.linalg.eigh} and then frozen for the remainder of training and
evaluation. No gradient updates are performed against the tokenizer.

\paragraph{Signal model.}
At a bin width of $\Delta = 1$\,ms over a window of $W = 1$\,s, each neuron
contributes a non-negative integer count vector $\mathbf{x} \in
\mathbb{Z}_{\geq 0}^{T}$ with $T = W/\Delta = 1000$. Cross-neuron structure is
deliberately ignored: the same $\mathbf{W}$ is reused across all neurons in a
window and across the dataset, so the latent has shape $[N, k]$ for an input
of shape $[T, N]$ (one token per neuron per window, no time axis in the
latent).

\paragraph{Fit procedure.}
During the first training epoch, each batch of per-neuron windows
$\mathbf{X} \in \mathbb{R}^{B \times T}$ updates running counters: a count of
samples $n$, a sum $\mathbf{s} \in \mathbb{R}^{T}$, and an outer-product
accumulator $\mathbf{S} \in \mathbb{R}^{T \times T}$. Both accumulators are
maintained in \texttt{float64} to guard against drift across many batches. At
the end of the first epoch we form the empirical mean and covariance
\begin{equation}
    \boldsymbol{\mu}
    \;=\;
    \tfrac{1}{n}\,\mathbf{s},
    \qquad
    \mathbf{C}
    \;=\;
    \tfrac{1}{n}\,\mathbf{S}
    -
    \boldsymbol{\mu}\boldsymbol{\mu}^{\top},
\end{equation}
symmetrize $\mathbf{C}$ to suppress floating-point asymmetry, and take the
top-$k$ eigenvectors of $\mathbf{C}$ as the columns of $\mathbf{W}$. The
projection and mean are then frozen as buffers; the running accumulators are
discarded for inference.

\paragraph{Encoding and decoding.}
With $\mathbf{W}$ and $\boldsymbol{\mu}$ fixed,
\begin{equation}
    \mathbf{z}
    \;=\;
    (\mathbf{x} - \boldsymbol{\mu})\,\mathbf{W}
    \;\in\;\mathbb{R}^{k},
    \qquad
    \hat{\mathbf{x}}
    \;=\;
    \mathbf{z}\,\mathbf{W}^{\top}
    +
    \boldsymbol{\mu}
    \;\in\;\mathbb{R}^{T}.
    \label{eq:pca-fwd}
\end{equation}
Reconstructions are clamped to a small floor $\varepsilon = 10^{-8}$ to keep
the Poisson negative log-likelihood finite.

\begin{figure}[h!]
    \centering
    \includegraphics[width=\linewidth]{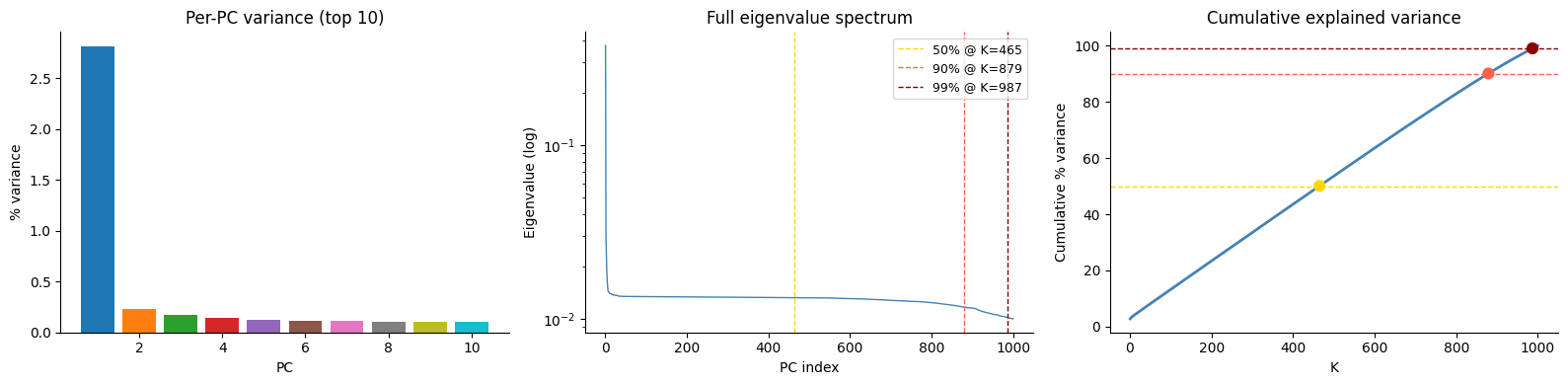}\\[0.6em]
    \includegraphics[width=0.84\linewidth]{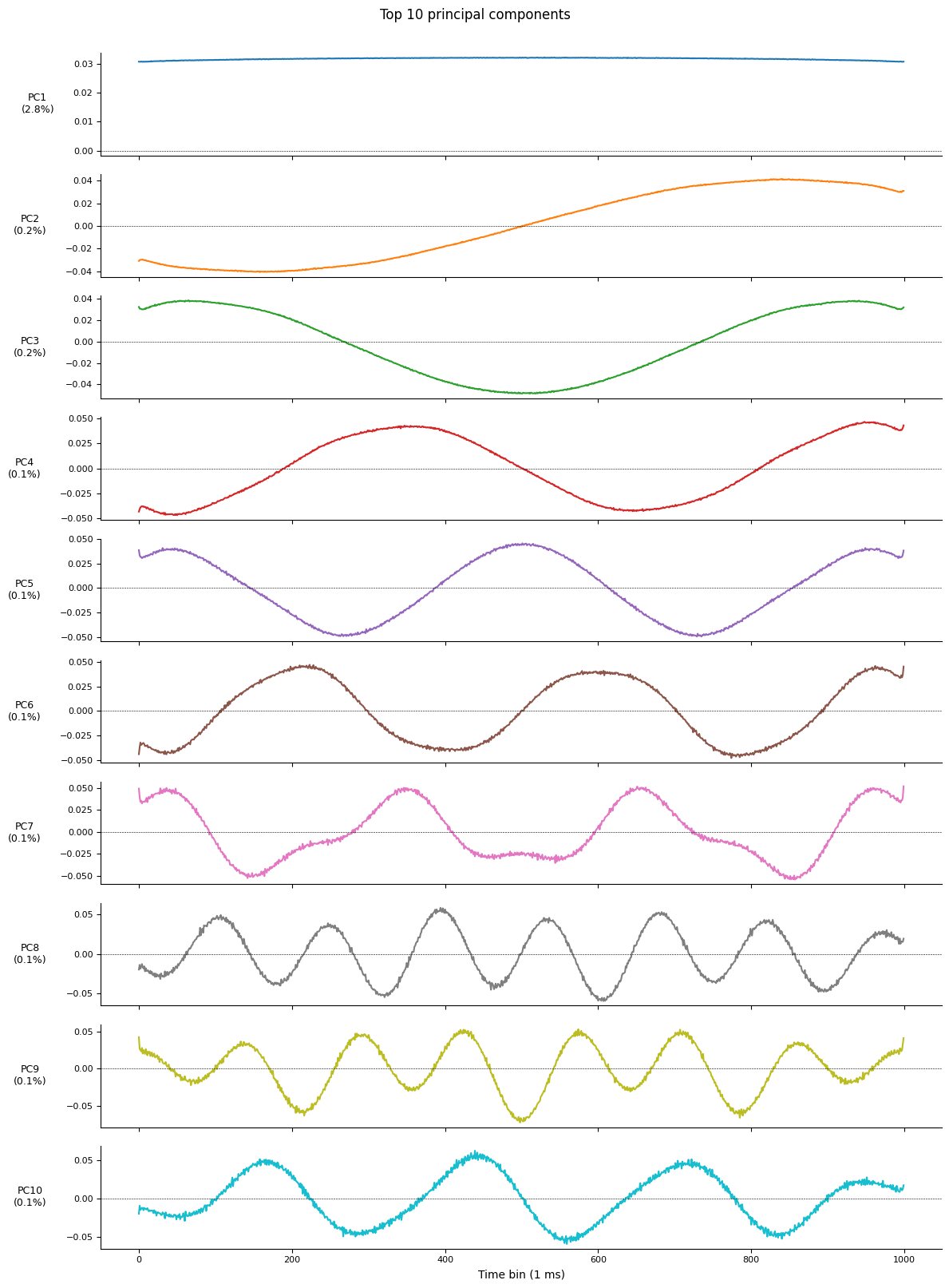}
    \caption{\textbf{PCA of $1$\,s windows of $1$\,ms-binned single-neuron
        spike counts.} \emph{Top row:} per-PC variance share for the top
        $10$ components (left); full eigenvalue spectrum on a log scale
        (center); cumulative explained variance, reaching $50\%$, $90\%$,
        and $99\%$ at $K = 465$, $879$, and $987$ respectively (right).
        \emph{Bottom:} the leading $10$ principal components plotted as
        functions of time over the $1000$-bin window. PC$1$ is flat and
        encodes the mean firing rate; PC$2$--PC$10$ are well-approximated
        by sinusoids at increasing frequency, recovering the discrete
        Fourier basis as expected for a (near-)stationary temporal
        covariance.}
    \label{fig:pca-components}
\end{figure}

\paragraph{Compression regime.}
The temporal downsampling ratio is $T/k$. We report two operating points,
$k = 1$ and $k = 10$, giving ratios of $1000{:}1$ and $100{:}1$ respectively
(in bits, Table~\ref{tab:main_results} reports a compression of $62\times$ for $k = 1$, since each latent value is counted as 16 bits). Because each neuron is encoded independently the
neuron count is preserved, and the total per-window latent is $N \cdot k$
continuous values for a $[T, N]$ input window.

\paragraph{What this baseline isolates.}
The PCA tokenizer is non-causal (the projection sees the full window), assumes
Gaussian observation noise (the Poisson likelihood is used only to score its reconstructions), and discards all cross-neuron structure. A Gaussian-MSE linear autoencoder
converges to the same principal subspace in the limit of perfect optimization, though not necessarily to the same basis. We include it to anchor the
benchmark at the low end of learned complexity and to separate gains
attributable to nonlinearity, causality, and cross-neuron mixing from those
attributable to the choice of bottleneck or training objective.

\paragraph{Interpreting the PCA components.}
The PCA tokenizer admits a directly interpretable linear decomposition: each
latent coordinate is the projection of the neuron's $1$\,s count vector onto
a fixed temporal basis fit on the training data. We fit one PCA per dataset
and inspect the leading components (Figure~\ref{fig:pca-components}). The
first component is essentially flat across the window and captures the mean
firing rate of the neuron over the $1$\,s segment. Subsequent components are
well-approximated by sinusoids at increasing frequency --- the discrete
Fourier basis recovered as eigenvectors of the (near-)stationary per-neuron
temporal covariance, a standard consequence of approximate translation
invariance in the time-bin statistics. The eigenvalue spectrum is
heavy-tailed: the top component carries ${\sim}2.8\%$ of the variance, and
reaching $50\%$ / $90\%$ / $99\%$ explained variance requires
$K = 465$ / $879$ / $987$ components respectively. PCA at small
$k \in \{1, 10\}$ therefore retains only the slowest temporal modes (DC plus
a few low-frequency sinusoids), which is precisely the regime in which we
expect learned, nonlinear, cross-neuron tokenizers to outperform.

\FloatBarrier

\subsection{Compressed Sensing Tokenizer}
\label{app:cs-tokenizer}

\begin{table}[H]
\centering
\small
\begin{tabular}{ll}
\toprule
\textbf{Property} & \textbf{Value} \\
\midrule
Operates over        & per-neuron (independent across units) \\
Latent type          & continuous \\
Output type          & non-negative integer spike counts \\
Window length        & fixed ($T \in \{100, 1000\}$ bins) \\
Neuron count         & variable (single matrix shared across units) \\
Trainable parameters & none \\
\bottomrule
\end{tabular}
\end{table}

The compressed sensing (CS) tokenizer is a training-free baseline. It produces
low-dimensional latents by linear measurement against a fixed random matrix
and reconstructs binned spike counts by sparse signal recovery. No parameters
are learned: the sensing matrix is a buffer, the decoder is an iterative
solver, and there is no objective optimized at fit time. This makes the CS
tokenizer a useful reference point that isolates what can be achieved purely
from the sparsity of single-neuron spike trains.

\paragraph{Signal model.}
We operate in the \emph{per-neuron} regime for this baseline. For a
window of duration $W \in \{100~\mathrm{ms}, 1000~\mathrm{ms}\}$ binned at
$\Delta = 1~\mathrm{ms}$, each neuron contributes a non-negative integer
count vector $\mathbf{x} \in \mathbb{Z}_{\geq 0}^{T}$ with
$T = W / \Delta \in \{100, 1000\}$. The CS tokenizer treats every neuron
independently: a single sensing matrix $\mathbf{A} \in \mathbb{R}^{m \times T}$
is reused across all neurons in a window and across the dataset. We use a
Gaussian sensing matrix with entries
$A_{ij} \sim \mathcal{N}(0, 1/m)$,
fixed by a global random seed, with $m \ll T$ the number of measurements
(latent dimension).

\paragraph{Encoding.}
Encoding is a single linear map applied per neuron:
\begin{equation}
    \mathbf{y} \;=\; \mathbf{A}\mathbf{x}
    \quad \in \mathbb{R}^{m},
    \qquad \mathbf{x} \in \mathbb{R}^{T}.
\end{equation}
For an input window of shape $[T, N]$ (time bins by neurons) the tokenizer
produces a latent of shape $[N, m]$: each neuron's full $T$-bin trace is
summarized by $m$ continuous measurements. The same matrix $\mathbf{A}$ acts
on every neuron, so the latent is unit-equivariant and the storage cost of
the tokenizer is $\mathcal{O}(mT)$ regardless of the number of neurons.

\paragraph{Decoding.}
Reconstruction targets a non-negative sparse signal $\hat{\mathbf{x}}$
consistent with the measurement $\mathbf{y}$. Because spike count vectors at
$\Delta = 1~\mathrm{ms}$ are very sparse (spike probability per bin is
typically $\ll 1$), we approximate non-negative basis pursuit by the
non-negative LASSO
\begin{equation}
    \hat{\mathbf{x}}
    \;=\;
    \arg\min_{\mathbf{x} \in \mathbb{R}^{T}_{\geq 0}}
    \tfrac{1}{2}\,\|\mathbf{A}\mathbf{x} - \mathbf{y}\|_{2}^{2}
    + \lambda\,\|\mathbf{x}\|_{1},
    \label{eq:cs-lasso}
\end{equation}
solved with FISTA~\citep{beck2009fast}. The Lipschitz constant for the
gradient step is $L = \lambda_{\max}(\mathbf{A}\mathbf{A}^{\top})$, giving a
fixed step size $\eta = 1/L$. The sparsity penalty is set per problem as a
fraction of the smallest $\lambda$ for which the trivial solution
$\mathbf{x} = \mathbf{0}$ is optimal:
\begin{equation}
    \lambda
    \;=\;
    \alpha \cdot \|\mathbf{A}^{\top}\mathbf{y}\|_{\infty},
    \qquad \alpha = 0.01,
\end{equation}
which makes the regularization scale with the measurement magnitude. Each
FISTA iteration is a non-negative soft-thresholding step combined with
Nesterov momentum:
\begin{align}
    \mathbf{v}_{k+1}
    &\;=\;
    \mathbf{z}_{k} - \eta\,\mathbf{A}^{\top}(\mathbf{A}\mathbf{z}_{k} - \mathbf{y}), \\
    \mathbf{x}_{k+1}
    &\;=\;
    \max\!\bigl(\mathbf{v}_{k+1} - \eta\lambda,\; \mathbf{0}\bigr),\\
    t_{k+1}
    &\;=\;
    \tfrac{1}{2}\bigl(1 + \sqrt{1 + 4 t_{k}^{2}}\bigr), \\
    \mathbf{z}_{k+1}
    &\;=\;
    \mathbf{x}_{k+1}
    + \tfrac{t_{k} - 1}{t_{k+1}}\bigl(\mathbf{x}_{k+1} - \mathbf{x}_{k}\bigr).
\end{align}
We run $K = 500$ iterations for every window and every neuron. Reconstructed
spike counts are then rounded to the nearest non-negative integer to match
the discrete output expected by downstream evaluation.

\paragraph{Compression regime.}
At $\Delta = 1~\mathrm{ms}$, single-neuron windows are highly sparse, which
places this configuration in the regime where compressed sensing is
informative: Gaussian random matrices satisfy the restricted isometry property
with high probability when $m = \mathcal{O}(s \log(T/s))$ for $s$-sparse
signals~\citep{candes2006near, donoho2006compressed},
and exact recovery of non-negative sparse vectors from noiseless Gaussian
measurements is well-studied~\citep{donoho2010counting}. We do not claim
exact recovery in our setting (signals are not strictly $s$-sparse, and
\eqref{eq:cs-lasso} is solved approximately), but the regime motivates the
parameter choices: $T \in \{100, 1000\}$ and $m \ll T$ such that the
information per measurement remains well above one bit for typical firing
rates.

\paragraph{What this baseline isolates.}
Since the tokenizer is untrained and acts on each neuron independently, any
performance it achieves on the downstream evaluation reflects only
(i) the sparsity of single-neuron spike trains at millisecond resolution and
(ii) the recovery properties of the LASSO solver. It does \emph{not} exploit
cross-neuron structure, temporal correlations beyond what is implicit in
single-neuron sparsity, or any task signal. Differences between trained
tokenizers and CS therefore quantify the contribution of these additional
inductive biases.

\paragraph{Sinusoidal sensing matrix.}
We additionally consider a deterministic alternative to the Gaussian sensing
matrix in which the rows of $\mathbf{A} \in \mathbb{R}^{m \times T}$ are
arranged as $\sin/\cos$ pairs at geometrically-spaced frequencies. For
$k = 0, 1, \ldots, \lceil m/2 \rceil - 1$ we set
\begin{equation}
    \omega_k \;=\; b^{-2k/m},
    \qquad
    A_{2k,\,t} \;=\; \tfrac{1}{\sqrt{m}}\sin(\omega_k t),
    \qquad
    A_{2k+1,\,t} \;=\; \tfrac{1}{\sqrt{m}}\cos(\omega_k t),
\end{equation}
for $t = 0, \ldots, T-1$, with base $b = 10^{4}$. When $m$ is odd, the last
row is an additional sine term at the highest frequency index. The
normalization by $1/\sqrt{m}$ matches the variance of the Gaussian
construction, so the latent magnitudes of the two variants are directly
comparable. This is exactly the sinusoidal positional encoding used in the
original Transformer~\citep{vaswani2017attention}, applied here as a sensing
operator. Concretely, the $k$-th measurement pair becomes the in-phase and
quadrature components of the input at frequency $\omega_k / (2\pi \Delta)$,
so the latent has a direct frequency-domain interpretation and is identical
across runs without requiring a random seed.

\paragraph{Why use it.}
Replacing the random Gaussian matrix with this deterministic Fourier-like
basis imposes structure on the latent space: nearby latent dimensions encode
adjacent frequencies, the latent ordering carries meaning, and runs are
exactly reproducible without seeding. The construction does not satisfy a
generic restricted isometry property --- rows are deterministic and highly
correlated across neighboring frequencies --- so the worst-case recovery
guarantees of the Gaussian regime no longer apply. Empirically, however, on
the $1$-ms-binned single-neuron windows used in this paper, the
LASSO/FISTA reconstruction in equation~\eqref{eq:cs-lasso} retains
reconstruction fidelity comparable to the Gaussian variant at $1000:1$, and somewhat lower at
$100:1$.

\subsection{Single-Neuron Convolutional-X (WanNeural)}
\label{app:wanneural}

\begin{table}[H]
\centering
\small
\begin{tabular}{ll}
\toprule
\textbf{Property} & \textbf{Value} \\
\midrule
Operates over        & per-neuron tokens (optional cross-neuron attention) \\
Latent type          & continuous (VAE) or discrete (FSQ / BSQ / VQ) \\
Output type          & spike probabilities (sigmoid) or firing rates (softplus) \\
Window length        & fixed (configurable; $1$\,s at $100{:}1$, $5$\,s at $1000{:}1$ in Table~\ref{tab:main_results}) \\
Neuron count         & fixed (processed independently or jointly) \\
Trainable parameters & encoder \& decoder weights, bottleneck parameters \\
\bottomrule
\end{tabular}
\end{table}

WanNeural is a convolutional autoencoder for neural spike data adapted from the Wan2.1/2.2 video VAE architecture~\citep{wan2025wan}. The 3-D (time $\times$ height $\times$ width) convolutional blocks of the original video model are reduced to 2-D (time $\times$ neurons), enabling the encoder-decoder to act on a $(B,T,N)$ tensor of binned spike counts or firing rates. Causality is enforced by left-padding all temporal convolutions, so each latent position depends only on past and current input bins. The architecture is modular: encoder depth, channel widths, temporal stride schedule, neuron kernel size, and bottleneck type are all independently configurable. In the default single-neuron operating mode, the neuron kernel is set to $k_n = 1$, which collapses cross-neuron mixing and processes every unit through the same shared weights --- analogous to depthwise-over-neurons convolution.

\paragraph{Signal model.}
Each input window is a tensor $\mathbf{X} \in \mathbb{R}^{T \times N}$ of binned activity at $\Delta = 1$\,ms resolution over a window of $W$\,s, giving $T = W / \Delta$ time bins and $N$ neurons. In the single-neuron mode, the $N$ axis is treated as a batch dimension: all neurons in a session share encoder/decoder weights, and the latent has shape $[T_{\mathrm{lat}}, N, d_z]$, one $d_z$-dimensional vector per compressed time step per neuron.

\paragraph{Encoder.}
An optional temporal patcher folds $p$ consecutive bins into the channel dimension, reducing the temporal axis from $T$ to $T/p$ before the first convolutional stage. The main encoder consists of $L$ stages; each stage applies $R$ residual blocks (CausalConv2d with kernel sizes $k_t$ along time and $k_n$ along neurons, RMSNorm, Swish activations) followed by optional Wan~2.2-style stage shortcuts and optional cross-neuron attention. Selected stages apply a strided causal convolution with kernel $k_{t,\mathrm{down}}$ and stride $s_{t,\mathrm{down}}$ to compress the time axis. Two mid-blocks with optional attention operate at the bottleneck resolution. Channel widths per stage follow a multiplier schedule $[m_0, m_1, \ldots, m_{L-1}]$ applied to a base dimension $d$; a head convolution maps the final feature map to $d_z$ (or $2 d_z$ for VAE) channels before the bottleneck.

\paragraph{Bottleneck.}
The codebase supports four interchangeable bottleneck families. The \textbf{VAE} bottleneck uses a diagonal Gaussian posterior with KL weight $\beta$, with an optional curvature penalty on the latent trajectory~\citep{wang2026temporal}. \textbf{FSQ} (finite scalar quantization~\citep{mentzer2023finite}) snaps each latent dimension to a per-axis lattice with straight-through gradients; no auxiliary loss is needed. \textbf{BSQ} (binary spherical quantization~\citep{zhao2024image}) projects to the unit sphere and binarizes, penalized by an entropy loss. The \textbf{VQPT} wrapper exposes any quantizer from the \texttt{vector-quantize-pytorch} library~\citep{wang2020vqpytorch} (VQ~\citep{vandenoord2017vqvae}, LFQ~\citep{yu2023language}, RVQ~\citep{zeghidour2021soundstream}, and variants) via a uniform interface.

\paragraph{Decoder.}
The decoder mirrors the encoder: it begins at the bottleneck resolution, applies $L$ up-stages with reversed channel widths, and inserts strided transposed causal convolutions at the positions corresponding to the encoder downsamples. All decoder parameters are independently overridable (e.g., different channel widths, attention placement, or stride schedule), defaulting to the encoder values. The output is passed through a sigmoid to produce spike probabilities (as in the Table~\ref{tab:main_results} models) or through a softplus with a small temperature parameter to produce non-negative firing-rate predictions.

\paragraph{Compression regime.}
Total temporal compression is $c = p \cdot s_{t,\mathrm{init}} \cdot s_{t,\mathrm{down}}^{n_\downarrow}$, where $n_\downarrow$ is the number of downsampled stages. Here $s_{t,\mathrm{init}}$ is the stride of the initial convolution (1 in all reported models). The $100{:}1$ models of Table~\ref{tab:main_results} use $p = 25$, $s_{t,\mathrm{down}} = 2$ and two downsampled stages ($c = 100$): a 1\,s window at 1\,ms resolution ($T = 1000$) is compressed to $T_\text{lat} = 10$ latent steps per neuron, each carrying a $d_z = 16$-dimensional code (160 latent values per neuron per window). The $1000{:}1$ Conv-VAE uses $p = 125$ and three downsampled stages ($c = 1000$) on a 5\,s window, giving $T_\text{lat} = 5$ steps with $d_z = 32$ (again 160 values).

\paragraph{Training protocol.}
The reconstruction loss depends on the decoder's output type. In the rate-output mode the primary loss is the Poisson negative log-likelihood between predicted rates and binned spike counts, optionally combined with a smoothed-rate MSE term via a multi-loss wrapper. In the binary spike-output mode (used by the Table~\ref{tab:main_results} models) the decoder targets $\{0,1\}$ bins directly and the primary loss switches to binary cross-entropy (or focal loss for highly imbalanced spike rates); Poisson NLL is not meaningful here. The KL term (VAE) or quantization loss (FSQ / BSQ / VQPT) is added with a configurable weight. We use AdamW with a cosine learning-rate schedule and mixed-precision training.

\paragraph{What this tokenizer provides.}
WanNeural extends the non-learned PCA and CS baselines with a fully learned nonlinear encoder-decoder while retaining the same per-neuron operating mode and comparable compression ratios. The configurable bottleneck (continuous vs.\ discrete, low vs.\ high bitrate) makes it straightforward to isolate the contribution of bottleneck type from that of encoder architecture. Enabling cross-neuron attention (e.g., \texttt{attn\_levels = [0, 1]}) or widening the neuron kernel ($k_n > 1$) converts the model from a single-neuron to a light population tokenizer without changing the rest of the pipeline. The $1000{:}1$ Conv-VAE of Table~\ref{tab:main_results} has such an attention block across neurons at the bottleneck resolution of its encoder and decoder.

\subsection{LDNS-Perceiver---a population-level extension of LDNS \citep{kapoor2024ldns}}
\label{app:perceiver-ldns}

\begin{table}[H]
\centering
\small
\begin{tabular}{ll}
\toprule
\textbf{Property} & \textbf{Value} \\
\midrule
Operates over        & session (population-level) \\
Latent type          & continuous (8 tokens of dimension 8 per 10\,ms time bin) \\
Output type          & Poisson firing rates, sampled to spikes \\
Window length        & variable \\
Neuron count         & variable \\
Trainable parameters & encoder \& decoder weights, latent token queries \\
\bottomrule
\end{tabular}
\end{table}

\begin{figure}
    \centering
    \includegraphics[width=0.9\linewidth]{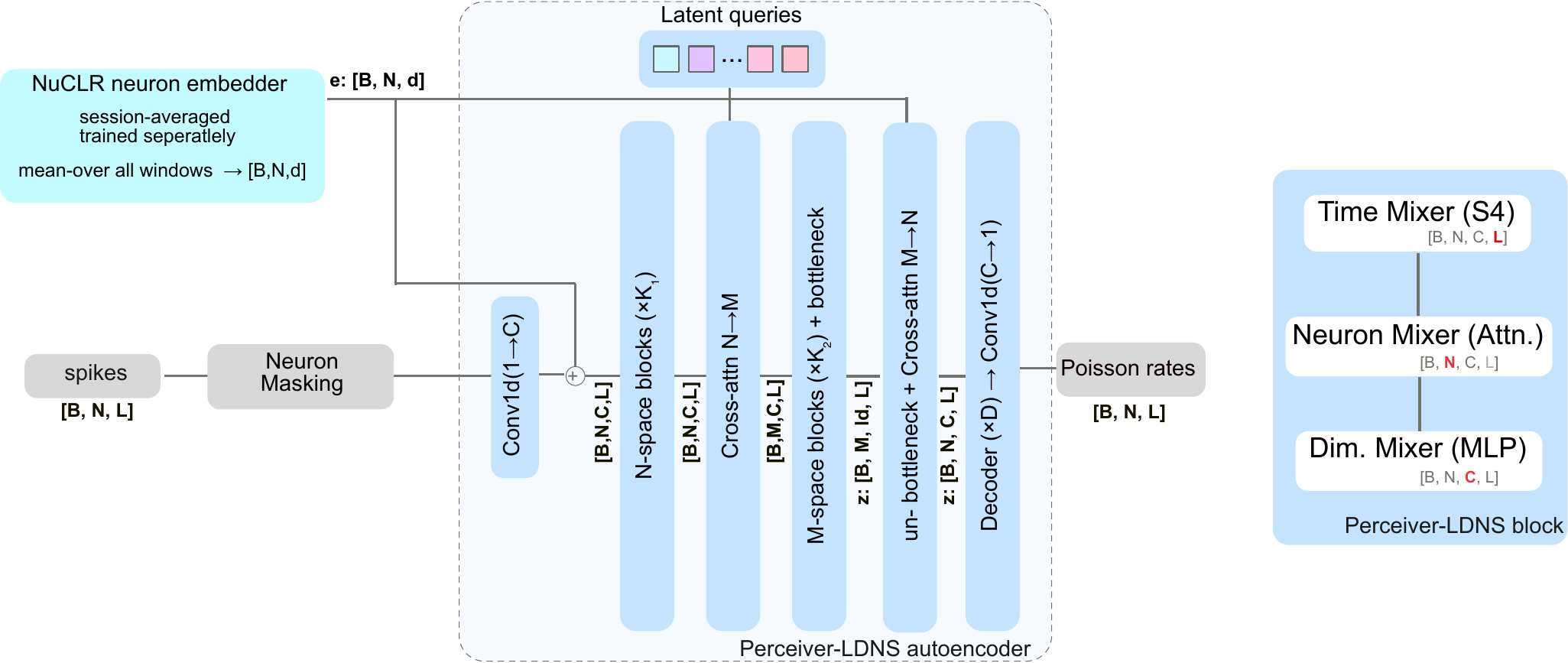}
    \caption{LDNS-Perceiver architecture}
    \label{fig:perceiver-arch}
\end{figure}

We use LDNS-Perceiver as the population-level tokenizer in our benchmark. The model keeps the autoencoding part of LDNS---a Poisson observation model, time-aligned continuous latents, and S4~\citep{gu2022efficiently} temporal mixing---but replaces the fixed-neuron autoencoder with a Perceiver-style architecture that supports variable neural populations. We use only the tokenizer; we do not use the diffusion stage of LDNS.

\paragraph{Input representation.}
The input is a window of binned spike counts together with fixed neuron embeddings. In the final model, we bin spikes at 10\,ms over 4\,s windows, giving 400 time bins. We assign each recorded unit a 256-dimensional NuCLR embedding \citep{aroraKnowThyselfKnowing2025}, extracted with a separately trained NuCLR model and kept frozen during tokenizer training. These embeddings provide neuron identity and session context, allowing the same tokenizer to operate on sessions with different recorded units.

\paragraph{Architecture.}
The encoder first processes the observed neurons in neuron space. At each time bin, neurons attend to one another, while S4 layers mix information along time. Compared to the LDNS autoencoder blocks, each block in this tokenizer has a time mixer, a self-attention based neuron mixer, and then an MLP. A Perceiver-style cross-attention layer then compresses the variable-size neuron set to a fixed set of latent tokens. Latent-space blocks further process these tokens with attention across latent tokens and temporal mixing along time. The final latent tensor has fixed size independent of the number of input neurons: in our final model, we use 8 latent tokens, each with an 8-dimensional bottleneck, at every 10\,ms time bin. 
The decoder reconstructs firing rates for all neurons from the latent tokens and the frozen neuron embeddings. We initialize decoder queries from the neuron embeddings and let them attend back to the latent tokens at each time bin. A final softplus output gives nonnegative rates for the Poisson observation model. See Figure~\ref{fig:perceiver-arch} for the full architecture and Table~\ref{tab:perceiver-ldns-settings} for its settings.

\paragraph{Training objective and details.}
During training, we hide 50\% of the real neurons from the encoder and reconstruct all real neurons in the session. This makes the latent tokens carry population-level information rather than a direct copy of each input neuron. We compute the reconstruction loss as the Poisson negative log-likelihood over all valid real-neuron and time-bin positions. We train for 100k iterations on a single A100 GPU with an effective batch size of 3 (batch size 1 with gradient accumulation over 3 steps) using AdamW, with a learning rate of $5\times 10^{-4}$ after 1000 warm-up steps, cosine-decayed to $5\times 10^{-5}$.

Compared with LDNS, we modify the latent regularization in two ways. First, we replace the LDNS latent \(L_2\) penalty with SIGReg \citep{balestriero2025lejepa}, which regularizes the latent distribution toward a well-scaled isotropic Gaussian using sliced Epps--Pulley statistics. Second, we replace the first-order temporal-difference penalty with a second-order curvature penalty, \(\mathcal{L}_{\mathrm{curv}} = \sum_t \|z_{t+2} - 2z_{t+1} + z_t\|_2^2\), which discourages sharp bends in the latent trajectory while allowing slowly varying trends. The full objective is
\[
\mathcal{L}
=
\mathcal{L}_{\mathrm{Poisson}}
+
\beta_{\mathrm{SIG}} \mathcal{L}_{\mathrm{SIG}}
+
\beta_{\mathrm{curv}} \mathcal{L}_{\mathrm{curv}},
\]
where \(\mathcal{L}_{\mathrm{Poisson}}\) is the reconstruction loss over all valid real-neuron bins, \(\mathcal{L}_{\mathrm{SIG}}\) is the SIGReg latent-distribution regularizer, and \(\mathcal{L}_{\mathrm{curv}}\) is the second-order temporal curvature penalty.

\begin{table}[t]
  \centering
  \caption{\textbf{LDNS-Perceiver settings.} Configuration of the population tokenizer used in the main experiments.}
  \label{tab:perceiver-ldns-settings}
  \small
  \setlength{\tabcolsep}{6pt}
  \begin{tabular}{@{}ll@{}}
    \toprule
    Setting & Value \\
    \midrule
    Input window & 4.0\,s \\
    Bin size & 10\,ms \\
    Time bins per window & 400 \\
    Input / output & Spike counts / Poisson rates \\
    Neuron embeddings & Frozen NuCLR binned-deep embeddings, \(d_e=256\) \\
    Hidden width & 224 \\
    Latent tokens & 8 \\
    Bottleneck dimension & 8 \\
    Attention heads & 7 \\
    Encoder blocks & 3 neuron-space blocks, 1 latent-space block \\
    Decoder blocks & 5 \\
    Temporal mixer & Bidirectional S4 in the encoder \\
    Neuron masking & 50\% hidden from encoder \\
    Reconstruction loss & Poisson NLL \\
    SIGReg weight & \(10^{-6}\) \\
    Curvature penalty & Second-order, lag 1, weight \(10^{-4}\) \\
    Temporal masking & Disabled \\
    \bottomrule
  \end{tabular}
\end{table}

\paragraph{NuCLR embedding model.}
We extract the neuron embeddings with a binned-deep NuCLR model \citep{aroraKnowThyselfKnowing2025} trained on 10\,s windows with 20\,ms bins. The model outputs 256-dimensional embeddings and uses contrastive two-view training with temporal and spatiotemporal attention layers. We keep these embeddings fixed during LDNS-Perceiver training instead of training neuron descriptors jointly with the tokenizer.

% \drafttodo{Add architecture diagrams, optimizer settings, hyperparameter grids, and supplementary result tables here.}

%Tokenizer comparisons should use a shared protocol for preprocessing, hyperparameter search, compute budget, convergence criteria, and evaluation. The main paper should give enough detail to make the results reproducible, with full configs deferred to supplementary material or code release.

%All learned tokenizers share a preprocessing pipeline, hyperparameter-search protocol, compute budget, and convergence criterion. We run the same evaluation pass on each tokenizer using the four generalization regimes defined in Section~4.3. Hyperparameters are shared across datasets except where noted; this is a deliberate choice to test how much per-dataset tuning the design can absorb.

% \drafttodo{Describe Hydra configs, random seeds, model selection criterion, early stopping, compute budget, and whether hyperparameters are shared or tuned per dataset.}

\subsection{Wavelet-FSQ}
\label{app:cosmos-2d}

\begin{table}[H]
\centering
\small
\begin{tabular}{ll}
\toprule
\textbf{Property} & \textbf{Value} \\
\midrule
Operates over        & population (joint neuron $\times$ time, fixed unit count) \\
Latent type          & discrete (finite-scalar quantization) \\
Output type          & non-negative firing rates (Poisson observation model) \\
Window length        & fixed ($T = 1000$ bins, $1$\,s at $1$\,ms) \\
Neuron count         & fixed ($N = 32$) \\
Trainable parameters & encoder, decoder (wavelet filters and quantizer frozen) \\
\bottomrule
\end{tabular}
\end{table}

Wavelet-FSQ is a wavelet-based 2D population tokenizer adapted from the
video tokenizer of \citet{agarwal2025cosmos}. It treats each $1$-second
window of $1$\,ms-binned spikes as a single-channel image of shape
$(H = N, W = T)$ and compresses both axes jointly. The pipeline has four
stages: a 2D wavelet \emph{patcher} that performs a multi-level packet-style
discrete wavelet transform (DWT) along each axis; a 2D convolutional encoder
that further downsamples and lifts to a latent representation; a
finite-scalar quantization bottleneck; and a 2D convolutional decoder that
mirrors the encoder, followed by an inverse-DWT \emph{unpatcher}. Causality
is enforced along the time axis inside the encoder/decoder convolutions; the
wavelet boundary uses symmetric (reflection) extension and is therefore not
strictly causal at the window edge.

\paragraph{Wavelet patcher.}
The patcher performs $\log_{2}(\text{patch}_h)$ levels of packet DWT along
the neuron axis and $\log_{2}(\text{patch}_w)$ levels along the time axis,
mapping
\begin{equation}
    [B,\,1,\,N,\,T]
    \;\xrightarrow{\text{Patcher2D}}\;
    [B,\;\text{patch}_h\cdot\text{patch}_w,\;N/\text{patch}_h,\;T/\text{patch}_w].
\end{equation}
We use $\text{patch}_h = 4$ and $\text{patch}_w = 2$, which downsamples to
$[N/4, T/2]$ and spreads high-frequency content into eight additional
channels. We use Daubechies-4 (\texttt{db4}) wavelets in all reported runs.
Compared with the Haar wavelet of the original Cosmos video tokenizer, the
longer \texttt{db4} support better preserves the smooth temporal envelopes
of spike rates; we found longer wavelets (\texttt{sym4}) gave no further
improvement. Symmetric extension yields bit-exact perfect reconstruction
across the orthogonal wavelets we support and avoids the unphysical
periodicity assumption of circular mode; the boundary extension is local
($\sim L$ samples for a length-$L$ filter), and downstream encoder layers
remain strictly causal.

\paragraph{Encoder.}
The encoder consists of three resolution stages with channel widths
$[64, 128, 128]$ (\texttt{channels}~$=64$, \texttt{channels\_mult}~$=
[1, 2, 2]$) and two residual blocks per stage, followed by two ``mid''
residual blocks at the bottleneck resolution. Each stage may downsample $H$
and/or $W$ independently by a stride-$2$ causal Conv2d. The number of
strided stages per axis is fixed by the compression configuration: with
$\text{neuron\_compression} = 32$, $\text{patch}_h = 4$ the encoder applies
$\log_{2}(32/4) = 3$ neuron-axis downsamples (one per stage); with
$\text{temporal\_compression} = 2$, $\text{patch}_w = 2$ no additional
time-axis downsamples are applied beyond the wavelet itself. For the
$N = 32$ window this fully collapses the neuron axis ($N_\text{lat} = 1$),
and the time axis is compressed only by the wavelet, giving
$T_\text{lat} = T / 2 = 500$. The encoder output has shape
$[B,\,z_\text{ch},\,1,\,500]$ with $z_\text{ch} = 64$. Internal layers use
GroupNorm applied per time column (so normalization statistics do not leak
the future) and Swish activations; convolutions are causal along $W$
(left-pad only) and symmetric along $H$.

\paragraph{FSQ bottleneck.}
The bottleneck is a finite-scalar quantizer
(FSQ)~\citep{mentzer2023finite} with two latent dimensions of $32$ levels
each ($1024$ codes, $10$\,bits per latent position). The encoder feature is passed
through a bounded $\tanh$-style nonlinearity and snapped to a per-axis lattice with
straight-through gradients; we use no auxiliary commitment loss. We also report
Wavelet-AE, the continuous counterpart in Table~\ref{tab:main_results}, which uses the same
encoder and decoder with an unquantized bottleneck of four latent channels.
The codebase additionally supports residual-FSQ and VAE bottlenecks, which we do not report.

\paragraph{Decoder.}
The decoder mirrors the encoder. It begins with two mid residual blocks at
the bottleneck resolution, then applies three up-stages with the encoder's
channel widths in reverse, each with three residual blocks (one more than in the encoder stages) and a per-axis
$2\!\times$ upsample (repeat-interleave followed by causal Conv2d). A final
inverse-DWT \emph{unpatcher} restores the full $[N, T]$ shape, with optional
trimming when symmetric-mode reconstruction overshoots the original support.
The output is passed through a softplus and clamped to a floor
$\varepsilon = 10^{-8}$ to give per-neuron Poisson rate predictions at
$1$\,ms resolution.

\paragraph{Compression regime.}
Each $1$-second window is summarized by $T_\text{lat} = 500$ latent
positions, each carrying $10$\,bits, for a total of
${\sim}5{,}000$\,bits per window against an input of
$T \cdot N = 32{,}000$ binned spike counts.

\paragraph{Training objective.}
We minimize the Poisson negative log-likelihood between the reconstructed
rates and the binned spike-count target. Quantization uses straight-through
gradients with no commitment loss; no reconstruction-side regularization is
used. Table~\ref{tab:cosmos-2d-settings} lists the full configuration.

\paragraph{What this tokenizer isolates.}
Unlike single-neuron tokenizers that fold the neuron dimension into the
batch, Wavelet-FSQ mixes information across neurons within each
$\text{patch}_h = 4$ group via 2D convolutions, capturing local
inter-neuron correlations that purely per-neuron compression discards at
equal latent budget. Comparing it with Wavelet-AE tests whether a fully discrete
population latent at $10$\,bits per position can match the
reconstruction quality of an unquantized continuous bottleneck on the
same encoder/decoder.

\begin{table}[t]
  \centering
  \caption{\textbf{Wavelet-FSQ settings.} Configuration of the FSQ
    population tokenizer used in Table~\ref{tab:main_results}.}
  \label{tab:cosmos-2d-settings}
  \small
  \setlength{\tabcolsep}{6pt}
  \begin{tabular}{@{}ll@{}}
    \toprule
    Setting & Value \\
    \midrule
    \multicolumn{2}{l}{\emph{Data}} \\
    Input window & $1.0$\,s \\
    Bin size & $1$\,ms \\
    Time bins per window & $T = 1000$ \\
    Neurons per window & $N = 32$ \\
    Input / output & Rates / Poisson rates \\
    \midrule
    \multicolumn{2}{l}{\emph{Wavelet patcher}} \\
    $\text{patch}_h$ (neuron axis) & $4$ \\
    $\text{patch}_w$ (time axis) & $2$ \\
    Wavelet family & \texttt{db4} \\
    Boundary extension & symmetric (perfect reconstruction) \\
    \midrule
    \multicolumn{2}{l}{\emph{Encoder / decoder}} \\
    Base channels & $64$ \\
    Channel multipliers & $[1, 2, 2]$ \\
    Residual blocks per stage & $2$ (encoder), $3$ (decoder) \\
    Mid residual blocks & $2$ \\
    Bottleneck channels $z_\text{ch}$ & $64$ \\
    Neuron compression (total) & $32\times$ ($\text{patch}_h \cdot 2^3$) \\
    Temporal compression (total) & $2\times$ ($\text{patch}_w \cdot 2^0$) \\
    Latent shape $[T_\text{lat}, N_\text{lat}]$ & $[500,\,1]$ \\
    Causality (time axis) & strict in conv stack; symmetric at wavelet edge \\
    \midrule
    \multicolumn{2}{l}{\emph{Bottleneck (FSQ)}} \\
    Quantizer & FSQ (single stage) \\
    Levels & $[32, 32]$ ($1024$ codes) \\
    Capacity per token & $10$\,bits \\
    Gradient estimator & straight-through \\
    Commitment loss & none \\
    \midrule
    \multicolumn{2}{l}{\emph{Training}} \\
    Reconstruction loss & Poisson NLL on per-bin rates \\
    Auxiliary losses & none \\
    Output activation & softplus + floor at $10^{-8}$ \\
    \bottomrule
  \end{tabular}
\end{table}

\FloatBarrier
\newpage

\section{Downstream Decoding Details}
\label{app:downstream-decoding}

We evaluate learned representations using two downstream decoding paradigms.
The first measures linear decodability directly from frozen latents via
a regularized linear probe. The second assesses whether richer temporal
structure can be extracted by a lightweight sequence decoder, using a
POYO-based~\cite{azabou2023poyo} architecture trained on top of frozen tokenizer outputs.

\subsection{Linear probes}
\label{app:downstream-decoding-linear-probes}

To assess linear decodability, we train a regularized linear probe from frozen
latents to four behavioral targets: cursor position and velocity (regression),
and trial target identity and movement phase (classification). Latents are
aggregated into non-overlapping 200\,ms segments by concatenating consecutive
token vectors, and behavioral targets are aligned to the same grid. Regression
targets are maintained at 10\,ms resolution and predicted jointly across all
bins within a segment; classification labels are assigned by majority vote, and
segments are discarded when the majority falls below 75\,\% or the fraction of
missing values exceeds 25\,\%. The regularization strength is selected by
generalized cross-validation on the training fold.

Each session is evaluated with a 3-fold trial-grouped cross-validation, using
stratified grouping for classification. Regression employs asymmetric phase
filtering: training includes segments spanning the inter-trial, reach, return,
and random phases, while $R^2$ is evaluated only on bins falling within the
reach and random phases of valid trials. Classification applies the same phase
mask symmetrically to both folds. 

\FloatBarrier

\subsection{Temporal decoder---POYO~\cite{azabou2023poyo}}
\label{app:downstream-decoding-poyo}

We use the original POYO decoder~\cite{azabou2023poyo}, where each spike is represented by a unit embedding and a token sequence with corresponding timestamps. For comparison using tokenized representations, we replace spike embeddings with tokenizer-derived tokens using a lightweight adapter to map tokenizer outputs to the decoder dimensionality. We use the 1.3M-parameter POYO configuration with \(d = 64\), see Table~\ref{tab:poyo-hyperparams} for all hyperparameter settings.

\begin{table}[t]
  \centering
  \caption{\textbf{POYO hyperparameters.} Configuration of the behavioral decoder, 1.3M param variant.}
  \label{tab:poyo-hyperparams}
  \small
  \setlength{\tabcolsep}{6pt}
  \begin{tabular}{@{}ll@{}}
    \toprule
    Setting & Value \\
    \midrule
    \multicolumn{2}{@{}l}{\textit{General}} \\
    Input window & 1.0\,s \\
    Model dim & 64 \\
    Attention head dimension & 64 \\
    Cross-attention heads & 2 \\
    Self-attention heads & 8 \\
    Self-attention depth & 6 \\
    FFN dropout & 0.2 \\
    Linear dropout & 0.4 \\
    Attention dropout & 0.2 \\
    Embedding init scale & 0.02 \\
    Rotary $t_{\min}\,/\,t_{\max}$ & $10^{-4}$\,s\,/\,4.0\,s \\
    \midrule
    \multicolumn{2}{@{}l}{\textit{Tokenizer}} \\
    Cross-attention & POYO Decoder only \\
    \midrule
    \multicolumn{2}{@{}l}{\textit{Spike-times baseline only}} \\
    Cross-attention & POYO Encoder \& Decoder \\
    Latent tokens per step & 16 \\
    Latent step & 0.125\,s (8 latent steps per window) \\
    \bottomrule
  \end{tabular}
\end{table}

For tokenizer-based inputs, we remove the encoder cross-attention stage and train a decoder directly on frozen tokenizer outputs. This avoids giving the downstream model an additional learned aggregation stage on top of already-compressed representations. The spike-time baseline retains the standard POYO event-token interface with encoder cross-attention, as native spike times are variable-length event tokens rather than fixed tokenizer latents. Each output query is a session embedding at a target behavioral timestamp. Then we map the resulting hidden state to \(\mathbb{R}^2\) (corresponding to normalized cursor velocity) using a linear map.

We minimize the MSE loss between the target and predicted cursor velocities for decoder training. We use sessions from the Perich--Miller dataset, with cursor velocity targets divided by a fixed scale of 20, as in POYO. Performance is reported as stitched held-out \(R^2\), averaged across sessions. Tables~\ref{tab:poyo-single-session-trials}--\ref{tab:poyo-plotted-means} list the sessions, trial counts, tokenizer inputs and plotted means.
As in the original paper, we train using SparseLamb with weight decay \(10^{-4}\) and a OneCycle learning-rate schedule with cosine annealing. The warmup fraction is \(0.05\) with initial and final learning-rate divisors of 25 and 40, respectively. Training windows are sampled randomly from the training split, while validation and test use fixed one-second stitching windows with a half-window stride. All plotted comparisons use seeds 42 and 43. We train the single-session models for 3400 steps each, and multi-session models for 6800 steps each. Calibration fine-tuning is performed over 800 steps.

\begin{table}[h]
  \centering
  \caption{\textbf{Single-session sessions and scratch-training trial counts.} Eight Perich--Miller sessions are used in the single-session panels of Figure~\ref{fig:poyo-evals}.}
  \label{tab:poyo-single-session-trials}
  \small
  \setlength{\tabcolsep}{5pt}
  \begin{tabular}{l rr}
    \toprule
    Session & Half trials & Full trials \\
    \midrule
    \texttt{c\_20131217\_random\_target\_reaching} & 52 & 105 \\
    \texttt{c\_20151103\_center\_out\_reaching} & 51 & 102 \\
    \texttt{c\_20151104\_center\_out\_reaching} & 64 & 129 \\
    \texttt{j\_20160406\_center\_out\_reaching} & 45 & 91 \\
    \texttt{m\_20140214\_random\_target\_reaching} & 52 & 105 \\
    \texttt{m\_20150511\_center\_out\_reaching} & 99 & 199 \\
    \texttt{t\_20130820\_random\_target\_reaching} & 43 & 86 \\
    \texttt{t\_20130822\_random\_target\_reaching} & 66 & 133 \\
    \bottomrule
  \end{tabular}
\end{table}

\begin{table}[h]
  \centering
  \caption{\textbf{Multi-session POYO model sessions.}}
  \label{tab:poyo-fold-a}
  \small
  \setlength{\tabcolsep}{5pt}
  \begin{tabular}{p{0.31\linewidth}p{0.61\linewidth}}
    \toprule
    Model & Sessions \\
    \midrule
    Multi-session trained POYO model &
    \texttt{c\_20151103\_center\_out\_reaching},
    \texttt{m\_20140214\_random\_target\_reaching},
    \texttt{c\_20151104\_center\_out\_reaching},
    \texttt{t\_20130820\_random\_target\_reaching} \\
    
    Finetuned model on target session &
    \texttt{c\_20131217\_random\_target\_reaching},
    \texttt{m\_20150511\_center\_out\_reaching},
    \texttt{j\_20160406\_center\_out\_reaching},
    \texttt{t\_20130822\_random\_target\_reaching} \\
    \bottomrule
  \end{tabular}
\end{table}

All temporal-decoder runs use one-second input windows, \(d=64\), six transformer layers, eight self-attention heads, two cross-attention heads where applicable, \(d_{\mathrm{head}}=64\), feed-forward dropout \(0.2\), linear dropout \(0.4\), and attention dropout \(0.2\).

\begin{table}[h]
  \centering
  \caption{\textbf{Tokenizer input configuration used in the POYO evaluation.}}
  \label{tab:poyo-tokenizer-inputs}
  \small
  \setlength{\tabcolsep}{4pt}
  \begin{tabular}{l c c}
    \toprule
    Method & Token dim. & \# input tokens \\
    \midrule
    Baseline (spike times) & 64 & $\sim$329 tokens/s\\
    Single-unit tokenizer & 16 & $\sim$553 tokens/s\\
    Pop.-level tokenizer (LDNS-Perceiver) & 8 & $\sim$800 tokens/s \\
    \bottomrule
  \end{tabular}
\end{table}

\begin{table}[h]
  \centering
  \caption{\textbf{Plotted means in Figure~\ref{fig:poyo-evals}.}}
  \label{tab:poyo-plotted-means}
  \small
  \setlength{\tabcolsep}{5pt}
  \begin{tabular}{llrrr}
    \toprule
    Panel & Condition & Baseline & Single-unit & Pop.-level \\
    \midrule
    (a) & Half-data test \(R^2\) & 0.8234 & 0.8024 & 0.7898 \\
     & Full-data test \(R^2\) & 0.8476 & 0.8365 & 0.8261 \\
    \midrule
    (b) & Finetune using 4 trials & 0.4618 & 0.4006 & 0.4015 \\
     & Finetune using 8 trials & 0.5911 & 0.5312 & 0.5306 \\
     & Finetune using 16 trials & 0.7002 & 0.6650 & 0.6752 \\
     & Finetune using 32 trials & 0.7709 & 0.7417 & 0.7512 \\
     & Finetune using 86 trials & 0.8187 & 0.7820 & 0.7966 \\
    \midrule
    (c) & Multi-session model validation \(R^2\) & 0.8497 & 0.8654 & 0.8397 \\
    \midrule
    (d) & Finetune using 4 trials & 0.4549 & 0.4791 & 0.4253 \\
     & Finetune using 8 trials & 0.6230 & 0.6214 & 0.5870 \\
     & Finetune using 16 trials & 0.7215 & 0.7431 & 0.6997 \\
     & Finetune using 32 trials & 0.7745 & 0.7842 & 0.7578 \\
     & Finetune using 86 trials & 0.8184 & 0.8309 & 0.7965 \\
    \bottomrule
  \end{tabular}
\end{table}

\subsection{Temporal decoding across the tokenizer zoo}
\label{app:poyo-tokenizer-zoo}

\begin{figure}[t]
    \centering
    \includegraphics[width=\linewidth]{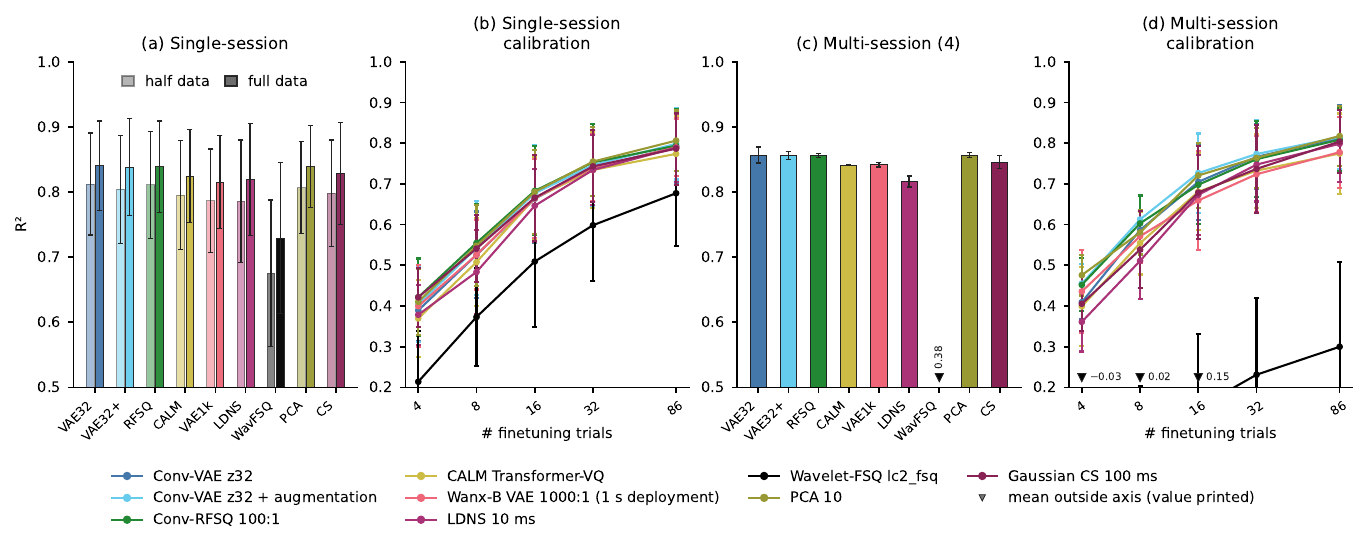}
    \caption{\textbf{POYO-style temporal decoding across nine
    frozen tokenizers.}
    \textbf{(a)} Single-session test $R^2$ using half or all
    available training data, averaged over eight sessions and
    two seeds ($n=16$ runs per bar).
    \textbf{(b)} Single-session calibration from each of two
    base sessions separately to the other seven sessions,
    averaged over base--target pairs and seeds
    ($n=28$ runs per trial budget).
    \textbf{(c)} Four-session base-model validation $R^2$,
    averaged over two seeds ($n=2$ runs per bar).
    \textbf{(d)} Calibration of the multi-session model to four
    additional sessions ($n=8$ runs per trial budget).
    Error bars indicate sample standard deviations across runs.
    Downward triangles indicate means below the displayed axis
    range, with numerical values shown beside the markers;
    some error bars extend beyond the displayed limits.}
    \label{fig:poyo-tokenizer-zoo}
\end{figure}

We extend the temporal-decoding evaluation to nine frozen tokenizers:
Conv-VAE (100:1, z32, SIGReg 1e-3), the same model trained with input jitter, Conv-RFSQ,
Transformer-VQ, Conv-VAE (1000:1), LDNS, Wavelet-FSQ, PCA, and Gaussian compressed sensing
(Figure~\ref{fig:poyo-tokenizer-zoo}). The LDNS model in this
comparison is the fixed-population LDNS baseline, rather than the
LDNS-Perceiver model used in the main three-method comparison.
We use the same POYO decoder architecture, training schedule, and
evaluation protocol described above. Single-session calibration
starts from each of two base sessions separately and adapts to
the other seven sessions. The multi-session training and calibration
sessions are listed in Table~\ref{tab:poyo-fold-a}.
Conv-VAE (1000:1) uses a one-second tokenizer window.

For fixed-population encoders, we use the all-unit input handoff
(\texttt{tokenizer.full\_window\_units=true}), which retains silent
neurons and preserves session-local channel indexing before
fixed-width input adaptation. In contrast, selecting only neurons
that spike within each window can change channel assignments across
windows, making this active-unit handoff unsuitable for encoders
that process neuron channels positionally. All applicable runs in
Figure~\ref{fig:poyo-tokenizer-zoo} use the all-unit handoff.

\section{Correlation of latent space quality with downstream metrics}
\label{app:latent_vs_downstream}

\begin{figure}[t]
    \centering
    \includegraphics[width=0.99\linewidth]{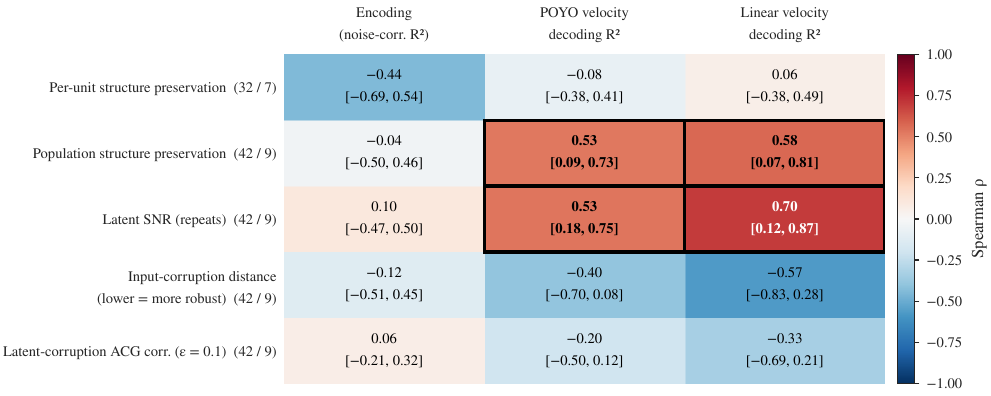}
    \caption{\textbf{Correlation of latent space quality axes with downstream metrics.}  Each cell gives the Spearman rank correlation across tokenizers (excluding tokenizers with 1s token bins) between a latent-quality metric (rows) and encoding (noise-corrected $R^2$) or cursor-velocity decoding (POYO, linear) $R^2$ (columns).
  Intervals show 95\% CI from 10,000 bootstrap resamples of tokenizer families. Row labels: number of tokenizers / number of tokenizer families. Boxed and bold: CI excludes 0 (uncorrected).}   
    \label{fig:appdx_latent_vs_downstream_heatmap}
\end{figure}

\begin{figure}[t]
    \centering
    \includegraphics[width=0.90\linewidth]{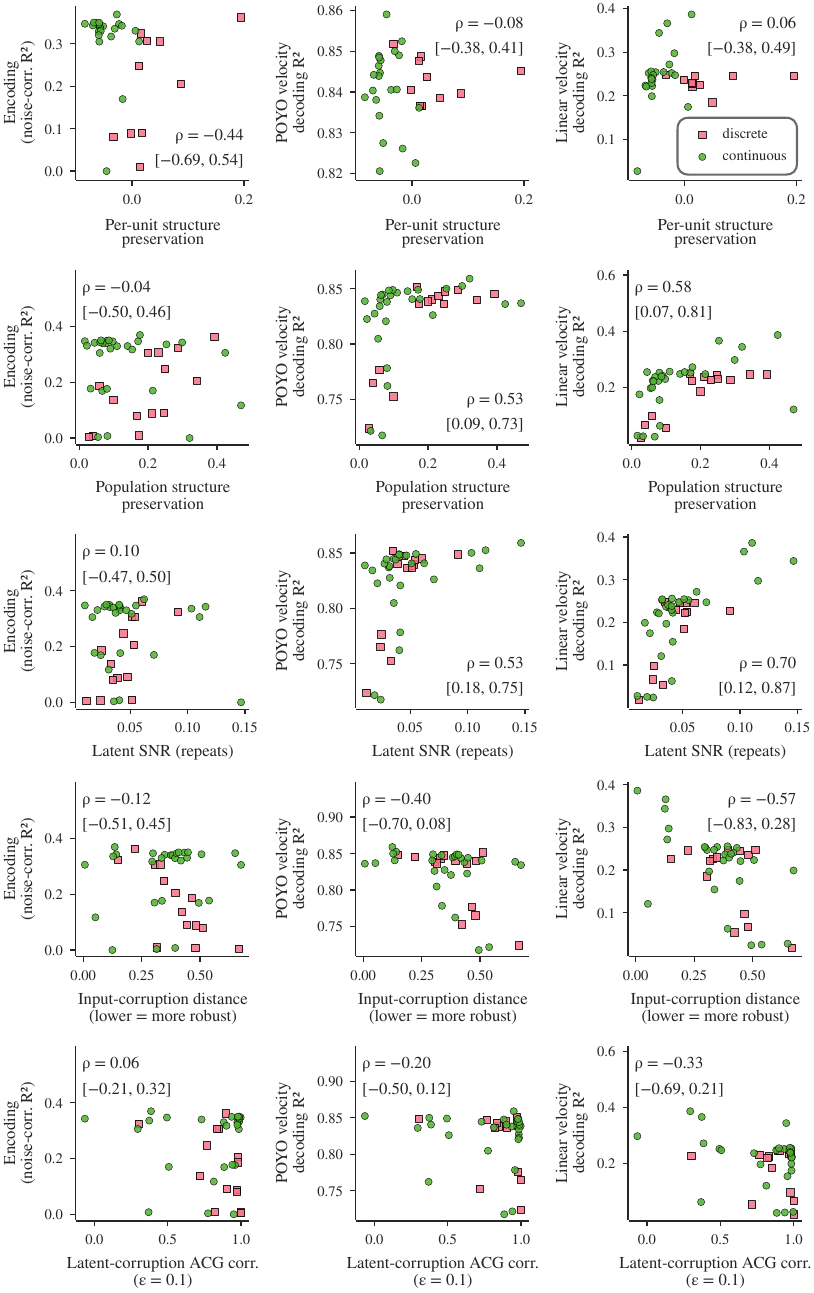}
    \caption{\textbf{Latent-quality metrics vs. downstream performance, per tokenizer.} One dot is one tokenizer, tokenizers with 1s token bin sizes are excluded. $\rho$: Spearman rank correlation; brackets: 95\% CI from 10,000 bootstrap resamples over tokenizer families. Latent quality metric values are not Poisson-normalized here.}   
    \label{fig:appdx_latent_vs_downstream_scatter}
\end{figure}

We include latent space metrics for diagnostics and as potential proxies for downstream evaluations, which are costly to evaluate. As figures \ref{fig:appdx_latent_vs_downstream_heatmap} and \ref{fig:appdx_latent_vs_downstream_scatter} show, there is indication that SNR of latents, and to a lesser degree population RSA, correlate with motor velocity decoding, in particular linear decoding. Given that our data contains only 12 families, not all of which can be evaluated on all latent quality metrics, results are preliminary and not multiple-comparison corrected. Whether these correlations hold up will show as more tokenizers are added to the evaluation.
No latent space metric correlates with encoding.

% \begin{figure}[t]
%     \centering
%     \includegraphics[width=0.5\linewidth]{figures/metric_analysis/latent_vs_X_scatterplots/snr_vs_input_rob_200ms_familyboot.pdf}
%     \caption{\textbf{Input robustness predicts SNR at higher (worse) input robustness values.} Same plot type as scatterplots in fig.~\ref{fig:appdx_latent_vs_downstream_scatter}. }
%     \label{fig:appdx_snr_vs_input_rob}
% \end{figure}

\FloatBarrier

\section{Downstream Encoding Details}
\label{app:downstream-encoding}

Here we describe the architecture, training procedure, and evaluation
used for the V4 image-encoding experiments on the Cowley dataset \citep{cowleyCompactDeepNeural2026}. The central
point is that the rate-like prediction of a tokenizer-based encoder is the
conditional mean of the decoded response under the model's predicted token
distribution. This is the direct analog of reading the rate parameter from a
Poisson encoder, and is the quantity that should be compared to repeat-averaged
PSTHs.

\subsection{Architecture}
\label{app:downstream-encoding-architecture}

All V4 experiments use the same architecture family. A Hiera vision backbone~\citep{ryali2023hiera}
\(\Phi\) maps the static image \(x \in \mathbb{R}^{C \times H \times W}\) to a
spatial token grid. We use a data-driven (trained from scratch on the encoding task) small Hiera backbone (about 4.7M parameters) at
\(112 \times 112\) resolution. A session-specific readout
\(\mathrm{Read}_s\) maps these features to predictions for the \(N_s\) units in
session \(s\). The readout uses \(N_s\) learnable queries
\(Q_s \in \mathbb{R}^{N_s \times d}\), cross-attends to the spatial features,
and then applies an MLP and temporal projection to produce \(T\) output bins.
The backbone is shared across sessions, while the readout is session-specific.
We only train on one subject (``Pepe'') to reduce training times for the downstream evaluation.

For tokenizer-based variants the readout emits a conditioning tensor
\(h_\theta(x) \in \mathbb{R}^{N_s \times T_z \times d_z}\), where \(T_z\) is
the number of latent time steps. A loss-specific head turns this tensor into
the predictive object used for training and evaluation:
\begin{itemize}
    \item \texttt{poisson}: a softplus output gives a
    per-bin firing rate directly; no tokenizer is used.
    \item \texttt{cross\_entropy}: a linear classifier produces logits over a
    vocabulary of size \(V\). For residual VQ with \(Q\) codebooks, the head
    produces \(Q V\) logits, interpreted as \(Q\) per-stage categorical
    distributions.
    \item \texttt{flow\_matching}, diffusion, and energy-based losses:
    \(h_\theta(x)\) conditions a sampler in continuous latent space.
    \item \texttt{mse} and \texttt{smooth\_l1}: a deterministic regression head
    predicts the tokenizer's continuous latent target.
\end{itemize}
For all tokenizer variants, the tokenizer encoder \(E\) and decoder \(D\) are
trained separately and then frozen. The downstream model sees the response
through \(E\) during training and through \(D\) during evaluation.

\subsection{Training}
\label{app:downstream-encoding-training}

The encoder is trained with AdamW, cosine annealing, a linear warmup over the
first 1000 steps, a backbone learning-rate multiplier (1 here, since the backbone is trained from scratch), mixed
\texttt{bf16} precision, and gradient clipping at 1.0. For tokenizer variants, the
tokenizer is loaded once and kept frozen, and the loss is computed in latent
space. A tokenizer-based training step computes \(h_\theta(x)\), encodes the
response as \(z = E(y)\), and minimizes the loss-specific likelihood or
divergence between the predictive distribution and the encoded target. The
Poisson variant instead minimizes the likelihood directly on the binned
response \(y\).

\subsection{From Single-Trial Likelihoods to Rate Predictions}
\label{app:downstream-encoding-rate-predictions}

Poisson and tokenizer-based encoders can both be viewed as conditional
generative models of neural responses. A PSTH is not observed on an individual
trial; it is a conditional expectation under the response distribution. Thus,
when a model predicts a distribution over latent response tokens, the
Poisson-rate analog is the expectation of the decoded response under that
predicted distribution.

\subsubsection{Poisson Encoders Learn the Conditional Mean}
\label{app:downstream-encoding-poisson-mean}

Let \(x\) denote the stimulus and
\(y \in \mathbb{N}^{T}\) a binned spike train. Under the Poisson observation
model
\begin{equation}
p_\theta(y \mid x)
= \prod_{t=1}^{T}
\mathrm{Poisson}\!\left(y_t;\lambda_{\theta,t}(x)\right),
\end{equation}
the per-stimulus negative log-likelihood, up to constants, is
\begin{equation}
\mathcal{L}_{\mathrm{Poisson}}(\lambda;x,y)
= \sum_t \left[\lambda_t(x) - y_t \log \lambda_t(x)\right].
\end{equation}
For fixed \(x\), the population risk is
\begin{equation}
\mathcal{R}(\lambda \mid x)
= \mathbb{E}_{y \sim p(y\mid x)}
\left[
\sum_t \lambda_t(x) - y_t \log \lambda_t(x)
\right].
\end{equation}
Differentiating with respect to \(\lambda_t(x)\) gives
\begin{equation}
\frac{\partial \mathcal{R}}{\partial \lambda_t(x)}
= 1 - \frac{\mathbb{E}[y_t \mid x]}{\lambda_t(x)}.
\end{equation}
The population minimizer is therefore
\begin{equation}
\lambda_t^*(x) = \mathbb{E}[y_t \mid x].
\end{equation}
Maximum-likelihood Poisson training on single trials, therefore, learns a
conditional mean response, i.e.\ the PSTH up to the chosen binning.

\subsubsection{Tokenizer Encoders Need Distributional Losses}
\label{app:downstream-encoding-tokenizer-losses}

In the tokenized formulation, \(z = E(y)\) is a latent response representation
and \(D\) maps latents back to response space. The downstream model should
represent \(q_\theta(z\mid x)\), an approximation to the conditional law of
encoded responses.

For a discrete tokenizer with \(z \in \{1,\ldots,V\}\), cross-entropy training
on single-trial samples minimizes
\begin{equation}
\mathcal{R}(q \mid x)
= -\mathbb{E}_{z \sim p(z\mid x)} \log q(z\mid x)
= H(p(\cdot\mid x))
+ \mathrm{KL}\!\left(p(\cdot\mid x)\,\|\,q(\cdot\mid x)\right).
\end{equation}
Since the entropy term is independent of \(q\), the population minimizer is
\begin{equation}
q^*(z\mid x) = p(z\mid x).
\end{equation}
For residual VQ, the implemented objective uses independent per-stage
cross-entropies and therefore recovers the best product-of-marginals
approximation under the model factorization
\(\prod_q q_\theta^{(q)}(z^{(q)}\mid x)\).

By contrast, deterministic regression on continuous latents recovers only a
point estimate. For an MSE objective,
\begin{equation}
\mathcal{R}(h_\theta \mid x)
= \mathbb{E}_{z \sim p(z\mid x)}
\left\|h_\theta(x)-z\right\|_2^2,
\end{equation}
whose minimizer is
\begin{equation}
h_\theta^*(x) = \mathbb{E}[z\mid x].
\end{equation}
Decoding this point estimate yields \(D(\mathbb{E}[z\mid x])\), while the
desired conditional mean response is
\begin{equation}
\mathbb{E}[y\mid x]
\approx
\mathbb{E}_{z\sim p(z\mid x)}[D(z)].
\end{equation}
These quantities are generally different for non-linear decoders. Thus MSE,
smooth-\(\ell_1\), and related deterministic losses collapse posterior
variation before decoding, which is the wrong order of operations for PSTH
prediction.

Continuous distributional losses avoid this collapse. Flow matching~\citep{lipman2023flow}, for
example, trains a conditional vector field \(v_\theta(z,t\mid x)\) to transport
a simple prior \(p_0(z)\) to \(p(z\mid x)\). With linear interpolation
\(z_t=(1-t)z_0+t z_1\), a common objective is
\begin{equation}
\mathcal{L}_{\mathrm{flow}}(\theta)
=
\mathbb{E}_{t,z_0,z_1}
\left\|
v_\theta(z_t,t\mid x) - (z_1-z_0)
\right\|_2^2,
\quad
z_0 \sim p_0,\; z_1 \sim p(z\mid x).
\end{equation}
Under the standard assumptions for the sampler and model class, the trained
ODE produces samples from \(q_\theta(z\mid x)\) rather than a delta at the
latent mean. Diffusion and energy-based samplers play the same role: they
preserve a predictive latent distribution that can be marginalized through
the decoder.

\subsubsection{The Principled Rate Analog}
\label{app:downstream-encoding-rate-analogue}

When \(q_\theta(z\mid x)\) approximates the conditional latent distribution,
the rate-like prediction is
\begin{equation}
\hat r_\theta(x)
=
\mathbb{E}_{z \sim q_\theta(z\mid x)}[D(z)].
\end{equation}
For a discrete single-token latent this is the finite sum
\begin{equation}
\hat r_\theta(x)
=
\sum_{z=1}^{V} q_\theta(z\mid x)D(z),
\end{equation}
and for continuous latents it is approximated by Monte Carlo:
\begin{equation}
\hat r_\theta(x)
\approx
\frac{1}{K}\sum_{k=1}^{K}D(z_k),
\qquad
z_k \sim q_\theta(z\mid x).
\end{equation}
This follows from the law of total expectation,
\begin{equation}
\mathbb{E}[y\mid x]
=
\mathbb{E}_{z \sim p(z\mid x)}
\left[\mathbb{E}[y\mid z,x]\right].
\end{equation}
Identifying the deterministic decoder output \(D(z)\) with the expected
response represented by \(z\) gives
\begin{equation}
\hat r_\theta(x)
=
\mathbb{E}_{z \sim q_\theta(z\mid x)}[D(z)]
\approx
\mathbb{E}[D(E(y))\mid x]
\approx
\mathbb{E}[y\mid x].
\end{equation}
The Poisson model exposes the conditional mean directly as
\(\lambda_\theta(x)\); the tokenizer model exposes it indirectly through
\(q_\theta\) and \(D\).

\subsubsection{Argmax and Single-Sample Decoding}
\label{app:downstream-encoding-argmax}

Decoding the maximum-probability token,
\(D(\arg\max_z q_\theta(z\mid x))\), gives a MAP latent response, not a
conditional mean. Decoding one sampled token gives one model-generated trial.
Neither quantity is matched to a repeat-averaged PSTH, which is an empirical
expectation. The correct PSTH-like quantity averages over the model's full
predictive distribution.

\subsection{Loss-Mode-Specific Implementation}
\label{app:downstream-encoding-implementation}

The model architecture is shared across loss modes; only the evaluation-time
decoding changes.

\subsubsection{Direct-Rate Models}
\label{app:downstream-encoding-direct-rate}

For \texttt{poisson}, the model output is already the predicted rate:
\begin{equation}
\hat r_\theta(x) = \lambda_\theta(x).
\end{equation}

\subsubsection{Cross-Entropy with One Codebook}
\label{app:downstream-encoding-ce-q1}

For a single-codebook tokenizer, each latent position has a categorical
distribution over \(V\) tokens. Because our windowed tokenizer decoder is local
enough in time for this evaluation, we precompute, once per session, decoded
response windows for each position \(p\) and token \(k\):
\begin{equation}
R_{p,k}[t,n]
=
D(\text{token } k \text{ at position } p)[t,n].
\end{equation}
The expected rate is then computed by an \(\mathrm{einsum}\):
\begin{equation}
\hat r_\theta(x)[t,n]
=
\sum_{p=1}^{T_z}\sum_{k=1}^{V}
q_\theta(z_p=k\mid x) R_{p,k}[t,n],
\qquad
t \in \mathrm{window}(p).
\end{equation}
This is exact for the implemented window-local approximation and avoids
Monte Carlo variance.

\subsubsection{Cross-Entropy with Residual VQ}
\label{app:downstream-encoding-ce-rvq}

For \(Q>1\) residual codebooks, exact enumeration would require \(V^Q\) joint
tuples per latent position. In the \(V=1024, Q=4\) setting this is
intractable, so we sample joint codes from the model's per-stage product
distribution:
\begin{equation}
\hat r_\theta(x)
\approx
\frac{1}{K}\sum_{k=1}^{K}D(\mathbf{z}_k),
\qquad
z_k^{(q)} \sim q_\theta^{(q)}(\cdot\mid x).
\end{equation}
We use \(K=32\) samples in the reported experiments. The approximation error
comes from Monte Carlo variance and from the product factorization imposed by
the model.

\subsubsection{Flow Matching}
\label{app:downstream-encoding-continuous-sampling}

For continuous latent models, \(q_\theta(z\mid x)\) is implicit in an ODE. We estimate the decoded conditional mean by
\begin{equation}
\hat r_\theta(x)
\approx
\frac{1}{K}\sum_{k=1}^{K}D(z_k),
\qquad
z_k \sim q_\theta(z\mid x),
\end{equation}
with \(K=32\) samples per stimulus. For flow matching we use a five-step Heun ODE solver.
After sampling, we simply decode each sample and average the decoded responses.

\subsubsection{MSE and Smooth-\(\ell_1\)}
\label{app:downstream-encoding-mse}

The deterministic latent-regression heads emit one latent prediction and
therefore require no Monte Carlo. We decode this latent once, obtaining
\(D(h_\theta(x))\). Because this corresponds to decoding a point estimate
rather than marginalizing a latent distribution, we report it as a baseline
for the cost of mean-only latent regression.

\subsection{Metrics and Aggregation}
\label{app:downstream-encoding-metrics}

For each session we evaluate both single-trial validation data and PSTH
validation data, where each unique image is represented by its empirical
repeat-average. We compute three neuron-wise metrics when applicable:
\begin{enumerate}
    \item \emph{Rate Pearson}: cross-image Pearson correlation between
    predicted and observed time-summed responses.
    \item \emph{Temporal Pearson}: per-image Pearson correlation between
    predicted and observed response time courses, averaged over images.
    \item \emph{Noise-corrected \(R^2\)}: the Pospisil and Bair ER estimator~\citep{pospisil2021unbiased},
    computed on single-trial validation where repeats are available.
\end{enumerate}
For each metric and session, we report the mean and standard deviation across
neurons. Across sessions, we report the mean of session means and
\(\mathrm{SE}=\mathrm{std}(\text{session means})/\sqrt{K}\), treating sessions
as independent observations.

\subsection{Caveat: Tokenizer Fidelity}
\label{app:downstream-encoding-tokenizer-fidelity}

The marginalization argument recovers \(\mathbb{E}[y\mid x]\) only to the
extent that \(D(E(y)) \approx y\). If the tokenizer is lossy, the predicted
quantity is
\begin{equation}
\mathbb{E}[D(E(y))\mid x]
\end{equation}
rather than \(\mathbb{E}[y\mid x]\). Agreement with empirical PSTHs therefore
depends on the fidelity and inductive biases of the tokenizer-decoder pair. We
evaluate this directly by measuring held-out reconstructions through
\(D \circ E\); when the tokenizer preserves PSTH-relevant response structure,
the marginalized decoder prediction is a faithful estimate of the conditional
mean response.

\clearpage
\FloatBarrier

%%%%%%%%%%%%%%%%%%%%%%%%%%%%%%%%%%%%%%%%%%

\section{Additional Results Tables}

% \subsection{Latent structure and robustness}
\begin{table}[H]
\centering
\setlength{\tabcolsep}{4pt}
\caption{\textbf{Latent space structure and robustness metrics for a subset of tokenizers:} Pop. RSA: Population RSA (unnormalized), SNR (unnormalized), robustness to window shifts of the input and to jittering a fraction of spikes by a small amount. -- indicates missing data.}
\label{tab:latent_metrics}
\begin{tabular}{lrrrr}
\toprule
Tokenizer & \makecell{Pop.\\RSA} & SNR & \makecell{Shift-\\robustness} & \makecell{Jitter-\\robustness} \\
\midrule
Poisson (100 ms) & 0.540 & 0.412 & 0.983 & 0.999 \\
PCA (1000:1) & 0.520 & -- & 0.999 & 1.000 \\
C.S. (100:1) & 0.017 & 0.010 & 0.370 & 0.827 \\
Transformer-VQ (100:1) & 0.342 & 0.053 & 0.525 & 0.856 \\
Wavelet-FSQ & 0.099 & 0.033 & 0.470 & 0.916 \\
Conv-RFSQ (100:1) & 0.286 & 0.091 & 0.724 & 0.977 \\
Conv-VAE (100:1) & 0.110 & 0.041 & 0.505 & 0.874 \\
LDNS & 0.055 & 0.036 & 0.617 & 0.921 \\
LDNS-Perceiver & 0.470 & 0.031 & 0.899 & 0.987 \\
\bottomrule
\end{tabular}
\end{table}

\begin{table}[H]
  \centering
  \caption{\textbf{Population-tokenizer reconstruction (10\,ms bins).}
  LDNS-Perceiver outputs firing rates; the metrics are computed on the decoded
  rates directly, without Poisson sampling. Columns: rate-matched Poisson floor
  (rates estimated from each session in 10\,ms bins); LDNS-Perceiver. ``Raw''
  columns report the raw metric value; ``Norm.'' columns are normalized
  scores (Poisson floor $=0$, perfect-reconstruction ceiling $=1$, higher is
  better in all cases; for firing\_rate\_err., $0$ is instead every unit
  given the population's mean rate). Single-neuron-resolution metrics (spike-detection F1,
  temporal precision, ISI) are omitted: they need spike times, which a 10\,ms
  rate output does not provide. Sessions: the 20 public test sessions (7 held-out-time, 13 held-out-data).}
  \label{tab:results-recon-pop-spike}
  \scriptsize
  \setlength{\tabcolsep}{4pt}
  \begin{tabular}{l c rr r}
    \toprule
    & & \multicolumn{2}{c}{Raw} & {Norm.} \\
    \cmidrule(lr){3-4} \cmidrule(lr){5-5}
    Metric & Dir. & Poisson & LDNS-Perceiver & LDNS-Perceiver \\
    \midrule
      binned\_rate\_diff.\ 10ms & $\downarrow$ & 7.177 & 7.270 & $-$0.013 \\
      binned\_rate\_diff.\ 50ms & $\downarrow$ & 5.592 & 4.482 & 0.199 \\
      smooth.\ SE 10ms & $\downarrow$ & 0.992 & 0.817 & 0.176 \\
      smooth.\ SE 50ms & $\downarrow$ & 0.993 & 2.955 & $-$1.976 \\
      smooth.\ SE 100ms & $\downarrow$ & 0.993 & 5.427 & $-$4.464 \\
      spike\_count\_corr.\ 10ms & $\uparrow$ & 0.690 & 0.865 & 0.564 \\
      spike\_count\_corr.\ 50ms & $\uparrow$ & 0.715 & 0.864 & 0.523 \\
      firing\_rate\_err. & $\downarrow$ & 0.102 & 1.578 & 0.782 \\
      acg\_corr.\ 10ms & $\uparrow$ & 0.804 & 0.879 & 0.383 \\
      fano\_err.\ 10ms & $\downarrow$ & 0.989 & 0.418 & 0.577 \\
      fano\_err.\ 50ms & $\downarrow$ & 0.989 & 0.528 & 0.466 \\
      ccg\_corr.\ 10ms & $\uparrow$ & 0.582 & 0.812 & 0.551 \\
      mut.\ info.\ 10ms & $\uparrow$ & 0.534 & 0.686 & 0.325 \\
      mut.\ info.\ 50ms & $\uparrow$ & 0.443 & 0.645 & 0.362 \\
      part.\ ratio err.\ 10ms & $\downarrow$ & 10.566 & 10.150 & 0.039 \\
      subspace overlap 10ms & $\uparrow$ & 0.691 & 0.636 & $-$0.181 \\
    \bottomrule
  \end{tabular}
\end{table}

% =============================================================================
% Table 3: Downstream behavior decoding (latents -> behavior)
% =============================================================================
\begin{table}[t]
  \centering
  \caption{\textbf{Downstream behavior decoding.} Targets are predicted from tokenizer latents using a linear probe (regression, classification) or a temporal decoder. The reported metric is the mean $R^2$ for regression and classification accuracy for classifiers. POYO scores are the 12-session means of Table~\ref{tab:main_results} (Appendix~\ref{sec:metrics-details}), not the eight-session protocol of Figure~\ref{fig:poyo-evals}. Implementation details in Sections \ref{app:downstream-decoding-linear-probes} and \ref{app:downstream-decoding-poyo}.
}
  \label{tab:results-decoding}
  \small
  \setlength{\tabcolsep}{4pt}
  \resizebox{\linewidth}{!}{%
  \begin{tabular}{l l c c c c c c c c c}
    \toprule
Metric & Score & Dir.
& \makecell{Conv-FSQ\\(100:1, 6 dims)} & \makecell{Transformer-VQ\\(100:1)} & \makecell{C.S.\\(100:1)} & \makecell{Conv-VAE\\(100:1)} & Wavelet-FSQ & LDNS-Perceiver & \makecell{Conv-RFSQ\\(100:1)} \\
\midrule

\multicolumn{10}{l}{\emph{Linear probe}} \\
\quad Cursor position  & $R^2$ & $\uparrow$
& 0.14 & 0.19 & 0.00 & 0.17 & 0.01 & 0.05 & 0.16 \\
\quad Cursor velocity  & $R^2$ & $\uparrow$
& 0.22 & 0.25 & 0.03 & 0.26 & 0.05 & 0.12 & 0.23 \\
\quad Trial target ID  & acc.  & $\uparrow$
& 0.35 & 0.35 & 0.20 & 0.37 & 0.20 & 0.23 & 0.37 \\
\quad Movement-phase   & acc.  & $\uparrow$
& 0.80 & 0.82 & 0.69 & 0.82 & 0.78 & 0.75 & 0.81 \\

\midrule
\multicolumn{10}{l}{\emph{Temporal decoder (POYO)}} \\
\quad Cursor velocity  & $R^2$ & $\uparrow$
& 0.84 & 0.84 & 0.84 & 0.85 & 0.71 & 0.84 & 0.85 \\

    \bottomrule
  \end{tabular}}
\end{table}
% =============================================================================
% Table 4: Downstream encoding (forward modelling)
% =============================================================================
\begin{table}[t]
  \centering
  \caption{\textbf{Downstream encoding.} Forward modeling of neural activity from images using tokenizer latents as the prediction target, scored by reconstruction quality on held-out images and responses.}
  \label{tab:results-encoding}
  \small
  \setlength{\tabcolsep}{3pt}
  \resizebox{\linewidth}{!}{%
  \begin{tabular}{l l c c c c c c c c c c c}
    \toprule
    Metric & Score & Dir. & Poisson @ 10 ms & \makecell{Conv-FSQ\\(100:1, 6 dims)} & \makecell{Transformer-VQ\\(100:1)} & \makecell{C.S.\\(100:1)} & \makecell{Conv-VAE\\(100:1)} & \makecell{Conv-RFSQ\\(100:1)} & Wavelet-FSQ & Wavelet-AE & LDNS & LDNS-Perceiver \\
    \midrule
    \multicolumn{13}{l}{\emph{Encoding-decoding roundtrip}} \\
    \quad Roundtrip temporal correlation & Pearson $r$ & $\uparrow$ & 1.0 & 0.872 & 0.749 & 0.965 & 0.989 & -- & -- & -- & -- & 0.545 \\
    \quad Roundtrip rate correlation & Pearson $r$ & $\uparrow$ & 1.0 & 0.950 & 0.828 & 0.992 & 0.994 & -- & -- & -- & -- & 0.566 \\
    \midrule
    \multicolumn{13}{l}{\emph{Stimulus $\rightarrow$ neural encoding (\citealp{cowleyCompactDeepNeural2026})}} \\
    \quad PSTH temporal correlation & Pearson $r$ & $\uparrow$ & \textbf{0.462} & 0.368 & 0.361 & \underline{0.369} & 0.323 & 0.187 & 0.059 & 0.086 & 0.268 & 0.366 \\
    \quad PSTH rate correlation & Pearson $r$ & $\uparrow$ & 0.236 & 0.005 & 0.315 & \textbf{0.403} & \underline{0.380} & 0.377 & 0.201 & 0.255 & $-$0.001 & 0.198 \\
    \quad Noise-corrected response fit & $R^2_{\mathrm{NC}}$ & $\uparrow$ & 0.128 & 0.010 & 0.205 & \textbf{0.347} & \underline{0.329} & 0.325 & 0.137 & 0.176 & 0.003 & 0.117 \\
    \bottomrule
  \end{tabular}}
    \vspace{2pt}
  \footnotesize
  \emph{Note. Tokenizer-decoded responses are re-sampled at 100 Hz prior to metric computation, except the Wavelet tokenizers', which are scored at the data's native 1\,ms resolution.}
\end{table}
%%%%%%%%%%%%%%%%%%%%%%%%%%%%%%%%%%%%%%%%%%%%%%%%%%%%%%%%%%%%

\begin{table}[t]
  \centering
  \caption{\textbf{Reconstruction quality across generalization regimes.} All
  tokenizers produce per-neuron tokens (some also attend across neurons; Appendix~\ref{app:tokenizer-details}). HR: convolutional
  tokenizers at $100:1$ (1\,tok/100\,ms); LR: Conv-VAE at $1000:1$ (1\,tok/1\,s,
  5$\times$ longer context). RFSQ/RLFQ:
  residual FSQ/LFQ; BSQ: binary spherical quantization. Regimes:
  xT = held-out time within session (7 sessions); xS = held-out session, same
  subject (6); xSub = held-out subject, same source (4); xSrc = held-out source (3). Pearson: spike-count correlation (1\,ms bins).
  Italicized LR-ACG values are negative because one token per second cannot
  capture sub-second ISI structure. \textbf{Bold}: best in row.}
  \label{tab:results-recon}
  \scriptsize
  \setlength{\tabcolsep}{2.5pt}
  \begin{tabular}{ll ccccc c}
    \toprule
    & & \multicolumn{5}{c}{HR ($100:1$)} & LR ($1000:1$) \\
    \cmidrule(lr){3-7}
    Metric & Reg.
      & Conv-VAE & Conv-RFSQ & Conv-RLFQ & Conv-FSQ & Conv-BSQ & Conv-VAE \\
    \midrule
    \multirow{4}{*}{Pearson $\uparrow$}
      & xT   & \textbf{0.99} & 0.93 & 0.94 & 0.87 & 0.90 & 0.42 \\
      & xS   & \textbf{0.99} & 0.93 & 0.94 & 0.88 & 0.91 & 0.43 \\
      & xSub & \textbf{0.99} & 0.96 & 0.97 & 0.92 & 0.94 & 0.46 \\
      & xSrc & \textbf{0.98} & 0.88 & 0.90 & 0.80 & 0.85 & 0.39 \\
    \midrule
    \multirow{4}{*}{ACG $\uparrow$}
      & xT   & \textbf{0.98} & 0.72 & 0.87 & 0.61 & 0.77 & \textit{$-$0.17} \\
      & xS   & \textbf{0.97} & 0.71 & 0.87 & 0.61 & 0.75 & \textit{$-$0.23} \\
      & xSub & \textbf{0.99} & 0.83 & 0.94 & 0.79 & 0.87 & \textit{$-$0.17} \\
      & xSrc & \textbf{0.99} & 0.59 & 0.89 & 0.51 & 0.74 & \textit{$-$0.27} \\
    \midrule
    \multirow{4}{*}{Fano err.\ $\downarrow$}
      & xT   & $\mathbf{3.6{\times}10^{-4}}$ & $2.6{\times}10^{-3}$ & $2.0{\times}10^{-3}$ & $3.3{\times}10^{-3}$ & $1.6{\times}10^{-3}$ & $9.6{\times}10^{-3}$ \\
      & xS   & $\mathbf{3.0{\times}10^{-4}}$ & $2.7{\times}10^{-3}$ & $2.0{\times}10^{-3}$ & $3.4{\times}10^{-3}$ & $1.7{\times}10^{-3}$ & $8.8{\times}10^{-3}$ \\
      & xSub & $\mathbf{6.7{\times}10^{-5}}$ & $8.5{\times}10^{-4}$ & $6.2{\times}10^{-4}$ & $1.1{\times}10^{-3}$ & $5.2{\times}10^{-4}$ & $4.1{\times}10^{-3}$ \\
      & xSrc & $\mathbf{4.1{\times}10^{-4}}$ & $4.4{\times}10^{-3}$ & $3.3{\times}10^{-3}$ & $5.6{\times}10^{-3}$ & $2.9{\times}10^{-3}$ & $1.2{\times}10^{-2}$ \\
    \midrule
    \multirow{4}{*}{Spk-F1 $\uparrow$}
      & xT   & \textbf{0.98} & 0.91 & 0.92 & 0.86 & 0.89 & 0.36 \\
      & xS   & \textbf{0.98} & 0.93 & 0.94 & 0.88 & 0.91 & 0.38 \\
      & xSub & \textbf{0.99} & 0.95 & 0.96 & 0.92 & 0.94 & 0.42 \\
      & xSrc & \textbf{0.98} & 0.88 & 0.90 & 0.81 & 0.86 & 0.34 \\
    \midrule
    \multirow{4}{*}{Subsp.\ ov.\ $\uparrow$}
      & xT   & \textbf{0.94} & 0.73 & 0.80 & 0.71 & 0.82 & 0.51 \\
      & xS   & \textbf{0.91} & 0.79 & 0.83 & 0.74 & 0.79 & 0.63 \\
      & xSub & \textbf{0.94} & 0.82 & 0.83 & 0.82 & 0.85 & 0.76 \\
      & xSrc & \textbf{0.86} & 0.77 & 0.85 & 0.76 & 0.85 & 0.75 \\
    \bottomrule
  \end{tabular}
\end{table}

\clearpage
\FloatBarrier

% DRAFT generated by scripts/make_deliverables.py from validated results (see results/PROVENANCE.md).
% Requires booktabs, makecell and graphicx (loaded by the paper). Do not edit numbers by hand; regenerate instead.
\section{Inference cost}
\label{app:inference-cost}

We evaluate tokenizer inference cost at a matched window duration and population size: a 1\,s window of
$N=1000$ neurons, with batch size 1. The encoder maps binned spike counts to tokens; the decoder reconstructs spike
counts from those tokens, including any conversion to latent vectors. We use zero-filled synthetic spike counts
and retain each checkpoint's architecture and input bin size: 1\,ms for all rows except LDNS and LDNS-Perceiver,
which use 10\,ms bins. For PCA and compressed sensing, each token spans the full evaluated window, so its
duration is adjusted with the window length. Other tokenizers retain their checkpoint token durations.
Several measurements differ from the training window or population size, and shape-dependent weights are
re-initialized where required. These measurements characterize computational cost,
not reconstruction accuracy at the modified input size.

We use PyTorch's \texttt{FlopCounterMode} to count matrix multiplications, convolutions, and attention.
We supplement these counts with shape-based estimates for real fast Fourier transforms (FFTs) in the
state-space layers of LDNS and LDNS-Perceiver and for pairwise distances in vector-quantizer lookup:
$2.5\,n\log_2 n$ operations per length-$n$ real FFT and $2nmd$ operations for distances between $n$ and $m$
vectors of dimension $d$, summed over all transforms or batches. Elementwise operations, including activations,
normalization, and softmax, are excluded. Reported GFLOP are the sum of counted and supplemental operations,
divided by $10^9$, for one input window. FLOP counts were identical across all repetitions.

We measure latency with one dedicated NVIDIA H100 80GB GPU per job, using fp32 with TF32 disabled, except that
LDNS-Perceiver retains its native bf16 attention kernel. Perceiver attention FLOPs are counted using a mathematically
equivalent fp32 attention implementation. Models run in evaluation mode with gradients disabled. After one untimed
encoder--decoder pass to populate model caches, each timed loop comprises three warm-up calls followed by ten
calls measured with CUDA synchronization. Table~\ref{tab:inference-cost} reports the median of 6 repetition means from
3 independent jobs, with two repetitions per job. These are steady-state, batch-size-one latencies;
they exclude checkpoint loading and do not measure throughput at larger batch sizes. The model-size column counts
scalar elements in model parameters and buffers.

\begin{table}[h]
\centering
\caption{\textbf{Tokenizer inference cost at a matched window duration and population size.}
Encoder and decoder costs are measured separately for a 1\,s window of 1000 neurons, batch size 1.
Input bins remain native: 1\,ms except for LDNS and LDNS-Perceiver (10\,ms).
GFLOP include counted matrix, convolution, and attention operations plus estimates for FFTs and vector-quantizer
distance calculations; elementwise operations are excluded. Latency is measured on an NVIDIA H100 80GB with TF32
disabled and is the median of six repetition means, each based on ten timed calls.
Computation uses fp32 except for LDNS-Perceiver's native bf16 attention kernel.}
%\small
\begin{tabular}{lrrrrr}
\toprule
Tokenizer & \makecell{Encoder\\GFLOP} & \makecell{Decoder\\GFLOP} & \makecell{Model size\\(elements)} & \makecell{Encoder\\latency (ms)} & \makecell{Decoder\\latency (ms)} \\
\midrule
\multicolumn{6}{l}{\textit{Single-neuron tokenizers}} \\[2pt]
PCA & 0.020 & 0.020 & 11.0\,k & 0.093 & 0.074 \\
C.S. & 0.040 & 40.041 & 20.0\,k & 0.044 & 35.26 \\
Conv-VAE ($100:1$) & 16.696 & 2.957 & 0.78\,M & 2.40 & 1.91 \\
Conv-RFSQ ($100:1$) & 16.596 & 31.340 & 1.75\,M & 3.25 & 4.05 \\
Transformer-VQ ($100:1$) & 189.082 & 71.854 & 14.27\,M & 6.95 & 2.90 \\
Conv-VAE ($1000:1$) & 8.300 & 11.750 & 4.54\,M & 4.70 & 5.12 \\
\midrule
\multicolumn{6}{l}{\textit{Population tokenizers}} \\[2pt]
LDNS & 0.391 & 0.165 & 3.16\,M & 5.41 & 1.93 \\
Wavelet-FSQ & 184.897 & 180.304 & 5.31\,M & 9.14 & 8.59 \\
LDNS-Perceiver & 476.596 & 307.274 & 3.95\,M & 21.10 & 25.11 \\
\bottomrule
\end{tabular}

\label{tab:inference-cost}
\end{table}

\begin{figure}[h]
\centering
\includegraphics{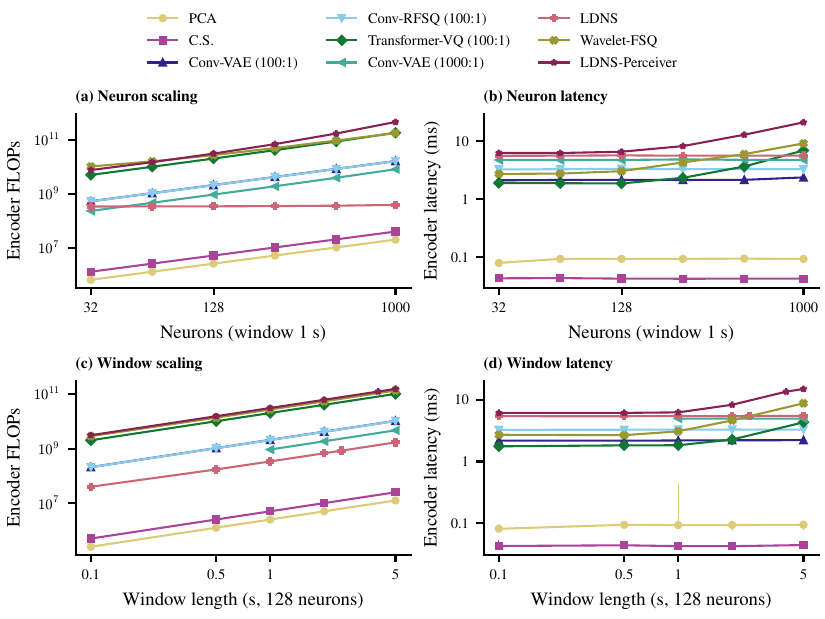}
\caption{\textbf{Encoder FLOPs and latency scale differently with input size.}
(a,b) Encoder FLOPs and latency versus neuron count at a 1\,s window; (c,d) versus window length at 128 neurons.
Both axes are logarithmic. Markers show median latency across six repetition means (ten calls each); bars span
their minimum and maximum. Conv-VAE ($1000:1$) is evaluated only for windows of at least 1\,s.}
\label{fig:inference-cost-scaling}
\end{figure}

\begin{figure}[h]
\centering
\includegraphics{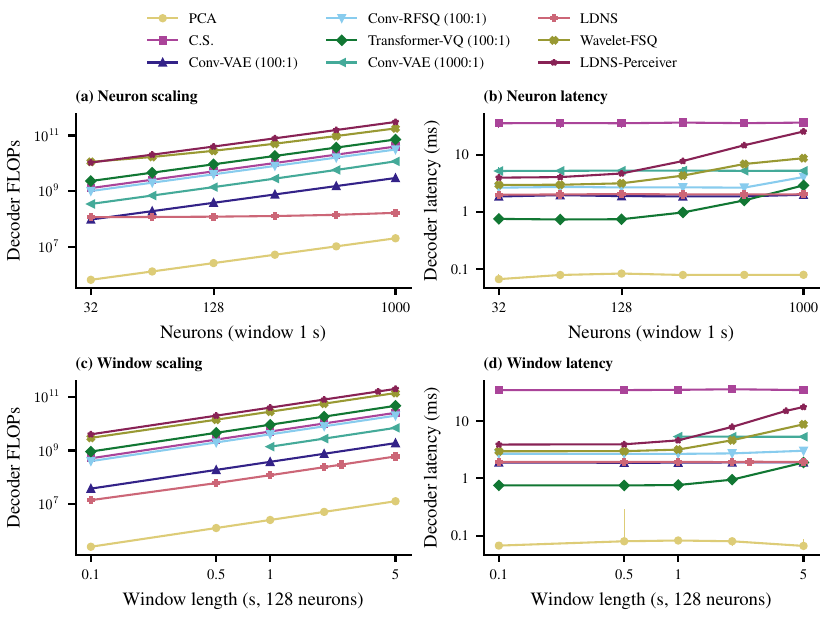}
\caption{\textbf{Decoder FLOPs and latency scale differently across tokenizers.}
(a,b) Decoder FLOPs and latency versus neuron count at a 1\,s window; (c,d) versus window length at 128 neurons.
Both axes are logarithmic. Markers show median latency across six repetition means (ten calls each); bars span
their minimum and maximum. Conv-VAE ($1000:1$) is evaluated only for windows of at least 1\,s.}
\label{fig:inference-cost-scaling-decoder}
\end{figure}

\paragraph{Scaling.} At a fixed 1\,s window, between 32 and 1000 neurons, the encoder FLOPs of PCA, C.S., Conv-VAE ($100:1$) and Conv-RFSQ ($100:1$) grow linearly with $N$. Transformer-VQ ($100:1$), Conv-VAE ($1000:1$) and LDNS-Perceiver grow faster than linearly (log-log slope between 512 and 1000 neurons: 1.12, 1.09, 1.48); for all three encoders, this reflects attention across neurons, including the middle attention block of Conv-VAE ($1000:1$). LDNS is nearly independent of $N$ (slope 0.04), since it projects the neurons to a fixed width in its first layer, and Wavelet-FSQ grows sub-linearly at small $N$ and close to linearly at large $N$ (slope 0.96 between 512 and 1000). At $N=128$, encoder and decoder FLOPs scale close to linearly with window length (endpoint log-log slopes 0.96--1.01) over the measured windows: 0.1--5\,s, except 1--5\,s for Conv-VAE ($1000:1$). Across the neuron or window sweeps, most tokenizers show little latency growth despite large increases in FLOPs. This is consistent with substantial per-call overhead at batch size 1; these measurements do not isolate its cause. Transformer-VQ ($100:1$), Wavelet-FSQ and LDNS-Perceiver show stronger latency growth in the measured range (Figures~\ref{fig:inference-cost-scaling} and~\ref{fig:inference-cost-scaling-decoder}).%% [2026-05-04 nrr] Population-tokenizer counterpart of the SN regime-split table

\clearpage
\newpage

\section{Compute resources}
\label{app:compute}

All experiments were run on internal GPU clusters using NVIDIA A100
(80\,GB) and L40S (48\,GB) accelerators, supplemented by a rented
cloud cluster of 8$\times$NVIDIA H100 GPUs. Approximate GPU-hour
totals are summarized in Table~\ref{tab:compute} and were estimated
from the project's W\&B run logs.

\paragraph{Tokenizer training.}
Tokenizer training runs spanned a wide range of compute footprints:
light convolutional tokenizers and non-learned baselines (PCA,
compressed sensing) train in $\approx$6\,h on a single A100 or L40S,
while transformer-based and population tokenizers (notably
LDNS-Perceiver and Wavelet-FSQ at high temporal resolution) train
for one to several days, putting the average run time across the
zoo at $\approx$50\,h per run on a single A100 or L40S GPU. The
reported zoo comprises $\approx$10 tokenizer families (12 sub-families
in the finer grouping of Appendix~\ref{app:metric-corr-analysis}) with
$\approx$10 hyperparameter and seed variants per family, totalling
$\approx$100 reported training runs; 47 of the resulting tokenizers
enter the final evaluation.

\paragraph{Downstream decoding.}
Each POYO temporal decoder run took $\approx$1\,h on a single A100
GPU; we report results across all decoders trained for the main
comparison (Figure~\ref{fig:poyo-evals}) and the tokenizer-zoo comparison
(Figure~\ref{fig:poyo-tokenizer-zoo}; run counts in the captions). The
cross-session calibration protocol additionally trains one multi-session
base model per tokenizer and seed for $\approx$12\,h each on A100 GPUs.

\paragraph{Downstream encoding.}
Each image-to-V4 encoder run (Hiera backbone with session-specific
readouts) took $\approx$10--15\,h on a single A100 GPU; we report
results across $\approx$30 trained encoders.

\paragraph{Evaluation pipeline.}
The full evaluation suite ran in $\approx$1\,h per tokenizer on a
single L40S GPU. Across the $\approx$60 tokenizer/checkpoint
combinations evaluated during the project, this totalled $\approx$60 GPU-hours.

\paragraph{Cloud compute.}
The project used a rented cloud cluster of 8$\times$NVIDIA H100 GPUs
for approximately three weeks ($\approx$4{,}030 GPU-hours), used
across the same experiment classes above (primarily for the larger
population-tokenizer training runs and for batched evaluation
sweeps).

\paragraph{Preliminary and unreported runs.}
Beyond the reported runs above, the project consumed additional
compute for architecture and hyperparameter exploration, failed
training runs, and intermediate development. We estimate
$\approx$200 additional tokenizer training runs and $\approx$60
additional encoding runs that did not make it into the reported
results.

\paragraph{Total.}
Aggregating reported experiments across A100, L40S, and H100 GPUs,
total reported compute is $\approx$9{,}500 GPU-hours. Including
preliminary and failed runs, total project compute is
$\approx$15{,}000 GPU-hours, as estimated from W\&B run logs.

\begin{table}[h]
\centering
\small
\caption{\textbf{Compute footprint.} Approximate GPU-hours per
experiment class. Internal cluster: NVIDIA A100 (80\,GB) and L40S
(48\,GB). Cloud: 8$\times$NVIDIA H100. Cloud compute is reported as
a single line item because the same experiment classes ran on both
internal and cloud hardware.}
\label{tab:compute}
\begin{tabular}{lrrr}
\toprule
Experiment class & Per run (h) & Runs & Total (GPU-h) \\
\midrule
Tokenizer training (reported) & 50 & 100 & 5{,}000 \\
Downstream decoding & 1 & 40 & 40 \\
Decoding calibration & 12 & 3 & 36 \\
Downstream encoding (reported) & 12 & 30 & 360 \\
Evaluation pipeline & 1 & 60 & 60 \\
Cloud cluster (8$\times$H100, 3 weeks) & -- & -- & $\approx$4{,}030 \\
\midrule
Reported subtotal & & & $\approx$9{,}500 \\
\midrule
Tokenizer training (exploratory) & 25 & 200 & 5{,}000 \\
Encoding (exploratory) & 8 & 60 & 480 \\
\midrule
Exploratory subtotal & & & $\approx$5{,}500 \\
\midrule
Project total & & & $\approx$15{,}000 \\
\bottomrule
\end{tabular}
\end{table}

\FloatBarrier
\clearpage
% \section{Correlation of latent space quality with downstream metrics}

% Figure \ref{fig:appdx_latent_vs_downstream_heatmap}

\section{Metric-level analysis}
\label{app:metric-corr-analysis}
\subsection{Pairwise correlation structure, clustering and reduced metric set}
\label{sec:appdx_metric_pairwise}
In total, we evaluate 39 metrics, after averaging scores over the granularities we evaluate them at, such as ACG correlation when binning spikes at different bin sizes.  %including versions at different granularities, such as ACG correlation when binning spikes at different bin sizes. This naturally leads to high correlation between some of the metrics. 

To simplify screening during development, we group metrics that rank tokenizers almost identically, then choose one metric as representative of each group, as detailed below (Section~\ref{sec:appdx_metric_analysis_details}).

\subsubsection{Method details: Determining metric groups and representatives}
\label{sec:appdx_metric_analysis_details}
We compare each pair of metrics, computing absolute Spearman rank correlation $|\rho|$ across tokenizers (min. 32 tokenizers, median 42 of 47, across pairs).  Error metrics are sign-flipped so that higher values always mean better results.

\textbf{Robust metric correlations:} The 47 %49
tokenizers belong to 12 sub-families. %, plus two Poisson baselines. 
Tokenizers in the same sub-family share an architecture and differ only in settings, so we require that a high correlation survives resampling of sub-families. For each pair with $|\rho| > 0.90$, we draw the 12 sub-families with replacement 1,000 times, include all tokenizers of the drawn sub-families, and recompute $|\rho|$. The lower 5th percentile, $|\rho_{_{5\%}}|$ is then used to define that pair's distance as $1 - |\rho_{_{5\%}}|$; $95\%$ of resamples have a correlation of at least that value. Pairs with $|\rho| \le 0.90$ are assigned distance 1 and can never be grouped.

\begin{figure}[p]
    \centering
    \begin{minipage}{0.90\textwidth}
      \centering
      \textbf{A}\\
      \includegraphics[width=\linewidth]{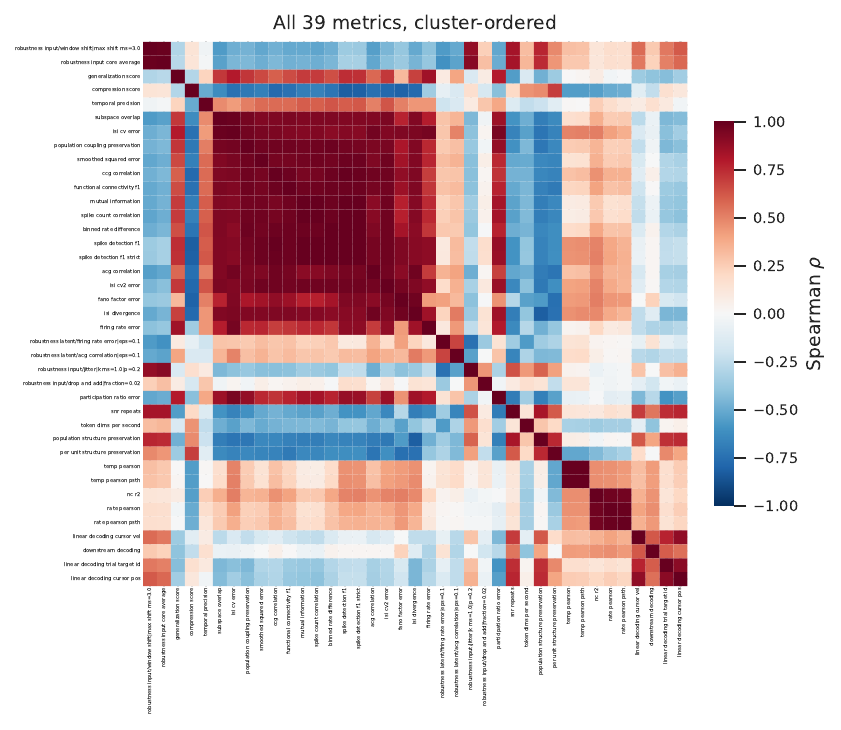}
    \end{minipage}

    \vspace{0.5em}
    \begin{minipage}{0.9\textwidth}
      \centering
      \textbf{B}\\
      \includegraphics[width=\linewidth]{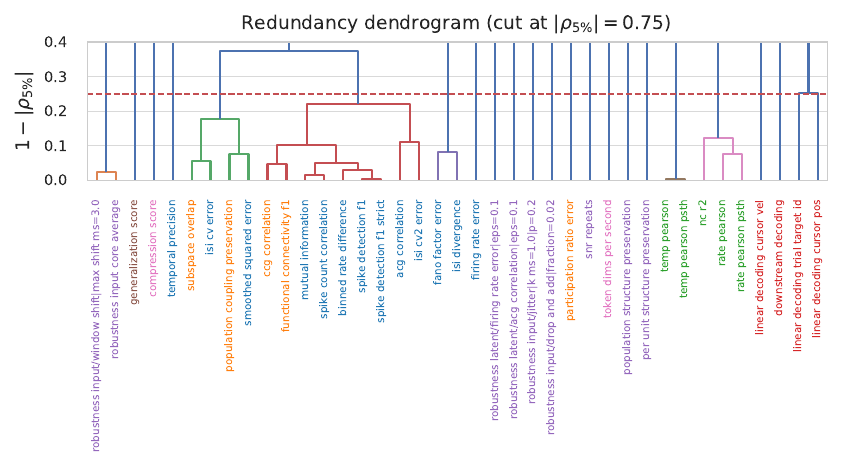}
    \end{minipage}
    \caption{\textbf{Metric-level correlation structure and redundancy groups. Best viewed with zoom.}
    \textbf{A} Spearman $\rho$ between all 39 metrics across tokenizers (pairwise complete;
    32--47 tokenizers per pair), with error metrics sign-flipped so that higher is better.
    Metrics are ordered as the leaves in B.
    \textbf{B} Hierarchical clustering (complete linkage) on the distance $1 - |\rho_{_{5\%}}|$, where $|\rho_{_{5\%}}|$ is the 5th percentile of $|\rho|$ over 1,000 resamples of the 12 tokenizer sub-families. Cutting at $|\rho_{_{5\%}}| = 0.75$ (dashed line) gives 23 groups; branches below the cut are colored by group. Pairs with $|\rho| \le 0.90$ were not resampled and have distance 1, so branches that leave the plot at the top are never grouped (the axis is shown up to 0.4). Label colors mark which of our 7 main categories a metric belongs to; they are independent of the branch colors.}
    \label{fig:metric-dendrogram}
  \end{figure}

\textbf{Grouping:} We use hierarchical clustering with complete linkage and distance $1 - |\rho_{_{5\%}}|$, iteratively merging the two clusters with minimum maximal distance between their elements. The dendrogram is shown in Fig.~\ref{fig:metric-dendrogram}. We cut the tree so that \emph{every} pair in a group has a lower bound $|\rho_{_{5\%}}|$ above $0.75$. This results in 23 groups, many with just one metric.

\textbf{Metric group representatives:} For each group, a metric for which we have scores for the largest number of tokenizers is selected as representative. Since there are usually multiple metrics in this position, we choose the one with highest average $|\rho|$ to the remaining metrics in the group. 

Table~\ref{tab:metric-clusters} shows each metric's group and the selected representative.

\subsubsection{Sensitivity analysis of metric groups}
\begin{figure}[t]
    \centering
    \includegraphics[width=0.6\textwidth]{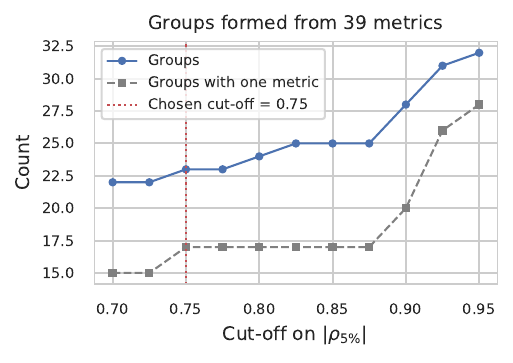}
    \caption{\textbf{Sensitivity of the metric groups to the cut-off.} Number of groups formed from the 39 metrics (solid) and number of groups with a single metric (dashed) for cut-offs on $|\rho_{_{5\%}}|$ from 0.70 to 0.95; the dotted line marks the chosen cut-off of 0.75.}
    \label{fig:metrics-floor-sensitivity}
  \end{figure}

Figure~\ref{fig:metrics-floor-sensitivity} shows how sensitive our grouping into 23 groups in Section~\ref{sec:appdx_metric_pairwise} is to varying the cut-off on the lower bound $|\rho_{_{5\%}}|$. 
 A stricter cut-off can only split groups, so the number of groups rises, from 22 at 0.70 to 32 at 0.95; it is the same (23) at %0.725, 
 0.75 and 0.775.

\subsubsection{Independent dimensions among metrics}
\begin{figure}[t]
  \centering
  \includegraphics[width=0.8\textwidth]{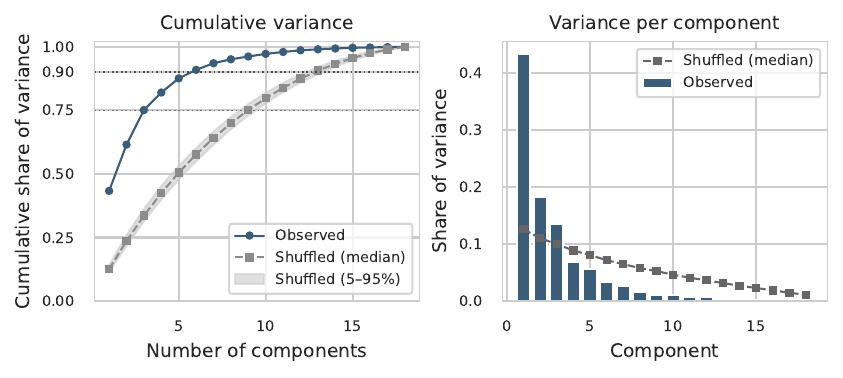}
  \caption{\textbf{Independent dimensions among metrics.} Cumulative variance explained by principal components of the rank-transformed scores of 18 of the 23 representative metrics (47 tokenizers), compared with scores shuffled across tokenizers separately per metric.}
  \label{fig:metrics-dimensionality}
\end{figure}

We use the 18 of the 23 representative metrics that have values for all 47 tokenizers. We apply principal component analysis to the scores after replacing each metric’s scores with their ranks across tokenizers. Ranks allow the analysis to use the same Spearman correlations as the analyses in the remainder of this section and prevent a tokenizer with extreme values from dominating a component.
It takes 2, 3 and 6 components to explain 50\%, 75\% and 90\% of the variance. After shuffling each metric’s scores across tokenizers separately, it takes 5, 9 and 13 components to explain those same amounts (Fig.~\ref{fig:metrics-dimensionality}). 
  
  \begin{figure}[t]
    \centering
    \includegraphics[width=0.9\textwidth]{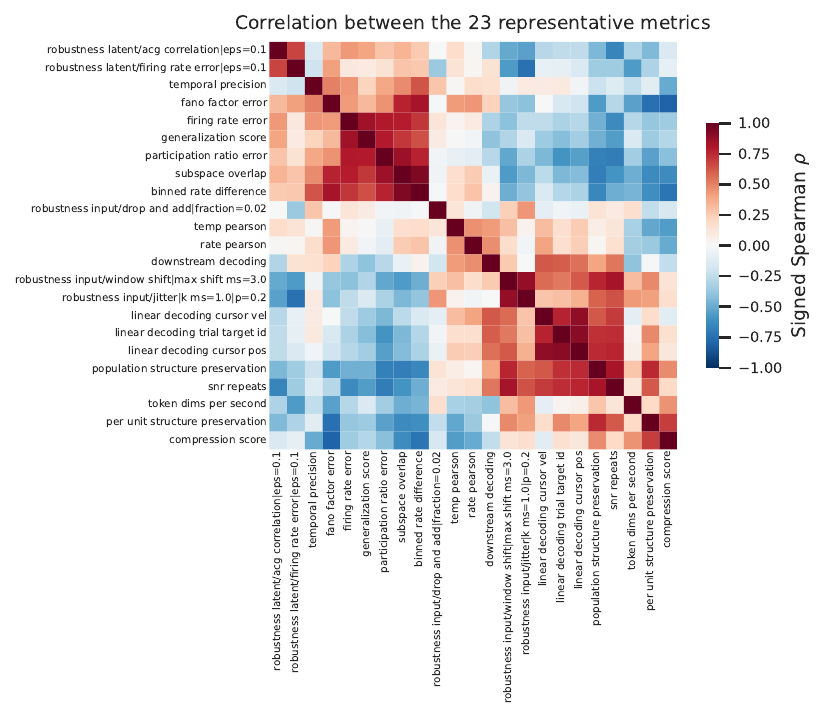}
    \caption{\textbf{Correlation between the 23 representative metrics.} Signed Spearman $\rho$,
    ordered as the leaves of the dendrogram in Fig.~\ref{fig:metrics-representatives-dendrogram}.}
    \label{fig:metrics-representatives-heatmap}
  \end{figure}

  % \begin{figure}[t]
  %   \centering
  %   \includegraphics[width=0.9\textwidth]{figures/metric_analysis/metrics_representatives_dendrogram.pdf}
  %   \caption{\textbf{Clustering of the 29 representative metrics.} Average linkage on $1-\rho$
  %   (signed Spearman); the dashed line marks the cut at $\rho = 0.75$ (height 0.25), and branches
  %   below it are coloured by cluster.}
  %   \label{fig:metrics-representatives-dendrogram}
  % \end{figure}

\subsection{Correlation structure of representatives}

To test whether the 23 representatives (Fig.~\ref{fig:metrics-representatives-heatmap}) form broader themes, we cluster them again using average linkage on $1-\rho$, where $\rho$ is the signed Spearman correlation across tokenizers, and cut the clusters at $\rho = 0.75$ (Fig.~\ref{fig:metrics-representatives-dendrogram}).  Unlike the grouping into representatives, which requires the resampled lower bound $|\rho_{_{5\%}}|$ to exceed 0.75, this cut uses the plain correlation $\rho$ and is therefore much looser. This gives 16 clusters; only 5 contain more than one metric. We repeat the clustering on 300 resamples of the tokenizer sub-families. %Only one multi-metric cluster is stable: every pair of its metrics is in the same cluster in at least 67\% of resamples. It contains spike count correlation, ACG correlation and population coupling; the weakest pair has $\rho = 0.93$. Thus, beyond near-duplicates, the representatives capture largely distinct aspects of the tokenizers.
Four of the five nontrivial clusters are fairly stable: on average, a pair of their metrics is grouped together in at least two-thirds of the resamples. %Those are: Firing rate error with generalization (which is based on reconstruction metrics, therefore some correlation with reconstruction metrics is expected). The three reconstruction metrics participation ratio error, subspace overlap and binned rate difference form another cluster, and two of the three input robustness metrics cluster (window shift and jitter), as do the three linear decoding metrics.

How well do these clusters map onto our predefined categories? This is shown in Table~\ref{tab:metrics-clusters-vs-categories}. Population and single neuron reconstruction metrics are partially correlated, leading to one cluster crossing these two categories. Firing rate error correlating with generalization indicates that reconstruction score generalization is correlated with plain reconstruction scores.
%the only other mixed cluster contains a latent quality metric. This is an unstable cluster, however. So generally, with our set of tokenizers, we cannot clearly separate population from single-neuron reconstruction. The other categories separate well.

\begin{figure}[t]
    \centering
    \includegraphics[width=0.8\textwidth]{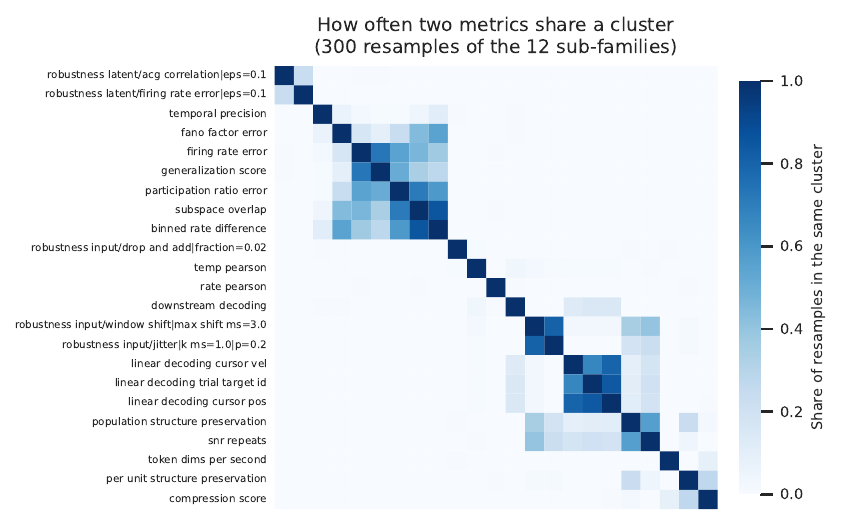}
    \caption{\textbf{How often two representative metrics share a cluster.} Each cell shows the share of 300 resamples of the 12 tokenizer sub-families where a   pair of metrics falls in the same cluster. For each resample, the clustering in Fig.~\ref{fig:metrics-representatives-dendrogram} is redone; metrics are ordered
  as its leaves. Dark blocks show groups that recur; pale areas show metrics without stable groups.}
    \label{fig:metrics-representatives-stability}
  \end{figure}

\begin{figure}[t]
    \centering
    \includegraphics[width=0.9\textwidth]{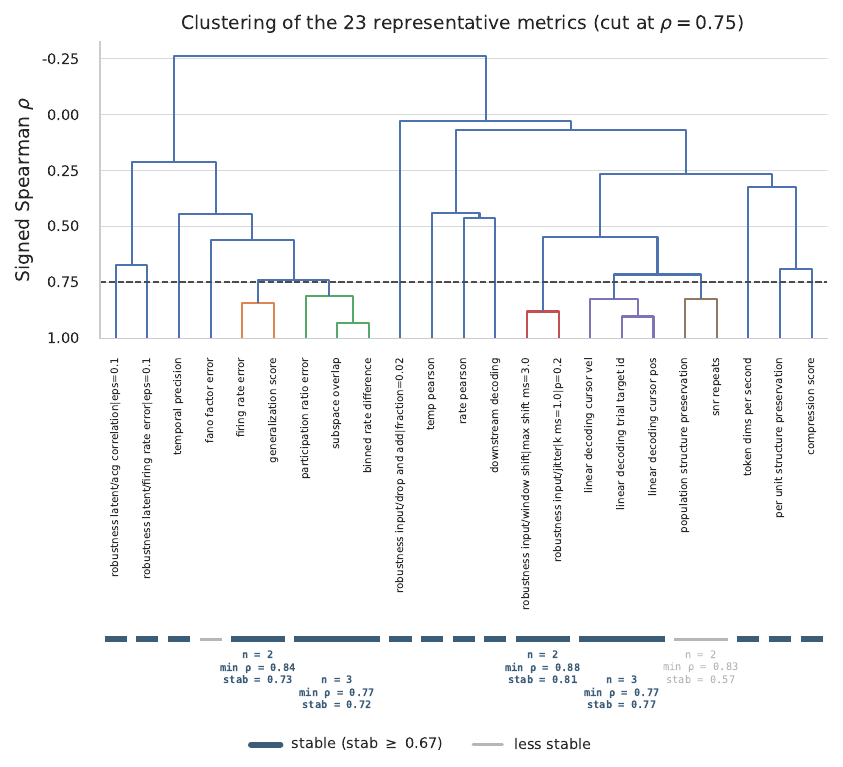}
    \caption{\textbf{Clustering of the 23 representative metrics.} Average linkage on $1-\rho$ (signed Spearman); the dashed line marks the cut at $\rho = 0.75$.  Bars under the leaves show the clusters made by the cut. For each multi-metric cluster: $n$ = number of metrics, min $\rho$ = weakest pair (``chained'' if below the cut), stab = average share of resamples in which a pair of its metrics falls in the same cluster (heatmap values in Fig.~\ref{fig:metrics-representatives-stability}). Bars are thick and dark when stab $\geq 0.67$.}
    \label{fig:metrics-representatives-dendrogram}
  \end{figure}

\subsection{Correlation within and between metric categories}

\begin{figure}[t]
      \centering
      \includegraphics[width=0.8\textwidth]{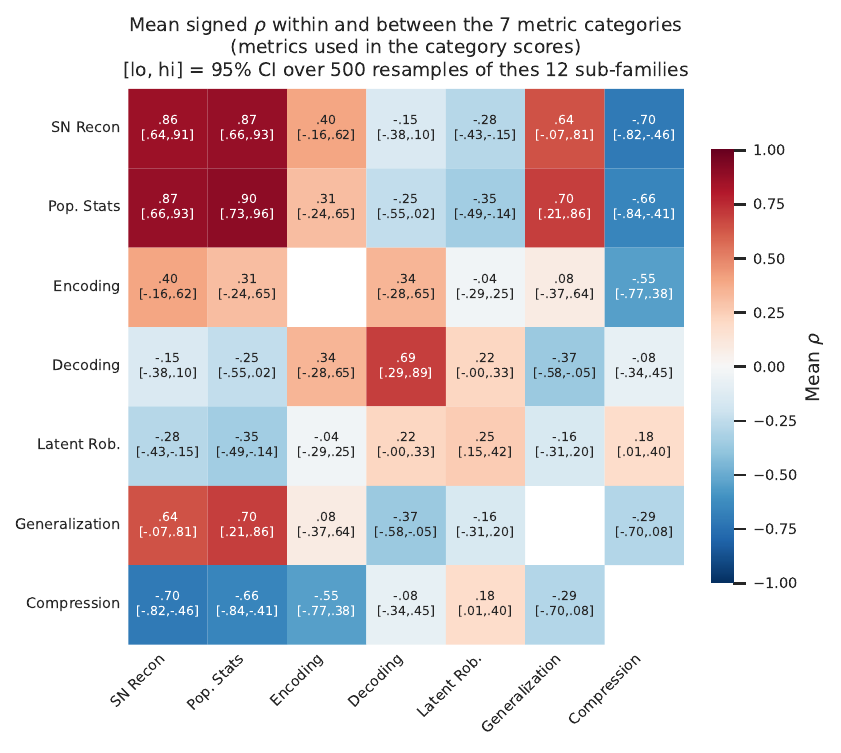}
      \caption{\textbf{Mean correlation within and between metric categories.}  Mean $\rho$ (Spearman) across metric pairs within one category (diagonal) or between two categories (off-diagonal), using the metrics included in the category scores. The three input-corruption metrics count as their mean in the latent structure and robustness score; in this figure, they remain three separate metrics in that category. Brackets show the 95\% interval for the mean across 500 resamples of the 12 tokenizer sub-families. Encoding, generalization and compression each use one metric, so they have no within-category value.}
      \label{fig:metrics-category-mean}
  \end{figure}

SN reconstruction metrics on average correlate similarly within themselves as with population statistics reconstruction metrics; population statistics and decoding metrics correlate most within themselves (Fig.~\ref{fig:metrics-category-mean}).
%Beyond that, metrics agree more with other metrics in their own category than with metrics in other categories when the category has several metrics (single-neuron reconstruction, population structure and decoding; Fig.~\ref{fig:metrics-category-mean}). This suggests that each  category groups metrics that measure related properties, while still measuring properties distinct from those in other categories. 
Latent structure and  robustness is the exception: their metrics agree only weakly with each other. This is by design: This category contains potential proxies for the remaining category scores -- we do not aim to have one coherent `latent space quality' metric, but instead to capture different latent space properties that might affect token quality as measured by the remaining metric categories.

\begin{figure}
      \centering
      \vspace{-10pt}  % trims space above
      \includegraphics[width=0.7\linewidth]{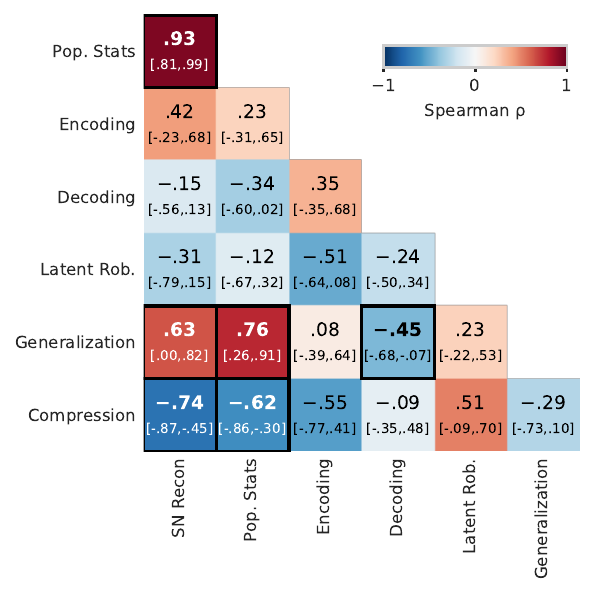}
      \caption{\textbf{Spearman rank correlation between main evaluation dimensions.} Brackets: 95\% CI from a cluster bootstrap over the 12 tokenizer sub-families; %and the Poisson baselines as one family; 
      bold: CI excludes 0 (uncorrected). With only 12 tokenizer sub-families, these correlations are preliminary and CIs are large.}
      \label{fig:metric-category-correlation}
      \vspace{-12pt}  % trims space below
\end{figure}

\newpage
\section{Code Availability}
\label{app:code-availability}

All code required to reproduce the experiments in this paper is released under two repositories:

\begin{itemize}
    \item \textbf{ntt} --- the tokenizer training framework, the 10-brainset benchmark config files, and the evaluation suite:\\
    \url{https://github.com/enigma-brain/ntt}
    
    \item \textbf{ntt-downstream} --- downstream evaluation of frozen \texttt{ntt} tokenizers on behavioral decoding (POYO) and stimulus$\rightarrow$response encoding (Cowley V4 image, Wingert auditory):\\
    \url{https://github.com/enigma-brain/ntt-downstream}
\end{itemize}

Both repositories use Hydra-driven CLIs on top of PyTorch Lightning. \texttt{ntt-downstream} installs \texttt{ntt} as a pinned git dependency (see \texttt{[tool.uv.sources]} in its \texttt{pyproject.toml}).

Pinned dependency versions, environment variables, and SLURM templates are documented in each repository's \texttt{docs/} directory.

Trained checkpoints of all tokenizers evaluated in this paper are provided at \url{https://huggingface.co/the-enigma-project/ntt-tokenizers}.

%%% Short table, metrics-clusters-vs-categories
\begin{table}[t]
\centering
\small
\begin{tabular}{rll}
\toprule
Cluster & Metric & Category \\
\midrule
1 & \textbf{subspace overlap} & Pop. Stats \\
 & binned rate difference & SN Recon \\
 & participation ratio error & Pop. Stats \\
\midrule
2 & \textbf{linear decoding cursor pos} & Decoding \\
 & linear decoding cursor vel & Decoding \\
 & linear decoding trial target id & Decoding \\
\midrule
3 & \textbf{firing rate error} & SN Recon \\
 & generalization score & Generalization \\
\midrule
4 & \textbf{robustness input/window shift\textbar{}max shift ms=3.0} & Latent Rob. \\
 & robustness input/jitter\textbar{}k ms=1.0\textbar{}p=0.2 & Latent Rob. \\
\midrule
5 & \textbf{population RSA} & Latent Rob. \\
 & snr repeats & Latent Rob. \\
\bottomrule
\end{tabular}
\caption{\textbf{Clusters of representative metrics and their categories.} The 5 clusters with more than one metric at the cut $\rho = 0.75$ (Fig.~\ref{fig:metrics-representatives-dendrogram}); the cluster's representative is in bold. Clusters 1 and 3 combine metrics from different categories.}
\label{tab:metrics-clusters-vs-categories}
\end{table}

%%%%% table
\newpage
\FloatBarrier

\begin{longtable}{llrr}
\caption{Metric redundancy pre-selection. Metrics are clustered on $1-|\rho|$ (Spearman, complete linkage) using the 5\% lower bound of a sub-family bootstrap rather than the point estimate: every pair inside a cluster satisfies $|\rho|>0.75$ in $95\%$ of 1000 resamples. 39 metrics reduce to 23 representatives (bold); $\rho_{\mathrm{rep}}$ is each metric's $|\rho|$ with its representative and $n$ its number of tokenizers.}\label{tab:metric-clusters}\\
\toprule
Cluster & Metric & $|\rho|$ w/ rep. & $n$ \\
\midrule
\endfirsthead
\toprule
Cluster & Metric & $|\rho|$ w/ rep. & $n$ \\
\midrule
\endhead
\multicolumn{4}{l}{\textit{Cluster 3 — 9 metrics, $\bar\rho=0.978\pm0.018$, $\rho^{-}_{\min}=0.78$}} \\
 & \textbf{binned rate difference} & — & 47 \\
 & \quad ccg correlation & 0.982 & 47 \\
 & \quad functional connectivity f1 & 0.984 & 47 \\
 & \quad acg correlation & 0.937 & 47 \\
 & \quad isi cv2 error & 0.972 & 35 \\
 & \quad mutual information & 0.978 & 47 \\
 & \quad spike count correlation & 0.981 & 47 \\
 & \quad spike detection f1 & 0.995 & 35 \\
 & \quad spike detection f1 strict & 0.996 & 35 \\
\midrule
\multicolumn{4}{l}{\textit{Cluster 2 — 4 metrics, $\bar\rho=0.957\pm0.029$, $\rho^{-}_{\min}=0.82$}} \\
 & \textbf{subspace overlap} & — & 47 \\
 & \quad population coupling preservation & 0.954 & 45 \\
 & \quad isi cv error & 0.987 & 35 \\
 & \quad smoothed squared error & 0.929 & 47 \\
\midrule
\multicolumn{4}{l}{\textit{Cluster 6 — 3 metrics, $\bar\rho=0.980\pm0.011$, $\rho^{-}_{\min}=0.88$}} \\
 & \textbf{rate pearson} & — & 42 \\
 & \quad nc r2 & 0.973 & 42 \\
 & \quad rate pearson psth & 0.988 & 42 \\
\midrule
\multicolumn{4}{l}{\textit{Cluster 1 — 2 metrics, $\bar\rho=0.992\pm0.000$, $\rho^{-}_{\min}=0.98$}} \\
 & \textbf{robustness input/window shift\textbar{}max shift ms=3.0} & — & 47 \\
 & \quad robustness input core average & 0.992 & 47 \\
\midrule
\multicolumn{4}{l}{\textit{Cluster 4 — 2 metrics, $\bar\rho=0.975\pm0.000$, $\rho^{-}_{\min}=0.92$}} \\
 & \textbf{fano factor error} & — & 47 \\
 & \quad isi divergence & 0.975 & 35 \\
\midrule
\multicolumn{4}{l}{\textit{Cluster 5 — 2 metrics, $\bar\rho=0.999\pm0.000$, $\rho^{-}_{\min}=1.00$}} \\
 & \textbf{temp pearson} & — & 42 \\
 & \quad temp pearson psth & 0.999 & 42 \\
\midrule
\multicolumn{4}{l}{\textit{Cluster 7 — 1 metric}} \\
 & \textbf{linear decoding trial target id} & — & 47 \\
\midrule
\multicolumn{4}{l}{\textit{Cluster 8 — 1 metric}} \\
 & \textbf{linear decoding cursor pos} & — & 47 \\
\midrule
\multicolumn{4}{l}{\textit{Cluster 9 — 1 metric}} \\
 & \textbf{downstream decoding} & — & 47 \\
\midrule
\multicolumn{4}{l}{\textit{Cluster 10 — 1 metric}} \\
 & \textbf{linear decoding cursor vel} & — & 47 \\
\midrule
\multicolumn{4}{l}{\textit{Cluster 11 — 1 metric}} \\
 & \textbf{single-neuron RSA} & — & 37 \\
\midrule
\multicolumn{4}{l}{\textit{Cluster 12 — 1 metric}} \\
 & \textbf{population RSA} & — & 47 \\
\midrule
\multicolumn{4}{l}{\textit{Cluster 13 — 1 metric}} \\
 & \textbf{token dims per second} & — & 47 \\
\midrule
\multicolumn{4}{l}{\textit{Cluster 14 — 1 metric}} \\
 & \textbf{snr repeats} & — & 42 \\
\midrule
\multicolumn{4}{l}{\textit{Cluster 15 — 1 metric}} \\
 & \textbf{participation ratio error} & — & 47 \\
\midrule
\multicolumn{4}{l}{\textit{Cluster 16 — 1 metric}} \\
 & \textbf{robustness input/drop and add\textbar{}fraction=0.02} & — & 47 \\
\midrule
\multicolumn{4}{l}{\textit{Cluster 17 — 1 metric}} \\
 & \textbf{robustness input/jitter\textbar{}k ms=1.0\textbar{}p=0.2} & — & 47 \\
\midrule
\multicolumn{4}{l}{\textit{Cluster 18 — 1 metric}} \\
 & \textbf{robustness latent/acg correlation\textbar{}eps=0.1} & — & 47 \\
\midrule
\multicolumn{4}{l}{\textit{Cluster 19 — 1 metric}} \\
 & \textbf{robustness latent/firing rate error\textbar{}eps=0.1} & — & 47 \\
\midrule
\multicolumn{4}{l}{\textit{Cluster 20 — 1 metric}} \\
 & \textbf{firing rate error} & — & 47 \\
\midrule
\multicolumn{4}{l}{\textit{Cluster 21 — 1 metric}} \\
 & \textbf{temporal precision} & — & 35 \\
\midrule
\multicolumn{4}{l}{\textit{Cluster 22 — 1 metric}} \\
 & \textbf{compression score} & — & 47 \\
\midrule
\multicolumn{4}{l}{\textit{Cluster 23 — 1 metric}} \\
 & \textbf{generalization score} & — & 47 \\
\bottomrule
\end{longtable}
%%% huge table end

\newpage
\section*{NeurIPS Paper Checklist}

\begin{enumerate}

\item {\bf Claims}
    \item[] Question: Do the main claims made in the abstract and introduction accurately reflect the paper's contributions and scope?
    \item[] Answer: \answerYes{} % Replace by \answerYes{}, \answerNo{}, or \answerNA{}.
    \item[] Justification: The abstract and introduction enumerate the paper's core contributions (defining spike tokenization as a modular component with explicit desiderata, introducing a multi-axis evaluation framework normalized between a Poisson floor and a perfect-reconstruction ceiling, collating a ten-dataset benchmark with nested generalization splits, and evaluating a zoo of continuous and discrete tokenizers within the framework) which are developed in the following section. The main empirical claim (the dissociation between reconstruction fidelity and downstream utility) is also clearly explored.

\item {\bf Limitations}
    \item[] Question: Does the paper discuss the limitations of the work performed by the authors?
    \item[] Answer: \answerYes{} % Replace by \answerYes{}, \answerNo{}, or \answerNA{}.
    \item[] Justification: The discussion section has a limitation paragraph covering key points to address in future work.

\item {\bf Theory assumptions and proofs}
    \item[] Question: For each theoretical result, does the paper provide the full set of assumptions and a complete (and correct) proof?
    \item[] Answer: \answerNA{} % Replace by \answerYes{}, \answerNo{}, or \answerNA{}.
    \item[] Justification: The paper does not include theoretical results.

    \item {\bf Experimental result reproducibility}
    \item[] Question: Does the paper fully disclose all the information needed to reproduce the main experimental results of the paper to the extent that it affects the main claims and/or conclusions of the paper (regardless of whether the code and data are provided or not)?
    \item[] Answer: \answerYes{} % Replace by \answerYes{}, \answerNo{}, or \answerNA{}.
    \item[] Justification: We extensively describe in the appendix the details of our experimental paradigms and metrics. We also release the code to run our metrics and experiments. 

\item {\bf Open access to data and code}
    \item[] Question: Does the paper provide open access to the data and code, with sufficient instructions to faithfully reproduce the main experimental results, as described in supplemental material?
    \item[] Answer: \answerYes{} % Replace by \answerYes{}, \answerNo{}, or \answerNA{}.
    \item[] Justification: The \texttt{ntt} repository (https://github.com/enigma-brain/ntt) contains the data preparation pipeline, the tokenizer training pipeline, and the evaluation suite; the downstream decoding and encoding code is released as \texttt{ntt-downstream} (Appendix~\ref{app:code-availability}). Code and trained tokenizer checkpoints are released under a permissive license. The tokenizer and metric interfaces are designed for community extension, with the goal of maintaining the benchmark as a living resource as new tokenizers and metrics are contributed.

\item {\bf Experimental setting/details}
    \item[] Question: Does the paper specify all the training and test details (e.g., data splits, hyperparameters, how they were chosen, type of optimizer) necessary to understand the results?
    \item[] Answer: \answerYes{} % Replace by \answerYes{}, \answerNo{}, or \answerNA{}.
    \item[] Justification: Training details, metric splits and computation are detailed in the appendix and available in the code.

\item {\bf Experiment statistical significance}
    \item[] Question: Does the paper report error bars suitably and correctly defined or other appropriate information about the statistical significance of the experiments?
    \item[] Answer: \answerYes{} % Replace by \answerYes{}, \answerNo{}, or \answerNA{}.
    \item[] Justification: Figures supporting the main empirical claims report variability explicitly. The variability factor is named explicitly when applicable. Headline summary tables report point-estimate aggregates to keep the multi-axis comparison readable; the per-session and per-seed values from which these aggregates are computed are produced by, and reproducible from, the released evaluation pipeline.

\item {\bf Experiments compute resources}
    \item[] Question: For each experiment, does the paper provide sufficient information on the computer resources (type of compute workers, memory, time of execution) needed to reproduce the experiments?
    \item[] Answer: \answerYes{} % Replace by \answerYes{}, \answerNo{}, or \answerNA{}.
    \item[] Justification: Per-experiment-class compute requirements (tokenizer training, downstream decoding and calibration protocols, downstream encoding, and the evaluation pipeline), GPU types (NVIDIA A100 80\,GB and L40S 48\,GB internally, and a rented cloud cluster of 8$\times$NVIDIA H100 GPUs), and an estimate of total project compute including preliminary and failed runs are reported in Appendix~\ref{app:compute}.
    
\item {\bf Code of ethics}
    \item[] Question: Does the research conducted in the paper conform, in every respect, with the NeurIPS Code of Ethics \url{https://neurips.cc/public/EthicsGuidelines}?
    \item[] Answer: \answerYes{} % Replace by \answerYes{}, \answerNo{}, or \answerNA{}.
    \item[] Justification: The research conforms with the NeurIPS Code of Ethics. Public neural-recording datasets are used under their original licenses and the institutional animal-care approvals of the original collecting groups (Section~\ref{sec:datasets}, Appendix~\ref{app:public-datasets}). The proprietary non-human mammalian dataset used in this work was collected at the authors' institution under approved animal-care protocols and the relevant government regulatory framework. The study involves no human-subjects research. Code, pretrained checkpoints, and the evaluation framework are released openly to support reproducibility and community extension.

\item {\bf Broader impacts}
    \item[] Question: Does the paper discuss both potential positive societal impacts and negative societal impacts of the work performed?
    \item[] Answer: \answerYes{} % Replace by \answerYes{}, \answerNo{}, or \answerNA{}.
    \item[] Justification: The paper introduces an evaluation framework for spike tokenizers, a methodological component upstream of any deployed system. Positive societal impacts include more reproducible and compute-efficient development of neural foundation models, with downstream applications in scientific neuroscience and clinical brain-computer interfaces such as motor neuroprosthetics. Brain decoding from invasive electrophysiology raises long-term privacy considerations, but the path from this work to consumer-facing privacy or surveillance concerns is indirect: the underlying recording technology requires surgically implanted electrodes and clinical infrastructure, and our contribution is an evaluation framework rather than a novel decoder, generative model, or recording method. Code, data preparation pipelines, and pretrained tokenizers are released openly under permissive licenses to maximize reproducibility and community use; we do not identify a direct misuse path that would warrant gated release.
    
\item {\bf Safeguards}
    \item[] Question: Does the paper describe safeguards that have been put in place for responsible release of data or models that have a high risk for misuse (e.g., pre-trained language models, image generators, or scraped datasets)?
    \item[] Answer: \answerNA{} % Replace by \answerYes{}, \answerNo{}, or \answerNA{}.
    \item[] Justification: We do not release language models, image generators, or scraped internet data, and the pretrained tokenizers have no meaningful function outside the specific scientific domain of invasive electrophysiology. Consistent with the broader-impacts statement, we do not identify a direct misuse path that would warrant gated release or usage restrictions.

\item {\bf Licenses for existing assets}
    \item[] Question: Are the creators or original owners of assets (e.g., code, data, models), used in the paper, properly credited and are the license and terms of use explicitly mentioned and properly respected?
    \item[] Answer: \answerYes{} % Replace by \answerYes{}, \answerNo{}, or \answerNA{}.
    \item[] Justification: All datasets, code, and pretrained components used in this work are credited by citation. Public neural recording datasets (Table~\ref{tab:datasets}) are referenced to their original publications and used under the licenses provided by the source repositories; detailed license information is available in the Croissant metadata file accompanying the submission. Code and methodological building blocks are cited at first use and used under their respective open-source licenses. The proprietary dataset is not released and is governed by the authors' institutional data-sharing policy.

\item {\bf New assets}
    \item[] Question: Are new assets introduced in the paper well documented and is the documentation provided alongside the assets?
    \item[] Answer: \answerYes{} % Replace by \answerYes{}, \answerNo{}, or \answerNA{}.
    \item[] Justification: The paper releases new assets: the \texttt{ntt} framework (data preparation pipeline for the ten-dataset benchmark, tokenizer training pipeline, and evaluation suite), the \texttt{ntt-downstream} code for the downstream decoding and encoding evaluations (Appendix~\ref{app:code-availability}), and trained tokenizer checkpoints. Each repository documents its usage and is released under a permissive license. Tokenizers trained on the public datasets can be reproduced from the released configurations. The proprietary dataset, and any tokenizer trained on it, are not part of the release.

\item {\bf Crowdsourcing and research with human subjects}
    \item[] Question: For crowdsourcing experiments and research with human subjects, does the paper include the full text of instructions given to participants and screenshots, if applicable, as well as details about compensation (if any)? 
    \item[] Answer: \answerNA{} % Replace by \answerYes{}, \answerNo{}, or \answerNA{}.
    \item[] Justification: No research with human subjects was conducted.

\item {\bf Institutional review board (IRB) approvals or equivalent for research with human subjects}
    \item[] Question: Does the paper describe potential risks incurred by study participants, whether such risks were disclosed to the subjects, and whether Institutional Review Board (IRB) approvals (or an equivalent approval/review based on the requirements of your country or institution) were obtained?
    \item[] Answer: \answerNA{} % Replace by \answerYes{}, \answerNo{}, or \answerNA{}.
    \item[] Justification: Not applicable.

\item {\bf Declaration of LLM usage}
    \item[] Question: Does the paper describe the usage of LLMs if it is an important, original, or non-standard component of the core methods in this research? Note that if the LLM is used only for writing, editing, or formatting purposes and does \emph{not} impact the core methodology, scientific rigor, or originality of the research, declaration is not required.
    %this research? 
    \item[] Answer: \answerNA{} % Replace by \answerYes{}, \answerNo{}, or \answerNA{}.
    \item[] Justification: LLMs are not an important or original component of the methods in this research.

\end{enumerate}

\end{document}